\documentclass[twocolumn,aps,nofootinbib]{revtex4}
\usepackage{graphicx}
\usepackage{epstopdf}
\usepackage{latexsym}
\usepackage{amssymb}
\usepackage{amsmath}
\usepackage{color}
\usepackage{float}
\usepackage{mathrsfs}
\usepackage{multirow}
\usepackage{tabularx,booktabs}
\usepackage{esvect}
\usepackage{ulem} 
\newcolumntype{Y}{>{\centering\arraybackslash}X}

\usepackage[center]{subfigure}
\usepackage{makecell}
\usepackage{ulem}
\usepackage[colorlinks=true]{hyperref} 
\usepackage{longtable}
\setcellgapes{4pt}
\begin{document}

  \renewcommand\arraystretch{2}
 \newcommand{\bq}{\begin{equation}}
 \newcommand{\eq}{\end{equation}}
 \newcommand{\bqn}{\begin{eqnarray}}
 \newcommand{\eqn}{\end{eqnarray}}
 \newcommand{\nb}{\nonumber}
 \newcommand{\lb}{\label}
 \newcommand{\cb}{\color{blue}}
    \newcommand{\cc}{\color{cyan}}
        \newcommand{\cm}{\color{magenta}}
\newcommand{\rc}{\rho^{\scriptscriptstyle{\mathrm{I}}}_c}
\newcommand{\rd}{\rho^{\scriptscriptstyle{\mathrm{II}}}_c} 
\newcommand{\PRL}{Phys. Rev. Lett.}
\newcommand{\PL}{Phys. Lett.}
\newcommand{\PR}{Phys. Rev.}
\newcommand{\CQG}{Class. Quantum Grav.}
\newcommand{\delete}[1] {\red{\sout{#1}}}

\newcommand{\ny}[1]{ {\bf{ny: #1}}}
\newcommand{\ky}[1]{\textcolor{magenta}{\bf{ky: #1}}}
\newcommand{\kent}[1]{\textcolor{magenta}{#1}}
\newcommand{\blue}[1]{ {#1}}

\title{Spherically symmetric static black holes in  Einstein-aether theory}

\author{Chao Zhang$^{1}$}
\email{C$\_$Zhang@baylor.edu}
\author{Xiang Zhao$^{1}$}
\email{Xiang$\_$Zhao@baylor.edu}
\author{Kai Lin$^{2, 3}$}
\email{lk314159@hotmail.com}
\author{Shaojun Zhang$^{4,5}$}
\email{sjzhang84@hotmail.com}
\author{Wen Zhao$^{6,7}$}
\email{wzhao7@ustc.edu.cn}
\author{Anzhong Wang$^{1}$ \footnote{Corresponding Author}}
\email{Anzhong$\_$Wang@baylor.edu}
 
\affiliation{$^1$ GCAP-CASPER, Physics Department, Baylor University, Waco, Texas, 76798-7316, USA}
\affiliation{$^2$ Hubei Subsurface Multi-scale Imaging Key Laboratory, Institute of Geophysics and Geomatics, 
	China University of Geosciences, Wuhan, Hubei, 430074, China}
\affiliation{$^3$ Escola de Engenharia de Lorena, Universidade de S\~ao Paulo, 12602-810, Lorena, SP, Brazil}
\affiliation{$^4$ {Institute for Theoretical Physics $\&$ Cosmology, Zhejiang University of Technology,} Hangzhou 310032, China}
\affiliation{$^5$ {United Center for Gravitational Wave Physics, Zhejiang University of Technology,} Hangzhou 310032, China}
\affiliation{$^6$ CAS Key Laboratory for Researches in Galaxies and Cosmology, Department of Astronomy, \\
	University of Science and Technology of China, Chinese Academy of Sciences, Hefei, Anhui 230026, China}
\affiliation{$^7$ School of Astronomy and Space Science, University of Science and Technology of China, Hefei 230026, China}

\date{\today}

\begin{abstract}

In this paper, we systematically study spherically symmetric static spacetimes in the framework of Einstein-aether  theory, and pay 
particular attention to the existence of black holes (BHs). In the theory, two additional gravitational modes (one scalar and one vector) 
appear, due to the presence of a timelike aether field. To avoid the vacuum gravi-\v{C}erenkov radiation, they must all propagate with 
speeds greater than or at least equal to the speed of light. In the spherical case, only the scalar mode is relevant, so BH horizons are 
defined by this mode, which are always inside or at most coincide with the metric (Killing) horizons. In the present studies we first clarify 
several subtle issues. In particular, we find that, out of the five non-trivial field equations,  only three are independent, so the problem is 
well-posed, as now generically there are only three unknown functions,   {$F(r), B(r), A(r)$, where $F$ and $B$ are metric
coefficients, and  $A$ describes the aether field.}  In addition,  the two second-order differential equations for $A$ and $F$ are independent 
of $B$, and once they are found, $B$ is given simply by an algebraic expression of $F,\; A$ and their derivatives.  To simplify the problem 
further, we explore the symmetry of field redefinitions, and work first with the redefined metric and aether field, and then obtain the physical 
ones by the inverse transformations. These clarifications significantly simplify the computational labor, which is important, as the problem is 
highly involved mathematically. In fact, it is exactly because of these, we find various numerical BH solutions with an accuracy that is at least 
two orders higher than previous ones. More important, these BH solutions are the only ones that satisfy the self-consistent conditions and 
meantime are consistent with all the observational constraints obtained so far. The locations of universal horizons  are also identified, together 
with several other observationally interesting quantities, such as the innermost stable circular orbits (ISCO), the ISCO frequency,  and the 
maximum redshift  $z_{max}$ of a photon emitted by a source orbiting the ISCO. All of these quantities are found to be quite  
{close to}  their relativistic limits.

 \end{abstract}


\maketitle
\section{Introduction}

\renewcommand{\theequation}{1.\arabic{equation}} \setcounter{equation}{0}

The detection of the first gravitational wave (GW) from the coalescence of two massive black holes (BHs) by  the advanced Laser Interferometer Gravitational-Wave Observatory (LIGO)  marked the beginning of a new era,  
the GW astronomy \cite{Ref1}. Following this observation, soon more than ten  GWs were  detected by the LIGO/Virgo scientific collaboration \cite{GWs,GWs19a,GWs19b}. More recently, 
about 50 GW  candidates  have been identified after LIGO/Virgo resumed operations on April 1, 2019, possibly including the coalescence of a neutron-star (NS)/BH binary. 
However, the details of these detections have not yet been released \cite{LIGO}. The outbreak of interest on GWs and BHs has further gained momentum after the detection of the 
shadow of the M87 BH  \cite{EHTa,EHTb,EHTc,EHTd,EHTe,EHTf}.

One of the remarkable observational results  is the discovery that the mass of  an individual BH in these binary systems can be  much larger than what was previously expected, 
both theoretically and observationally \cite{Ref4,Ref5,Ref6}, leading to the proposal and refinement of various formation scenarios \cite{Ref7,Ref8}. A consequence of this discovery 
is that the early inspiral phase may also be detectable by space-based observatories, such as as the Laser Interferometer Space Antenna (LISA)  \cite{PAS17}, TianQin \cite{TianQin}, Taiji \cite{Taiji}, 
and the Deci-Hertz Interferometer Gravitational wave Observatory (DECIGO) \cite{DECIGO}, 
for several years prior to their coalescence \cite{AS16, Moore15}. Such space-based detectors may be able to see many such systems, which will result in a variety of profound scientific
consequences. In particular, multiple observations with different detectors at different frequencies of signals from the same source can provide excellent opportunities to study the evolution 
of the binary in detail. Since different detectors observe at disjoint frequency bands, together they cover different evolutionary stages of the same binary system. Each stage of the evolution 
carries information about different physical aspects of the source.

As a result, multi-band GW detections will provide  an unprecedented opportunity to test  different theories of gravity in the strong field regime \cite{Ref17,Carson:2019rda,Carson:2019fxr,Carson:2019yxq,Gnocchi:2019jzp,Carson:2019kkh}. Massive systems will be observed by ground-based  detectors with high signal-to-noise ratios, after being tracked for years by space-based detectors in their inspiral phase. The two portions of signals can be combined to make precise tests for different theories of gravity. In particular, joint observations of binary black holes (BBHs) with a total mass larger than about $60$ solar masses by LIGO/Virgo and space-based detectors can potentially improve current bounds on dipole emission from BBHs by more than six orders of magnitude \cite{Ref17}, which will impose severe constraints on various theories of gravity \cite{Ref18}.

In recent works, some of the present authors generalized the post-Newtonian (PN) formalism to certain modified theories of gravity and applied it to the quasi-circular inspiral of compact binaries. In particular, we calculated in detail the waveforms, GW polarizations, response functions and energy losses due to gravitational radiation in Brans-Dicke (BD) theory \cite{Ref23},  and screened modified gravity (SMG) \cite{Tan18,Ref25,Ref25b} to the leading PN order, with which we then considered projected constraints from the third-generation detectors. Such studies have been further generalized to triple systems in Einstein-aether ($\ae$-) theory \cite{Kai19,Zhao19}. When applying such formulas to the first relativistic triple system discovered in 2014 \cite{Ransom14}, we studied the radiation power, and found that quadrupole emission has almost the same amplitude as that in general relativity (GR), but the dipole emission can be as large as the quadrupole emission. This can provide a promising window to place severe constraints on $\ae$-theory with multi-band GW observations \cite{Ref17,Carson:2019yxq}. 

More recently, we revisited the problem of a binary system of non-spinning bodies in a quasi-circular inspiral within the framework of $\ae$-theory \cite{Foster06,Foster07,Yagi13,Yagi14,HYY15,GHLP18}, and provided the explicit expressions for the time-domain and frequency-domain waveforms, GW polarizations, and response functions for both ground- and space-based detectors in the PN approximation \cite{Zhang20}. In particular, we found that, when going beyond  the leading order in the PN approximation, the non-Einsteinian polarization modes contain terms that depend on both the first and  second harmonics of the orbital phase. With this in mind, we  calculated analytically the corresponding parameterized post-Einsteinian parameters, generalizing the existing framework to allow for different propagation speeds among scalar, vector and tensor modes, without assuming the magnitude of its coupling parameters, and meanwhile allowing the binary system to have relative motions with respect to the aether field. Such results will particularly allow for the easy construction of Einstein-aether templates that could be used in Bayesian tests of GR in the future.

In this paper, we shall continuously work on GWs and BHs in the framework of  $\ae$-theory, but move to the ringdown phase, which consists  of the relaxation of the highly perturbed, newly formed merger remnant to its equilibrium state through the shedding of any perturbations in GWs as well as in matter waves. Such a remnant will typically be  a Kerr BH, provided that the binary system is massive enough and GR provides the correct description. This phase  can be well described as a sum of damped exponentials with unique frequencies and damping times - quasi-normal modes (QNMs) \cite{Berti18}.
 
 The information contained in QNMs provide the keys in revealing whether BHs are ubiquitous in our Universe, and more important whether  
 GR is the correct theory to describe the event even in the strong field regime. 
 In fact, in GR according to the no-hair theorem \cite{NHTs},  an isolated  and stationary BH is completely characterized by only three quantities, mass,  spin angular momentum and electric charge. 
 Astrophysically, we expect BHs to be neutral, so it must be described by
 the Kerr solution. Then,  the quasi-normal frequencies and damping times  will depend only on the mass and angular momentum of the final BH. Therefore, to extract the physics from
 the ringdown phase, at least two QNMs are needed. This will require the signal-to-noise  ratio (SNR) to be of the order 100 \cite{Berti07}. 
 Although such high SNRs are not achievable right now, it was shown that \cite{Berti16} they may be achievable once the advanced LIGO and Virgo reach their design sensitivities. In any case,  it is certain that 
they will be detected by the ground-based third-generation detectors,  such as Cosmic Explorer \cite{Dwyer15,BPA17}  or the Einstein Telescope \cite{MP10}, as well as the space-based detectors,  including
 LISA \cite{PAS17}, TianQin \cite{TianQin}, Taiji \cite{Taiji}, and DECIGO \cite{DECIGO}, as just mentioned  above.

In the framework of $\ae$-theory, BHs with rotations have not been found yet,   while   spherically symmetric BHs have been extensively
 studied in the past couple of years both analytically \cite{Eling2006-1, Oost2019, Per12, Dingq15, Ding16, Kai19b, Ding19, Gao2013, Chan2020, Leon2019, Leon2020,AA20} 
 and numerically \cite{Eling2006-2, Eling2007,Tamaki2008, BS11,Enrico11, Enrico2016, Zhu2019}. It was shown that they  {can also be formed }from gravitational collapse  
 \cite{Garfinkle2007}. Unfortunately,  in these studies, the parameter space has all been ruled out by current observations \cite{OMW18}.  
  Therefore, as a first step to the study of the  ringdown phase of a coalescing massive binary system, 
 in this paper we shall focus ourselves mainly on spherically symmetric static BHs in the parameter space that satisfies the self-consistent conditions and the current observations \cite{OMW18}. 
 As shown explicitly in \cite{Bhattacharjee2018},   spherically symmetric BHs in the new physically viable  phase space can be still formed  from the gravitational collapse of realistic matter.
 
 It should be noted that  the definition of BHs in $\ae$-theory is different from that given in GR. In particular,  in $\ae$-theory there are three gravitational modes, the scalar, vector and tensor, which will
 be referred to as the spin-0, spin-1 and spin-2 gravitons, respectively. Each of them moves in principle with a different speed, given, respectively, by \cite{JM04}
 \begin{eqnarray}
 \label{eq1.1}
 c_S^2 & = & \frac{c_{123}(2-c_{14})}{c_{14}(1-c_{13}) (2+c_{13} + 3c_2)}\,,\nonumber\\
 c_V^2 & = & \frac{2c_1 -c_{13} (2c_1-c_{13})}{2c_{14}(1-c_{13})}\,,\nonumber\\
 c_T^2 & = & \frac{1}{1-c_{13}},
\end{eqnarray}
where $c_i$'s   are the four dimensionless coupling constants of the theory, and $c_{ij} \equiv c_i + c_j,\; c_{ijk} \equiv c_i + c_j + c_k$.  The constants
$c_S,\; c_V$ and $c_T$ represent the speeds of the spin-0, spin-1 and spin-2 gravitons, respectively. In order to avoid the existence of the vacuum gravi-\v{C}erenkov radiation by matter
 such as cosmic rays \cite{EMS05}, we must require
 \bq
 \lb{eq1.2}
 c_S,\; c_V, c_T \ge c,
 \eq
 where $c$ denotes the speed of light. Therefore, as far as the gravitational sector is concerned, the horizon of a BH should be defined by the largest speed of the three different species of gravitons.
 However, in the spherically symmetric spacetimes, the spin-1 and spin-2 gravitons are not excited, and only the spin-0 graviton is relevant. Thus, the BH horizons in spherically symmetric spacetimes are defined by the metric  \cite{Jacobson},
 \bq
 \lb{eq1.3}
 g_{\mu\nu}^{(S)} \equiv g_{\mu\nu} -\left(c_S^2 -1\right)u_{\mu}u_{\nu}, 
 \eq
 where $u_{\mu}$ denotes the four-velocity of the aether field, which is always timelike and unity, $u_{\mu}u^{\mu} = -1$. 
 
 Because of the presence of the aether   in the whole spacetime,  it uniquely determines a preferred direction at each point of the spacetime.
 As a result, the Lorentz symmetry is locally violated in $\ae$-theory \cite{LZbreaking} \footnote{It should be noted that 
 the invariance under the Lorentz symmetry group is a cornerstone of modern physics and strongly supported by experiments and observations \cite{Liberati13}. 
Nevertheless, there are various  reasons to construct gravitational theories with broken Lorentz invariance (LI). For example, if  space and/or time at the Planck scale 
are/is discrete, as currently  understood  \cite{QGs}, Lorentz symmetry is absent at short distance/time scales and must be an emergent low energy symmetry. A 
concrete example of gravitational theories with broken LI is the Ho\v{r}ava theory of quantum gravity \cite{Horava}, 
 in which the LI  is broken via the anisotropic scaling between time and space in the ultraviolet (UV), $t \rightarrow b^{-z} t$, $x^{i} \rightarrow b^{-1} x^{i}, \; (i = 1, 2, ..., d)$, 
 where $z$ denotes the dynamical  critical exponent, and $d$ the spatial dimensions.  Power-counting renormalizability requires $z \ge d$ at short distances, while LI demands  $z = 1$. 
 For more details about Ho\v{r}ava gravity, see, for example, the review article \cite{Wang17}, and references therein.}.
 It must be emphasized  that   the breaking of Lorentz symmetry can have significant effects on the low-energy physics  through
the interactions between gravity and matter, no matter how high the scale of symmetry breaking is \cite{Collin04}, unless supersymmetry is invoked \cite{NP05}. 
In this paper, we shall not be concerned with this question. First, we consider $\ae$-theory as a low-energy effective theory, and second the constraints 
on the breaking of the Lorentz symmetry in the gravitational sector is much
weaker than that in the matter sector \cite{LZbreaking}. So, to avoid this problem, in this paper we simply assume that the matter sector still satisfies the Lorentz symmetry. Then, all the particles from the
 matter sector will travel with speeds less or equal to the speed of light. Therefore, for these particles, the Killing (or metric) horizons still serve as the boundaries. Once inside them, they will be trapped
 inside the metric horizons (MHs) forever, and never  be  able to escape to spatial  infinities. 

With the above in mind, in this paper we shall carry out a systematical study of spherically symmetric spacetimes in $\ae$-theory, clarify several subtle points, and then present numerically new BH
solutions that satisfy all the current observational constraints  \cite{OMW18}. In particular, we shall show that, among the five non-trivial field equations (three  {evolution} equations and two constraints), 
only three of them are independent. As a result, the system is well defined, since in the current case  there are only three unknown functions: two describe the spacetime, denoted by $F(r)$
 and $B(r)$  in Eq.(\ref{eq2.14}), and one describes the aether field, denoted by  $A(r)$ in Eq.(\ref{eq2.15}).

 An important result, born out of the above observations, is that the three independent equations can be divided into two groups, which decouple one from the other, that is, the equations for the two functions
 $A(r)$ and $F(r)$ [cf. Eqs.(\ref{eq2.22a}) and (\ref{eq2.22b})] are independent of the function $B(r)$. Therefore, to solve these three field equations, one can first solve Eqs.(\ref{eq2.22a}) and (\ref{eq2.22b})
 for  $A(r)$ and $F(r)$. Once they are found, one can obtain $B(r)$ from the third equation. It is even more remarkable, if the third equation is chosen to be the constraint $C^v = 0$, given by
 Eq.(\ref{eq2.23a}), from which one finds that  $B(r)$ is then directly given by the algebraic equation (\ref{Bcv}) without the need of any further integration. Considering the fact that the field equations are in general 
highly involved mathematically, as it can be seen from Eqs.(\ref{eq2.22a})-(\ref{eq2.23a}) and Eqs.(\ref{fns})-(\ref{nns}), this  is important, as it shall significantly simplify the computational labor, when we try to solve  these field equations. 
 
Another important step of solving the field equations  is Foster's discovery of the symmetry of the action, the so-called field redefinitions \cite{Foster05}: the action remains invariant under  the replacements,
\bq
 \lb{eq1.4}
 \left(g_{\mu\nu}, u^{\mu}, c_i\right) \rightarrow \left(\hat g_{\mu\nu}, \hat u^{\mu}, \hat c_i\right), 
 \eq
 where $\hat g_{\mu\nu}$, $\hat u^{\mu}$ and $\hat c_i$ are given by Eqs.(\ref{eq2.2}) and (\ref{eq2.3}) through the introduction of a free parameter $\sigma$. Taking the advantage of the arbitrariness of
 $\sigma$, we can choose it as $\sigma = c_S^2$, where $c_S^2$ is given by Eq.(\ref{eq1.1}).
  Then, the spin-0 and metric horizons for the metric
 $\hat g_{\mu\nu}$ coincide \cite{Eling2006-2,Tamaki2008,Enrico11}. Thus, instead of solving the field equations for  $(g_{\mu\nu}, u^{\mu})$, we first solve the ones for  $\left(\hat g_{\mu\nu}, \hat u^{\mu}\right)$,
  as in the latter the 
 corresponding    initial value problem can be easily imposed at  horizons. Once  $\left(\hat g_{\mu\nu}, \hat u^{\mu}\right)$ is found, using the inverse transformations, we can easily obtain 
 $(g_{\mu\nu}, u^{\mu})$.
 
With the above observations, we are able to solve numerically the field equations with very high accuracy, as to be shown below [cf. Table \ref{table1}]. In fact, the accuracy is significantly improved and
 in general at least two orders higher than the previous  works.
 
In theories with breaking Lorentz symmetry, another important quantity is the universal horizon (UH) \cite{BS11,Enrico11}, which is the causal boundary even for particles with infinitely large speeds.
 The thermodynamics of UHs and relevant physics have been extensively studied since then (see, for example, Section III of the  review article \cite{Wang17}, and references therein). In particular, 
 it was shown that such horizons can be formed from gravitational collapse of a massless scalar field  \cite{Bhattacharjee2018}. In this paper, 
 we shall also identify the locations of the UHs of our numerical new BH solutions. 
 
The rest of the paper is organized as follows:  Sec. II provides a brief review to $\ae$-theory, in which the introduction of the field redefinitions, the current observational constraints  on the 
four dimensionless coupling constants $c_i$'s of the theory, and the definition of the spin-0 horizons  (S0Hs) are given. 

In Sec. III, we systematically study spherically symmetric static spacetimes, and
show explicitly that among the five non-trivial field equations,  only three of them are independent, so the corresponding problem is 
well defined: three independent equations  for three unknown functions.  {Then, from these three independent equations we are able to obtain a three-parameter family of exact solutions for the special case 
$c_{13} = c_{14} = 0$, which depends in general on the coupling constant $c_2$. However,   requiring that the solutions be asymptotically flat makes the solutions independent of $c_2$, and the metric reduces
precisely to the Schwarzschild BH solution with a non-trivially coupling aether field [cf. Eq.(\ref{eqA6})], which is timelike over the whole spacetime, including the region inside the BH. }
To further simplify the problem, in this section 
we also explore the advantage of the field redefinitions \cite{Foster05}.  
 In particular,   we show step by step how to choose 
the initial values of the differential equations Eqs.(\ref{eq2.30a}) and (\ref{eq2.30b}) on S0Hs, and how to reduce the phase space from four dimensions, spanned by $(\tilde F_H, \; \tilde F'_H,\; \tilde A_H,\; \tilde A'_H$), to one dimension, spanned only by $\tilde A_H$.  So, finally the problem reduces to finding the values of $\tilde A_H$ that lead to asymptotically flat solutions  of the form (\ref{eq2.38}) \cite{Eling2006-2,Enrico11}.

 In Sec. IV, we spell out in detail the steps to carry out our numerical analysis. In particular,  as we show explicitly, Eq.(\ref{eq2.30c}) is not independent from other three differential equations. 
 Taking this advantage, we use it to monitor 
 our numerical errors [cf. Eq.(\ref{scC})]. To check our numerical code further, we reproduce the BH solutions obtained in  \cite{Eling2006-2,Enrico11}, but with an accuracy two orders higher than those obtained
in   \cite{Enrico11} [cf. Table I]. Unfortunately, all these BH solutions have been ruled out by the current observations \cite{OMW18}. So, in Sec. IV.B we consider cases that satisfy 
all the observational  {constraints and obtain} various new static BH solutions.

 Then,  in Sec. V, we present   the  physical metric and $\ae$-field for these viable new BH solutions, by using the inverse transformations from the effective fields to the physical ones. 
 In this section, we also show explicitly that  the physical fields, ${g}_{\mu\nu}$ and ${u}^{\mu}$,  are also asymptotically flat, provided that the effective fields $\tilde g_{\mu\nu}$ and $\tilde u^{\mu}$ are,
 which are related to $\hat g_{\mu\nu}$ and $\hat u^{\mu}$ via the coordinate transformations given by  Eq.(\ref{eq2.18}).
Then, we calculate explicitly the locations of the metric, spin-0 and universal horizons, as well as the locations of the  innermost stable circular orbits (ISCO),  the Lorentz gamma factor, 
 the gravitational radius, the orbital frequency  of the ISCO, the maximum redshift  of a photon emitted by a source orbiting the ISCO (measured at the infinity), the radii  of the circular photon orbit, and the impact parameter   of the circular photon orbit. All of them are given in Table \ref{table4}-\ref{table5}. In Table \ref{table6} we also calculate the differences of these quantities obtained in $\ae$-theory and GR. From  these results, we find  that the differences are very small, and  it is very hard to distinguish GR and $\ae$-theory through these quantities, as far as the cases considered in this paper
 are concerned.

Finally,  in Sec. VI we summarize  our main results and present some concluding remarks. There is also an appendix,  in which the coefficients of the  field equations  for both  ($g_{\mu\nu}, u^{\mu}$) and 
  ($\tilde g_{\mu\nu}, \tilde u^{\mu}$) are given.


\section{$\ae$-theory}
 \renewcommand{\theequation}{2.\arabic{equation}} \setcounter{equation}{0}

In $\ae$-theory, the fundamental variables of the gravitational  sector are \cite{JM01},
\bq
\lb{2.0a}
\left(g_{\mu\nu}, u^{\mu}, \lambda\right),
\eq
with the Greek indices $\mu,\nu = 0, 1, 2, 3$, and $g_{\mu\nu}$ is  the four-dimensional metric  of the {spacetime}
with the  {signature} $(-, +,+,+)$ \cite{Foster06,Garfinkle2007},  $u^{\mu}$  {is} the aether four-velocity, as mentioned above, 
and $\lambda$ is a Lagrangian multiplier, which guarantees that the aether  four-velocity  is always timelike and unity.
In this paper, we also adopt units so that  the   speed of light is one ($c=1$).
Then, the general action of the theory  is given  by  \cite{Jacobson},
\bq
\lb{2.0}
S = S_{\ae} + S_{m},
\eq
where  $S_{m}$ denotes the action of matter,  and $S_{\ae}$  the gravitational action of the $\ae$-theory, given, respectively, by
\bqn
\lb{2.1}
 S_{\ae} &=& \frac{1}{16\pi G_{\ae} }\int{\sqrt{- g} \; d^4x \Big[  {\cal{L}}_{\ae}\left(g_{\mu\nu}, u^{\alpha}, c_i\right)}\nb\\
 && ~~~~~~~~~~~~ + {\cal{L}}_{\lambda}\left(g_{\mu\nu}, u^{\alpha}, \lambda\right)\Big],\nb\\
S_{m} &=& \int{\sqrt{- g} \; d^4x \Big[{\cal{L}}_{m}\left(g_{\mu\nu}, u^{\alpha}; \psi\right)\Big]}.
\eqn
Here
$\psi$ collectively denotes the matter fields, $R$    and $g$ are, respectively, the  Ricci scalar and determinant of $g_{\mu\nu}$, and
\bqn
\lb{2.2}
 {\cal{L}}_{\lambda}  &\equiv&  \lambda \left(g_{\alpha\beta} u^{\alpha}u^{\beta} + 1\right),\nb\\
 {\cal{L}}_{\ae}  &\equiv& R(g_{\mu\nu}) - M^{\alpha\beta}_{~~~~\mu\nu}\left(D_{\alpha}u^{\mu}\right) \left(D_{\beta}u^{\nu}\right),
\eqn
 where $D_{\mu}$ denotes the covariant derivative with respect to $g_{\mu\nu}$, and  $M^{\alpha\beta}_{~~~~\mu\nu}$ is defined as
\bqn
\lb{2.3}
M^{\alpha\beta}_{~~~~\mu\nu} \equiv c_1 g^{\alpha\beta} g_{\mu\nu} + c_2 \delta^{\alpha}_{\mu}\delta^{\beta}_{\nu} +  c_3 \delta^{\alpha}_{\nu}\delta^{\beta}_{\mu} - c_4 u^{\alpha}u^{\beta} g_{\mu\nu}.\nb\\
\eqn
Note that here we assume that matter fields couple not only to $g_{\mu\nu}$ but also to the aether field $u^{\mu}$. However, in order to satisfy the severe observational constraints,    such a coupling in general is assumed to be absent  \cite{Jacobson}.

The four coupling constants $c_i$'s are all dimensionless, and $G_{\ae} $ is related to  the Newtonian constant $G_{N}$ via the relation \cite{CL04},
\bq
\lb{2.3a}
G_{N} = \frac{G_{\ae}}{1 - \frac{1}{2}c_{14}}.
\eq
 
The variations of the total action with respect to $g_{\mu\nu}$,  $u^{\mu}$   and $\lambda$ yield, respectively, the field equations,
 \bqn
 \lb{2.4a}
 R^{\mu\nu} - \frac{1}{2} g^{\mu\nu}R - S^{\mu\nu} &=& 8\pi G_{\ae}  T^{\mu\nu},\\
 \lb{2.4b}
  \AE_{\mu} &=& 8\pi G_{\ae}  T_{\mu}, \\
   \lb{2.4c}
  g_{\alpha\beta} u^{\alpha}u^{\beta} &=& -1, 
 \eqn
where   $R^{\mu \nu}$ denotes the Ricci tensor, and
 \bqn
 \lb{2.5}
  S_{\alpha\beta} &\equiv&
  D_{\mu}\Big[J^{\mu}_{\;\;\;(\alpha}u_{\beta)} + J_{(\alpha\beta)}u^{\mu}-u_{(\beta}J_{\alpha)}^{\;\;\;\mu}\Big]\nb\\
&& + c_1\Big[\left(D_{\alpha}u_{\mu}\right)\left(D_{\beta}u^{\mu}\right) - \left(D_{\mu}u_{\alpha}\right)\left(D^{\mu}u_{\beta}\right)\Big]\nb\\
&& + c_4 a_{\alpha}a_{\beta}    + \lambda  u_{\alpha}u_{\beta} - \frac{1}{2}  g_{\alpha\beta} J^{\delta}_{\;\;\sigma} D_{\delta}u^{\sigma},\nb\\
 \AE_{\mu} & \equiv &
 D_{\alpha} J^{\alpha}_{\;\;\;\mu} + c_4 a_{\alpha} D_{\mu}u^{\alpha} + \lambda u_{\mu},\nb\\
  T^{\mu\nu} &\equiv&  \frac{2}{\sqrt{-g}}\frac{\delta \left(\sqrt{-g} {\cal{L}}_{m}\right)}{\delta g_{\mu\nu}},\nb\\
T_{\mu} &\equiv& - \frac{1}{\sqrt{-g}}\frac{\delta \left(\sqrt{-g} {\cal{L}}_{m}\right)}{\delta u^{\mu}},
 \eqn
 with
\begin{equation}
 \lb{2.6}
J^{\alpha}_{\;\;\;\mu} \equiv M^{\alpha\beta}_{~~~~\mu\nu}D_{\beta}u^{\nu}\,,\quad
a^{\mu} \equiv u^{\alpha}D_{\alpha}u^{\mu}.
\end{equation}
From Eq.(\ref{2.4b}),  we find that
\bq
\lb{2.7}
\lambda = u_{\beta}D_{\alpha}J^{\alpha\beta} + c_4 a^2 - 8\pi G_{\ae}  T_{\alpha}u^{\alpha},
\eq
where $a^{2}\equiv a_{\lambda}a^{\lambda}$.

 It is easy to show that the Minkowski spacetime  is a solution of  $\ae$-theory, in which the aether is aligned along the time direction, $\bar{u}_{\mu} = \delta^{0}_{\mu}$. 
 Then, the linear perturbations around the Minkowski background show that the theory in general possess three types of excitations, scalar  (spin-0), vector (spin-1) and tensor (spin-2)
  modes  \cite{JM04}, with their squared  speeds given by Eq.(\ref{eq1.1}).

In addition, among the 10 parameterized post-Newtonian (PPN) parameters \cite{Will06, Will2018},  in $\ae$-theory the only two parameters 
that deviate from  GR are $\alpha_1$ and $\alpha_2$, which measure the preferred frame effects. In terms of the four dimensionless coupling 
constants $c_i$'s of the $\ae$-theory, they are given by \cite{FJ06},
\bqn
\lb{2.3aa}
\alpha_1 &=& -  \frac{8\left(c_1c_{14} - c_{-}c_{13}\right)}{2c_1 - c_{-}c_{13}}, \nb\\
\alpha_2 &=&   \frac{1}{2}\alpha_1  + \frac{\left(c_{14}- 2c_{13}\right)\left(3c_2+c_{13}+c_{14}\right)}{c_{123}(2-c_{14})},~~~~~~~~
\eqn
where $c_{-} \equiv c_1 - c_3$.   In the weak-field regime, using lunar laser ranging and solar alignment with the ecliptic, 
Solar System observations constrain these parameters to very small values \cite{Will06},
\bq
\lb{CD5}
\left| \alpha_1\right| \le 10^{-4}, \quad 
 \left|\alpha_2\right| \le 10^{-7}.
 \eq

Recently,   the combination of the  GW  event GW170817 \cite{GW170817}, observed by the LIGO/Virgo collaboration, and the event of the gamma-ray burst
GRB 170817A \cite{GRB170817} provides  a remarkably stringent constraint on the speed of the spin-2 mode, 
$- 3\times 10^{-15} < c_T -1 < 7\times 10^{-16}$,
which, together with Eq.(\ref{eq1.1}), implies that 
\bq
\lb{2.8a}
\left |c_{13}\right| < 10^{-15}.
 \eq
 
Requiring that the theory:  (a)  be self-consistent, such as free of ghosts and instability; and (b) satisfy all the observational constraints obtained so far, 
 it was found that   the parameter space of the theory is considerably restricted \cite{OMW18}. In particular,    $c_{14}$ and $c_2$  are restricted to
\bqn
\lb{c1234a}
&& 0 \lesssim c_{14} \lesssim 2.5 \times 10^{-5},\\
\lb{c1234ab}
&&  0 \lesssim c_{14} \lesssim c_2 \lesssim  0.095.
\eqn

The constraints on other parameters depend on the values of $c_{14}$. If dividing the above range into three intervals: (i) $ 0 \lesssim c_{14}\leq 2\times 10^{-7}
$;  (ii)  $2\times 10^{-7}< c_{14}\lesssim 2\times 10^{-6}$;  and (iii) $ 2\times 10^{-6}\lesssim c_{14}\lesssim 2.5\times 10^{-5}$, in the first and last intervals, one finds
  \cite{OMW18}, 
\bqn
\lb{c1234}
&&  \mbox{(i)} \;\;\;\;
  0 \lesssim c_{14}\leq 2\times 10^{-7}, \nb\\
&&   ~~~~~~~ c_{14} \lesssim c_2 \lesssim  0.095, \\
   \lb{c1234b}
&&  \mbox{(iii)}\;\; 
  2\times 10^{-6}\lesssim c_{14}\lesssim 2.5\times 10^{-5}, \nonumber\\
 & & ~~~~~~ 0 \lesssim c_2-c_{14} \lesssim 2\times 10^{-7}.
\eqn
In the intermediate regime  (ii) $ \; 2\times 10^{-7}< c_{14}\lesssim 2\times 10^{-6}$, in addition to the ones given by Eqs.(\ref{c1234a}) and (\ref{c1234ab}), 
the following constraints must be also satisfied,
\bq
\lb{c1234bb}
- 10^{-7} \le  \frac{c_{14}\left(c_{14} + 2c_2c_{14} - c_2\right)}{c_2\left(2-c_{14}\right)} \le 10^{-7}. 
\eq
Note that in writing Eq.(\ref{c1234bb}), we had set $c_{13} = 0$, for which the errors are of the order ${\cal{O}}\left(c_{13}\right) \simeq 10^{-15}$, which can be safely neglected for the current 
and forthcoming experiments. The  results in this intermediate interval of $c_{14}$ were  shown explicitly  by Fig.~1 in \cite{OMW18}. Note that in this figure, the physically valid region is restricted only  to
the half plane $c_{14} \ge 0$, as shown by Eq.(\ref{c1234a}).

Since the theory possesses three different modes, and all of them are moving in different speeds, in general these different modes define different horizons   \cite{Jacobson}.  
These horizons are the null surfaces of the effective metrics,
\bq
\lb{EHM}
g_{\alpha \beta}^{(A)}  \equiv g_{\alpha \beta} - \left(c_{A}^2 - 1\right) u_{\alpha}u_{\beta},
\eq
where $A = S, V, T$. If a BH is defined to be a region that traps all possible causal influences, it must be bounded by a horizon corresponding to the fastest speed. Assuming that the matter sector always satisfies the Lorentz symmetry, we can see that in the matter sector  the fastest speed  will be the speed of light. Then,  overall,  the fastest speed must be one of the three gravitational  modes. 

However, in the spherically symmetric case, the spin-1 and spin-2 modes are not excited, so only the spin-0 gravitons are relevant. Therefore,  in the present paper the relevant horizons for the gravitational sector are  the S0Hs \footnote{If we consider  Ho\v{r}ava gravity \cite{Horava} as the UV complete theory of  {the hypersurface-orthogonal  $\ae$-theory (the khronometric theory) \cite{BPSa,BPSb,Jacobson10,Jacobson14},} even in the gravitational sector, the relevant boundaries will be the UHs, once such a UV complete theory is taken into account \cite{Wang17}.}. In order to avoid the existence of the vacuum gravi-\v{C}erenkov radiation by matter such as cosmic rays \cite{EMS05}, we assume that $c_S \ge 1$, so that S0Hs are always inside or at most coincide with the metric horizons, the null surfaces defined by the metric $g_{\alpha \beta}$.  The equality  happens only when $c_S =  1$.  

\subsection{Field Redefinitions}

Due to the specific symmetry of the theory, Foster found that the action  ${{S}}_{\ae}\left(g_{\alpha\beta}, u^{\alpha}, c_i\right)$
 given by Eqs.(\ref{2.1})-(\ref{2.3}) does not change under the following field 
redefinitions   \cite{Foster05}, 
\bq
\lb{eq2.1}
\left(g_{\alpha\beta}, u^{\alpha}, c_i\right) \rightarrow \left(\hat{g}_{\alpha\beta}, \hat{u}^{\alpha}, \hat{c}_i\right),
\eq
where 
\bqn
\lb{eq2.2}
\hat{g}_{\alpha \beta}&=& g_{\alpha \beta}- (\sigma-1) u_{\alpha} u_{\beta},\; \;\; \hat{u}^{ \alpha} = \frac{1}{\sqrt{\sigma}} u^{\alpha}, \nb\\
\hat{g}^{\alpha \beta}&=& g^{\alpha \beta}-(\sigma^{-1}-1) u^{\alpha} u^{\beta},\; \;\;  \hat{u}_{ \alpha} = \sqrt{\sigma} u_{\alpha}, ~~~
\eqn
and 
\bqn
\lb{eq2.3}
 \hat{c}_{1} &=& \frac{\sigma}{2}\left[\left(1+\sigma^{-2}\right) c_{1}+\left(1-\sigma^{-2}\right) c_{3}-\left(1-\sigma^{-1}\right)^{2}\right], \nb\\
\hat{c}_{2} &=&\sigma\left(c_{2}+1-\sigma^{-1}\right), \nb\\
\hat{c}_{3} &=& \frac{\sigma}{2}\left[\left(1-\sigma^{-2}\right) c_{1}+\left(1+\sigma^{-2}\right) c_{3}-\left(1-\sigma^{-2}\right)\right], \nb\\
\hat{c}_{4} &=& c_{4} -\frac{\sigma}{2} \Bigg[\left(1-\frac{1}{\sigma}\right)^{2} c_{1}+\left(1-\frac{1}{\sigma^2}\right) c_{3} \nb\\
&& ~~~~~~~~~~~~ -\left(1-\frac{1}{\sigma}\right)^{2}\Bigg],
\eqn
 with $\sigma$ being a positive otherwise arbitrary constant.  Then,  the following useful relations between $c_i$ and $\hat c_i$ hold,
\bqn
\lb{eq2.8}
&&\hat c_2 = \sigma (c_2 + 1) -1, \quad \hat c_{14}  = c_{14}, \nb\\
&&   \hat c_{13}  = \sigma (c_{13}  - 1) + 1,  \quad \hat c_{123} = \sigma c_{123}, \nb\\
&& \hat c_- = \sigma^{-1}(c_- + \sigma -1).
\eqn

Note that $\hat{g}^{\alpha \beta} \hat{g}_{\beta \gamma} =\delta^\alpha_\gamma$ and $\hat{u}_\alpha\equiv \hat{g}_{\alpha \beta} \hat{u}^\beta$. Then, from Eq.(\ref{eq2.2}), we find that
\bq
\lb{eq2.4a}
\hat{g}_{\alpha \beta}\hat{u}^{ \alpha}\hat{u}^{\beta} = -1, \quad \hat{g} = \sigma g,
\eq
 where $\hat{g}$ is the determinant of $\hat{g}_{\alpha \beta}$. Thus, replacing $G_{\ae}$ and ${\cal{L}}_{\lambda}$ by $\hat G_{\ae}$ and $\hat {\cal{L}}_{\lambda}$ in Eq.(\ref{2.1}), where
\bq
\lb{eq2.4}
\hat{G}_{\ae}  \equiv \sqrt{\sigma} G_{\ae},\quad  \hat{\cal{L}}_{\lambda}  \equiv  \lambda \left(\hat{g}_{\alpha\beta} \hat{u}^{\alpha}\hat{u}^{\beta} + 1\right), 
\eq
we find that 
\bq
\lb{eq2.6}
 S_{\ae}\left(g_{\alpha\beta}, u^{\alpha}, c_i,  G_{\ae}, \lambda\right) 
 = \hat{S}_{\ae}\left(\hat{g}_{\alpha\beta}, \hat{u}^{\alpha}, \hat{c}_i, \hat{G}_{\ae},  \lambda\right).
 \eq
 As a result, when the matter field is absent, that is, ${\cal{L}}_{m} = 0$, the Einstein-aether vacuum field equations take the same forms 
 for the fields $\left(\hat{g}_{\alpha \beta}, \hat{u}^{ \alpha}, \hat{c}_i, \lambda\right)$,  
 \bqn
 \lb{2.4aa}
 \hat R^{\mu\nu} - \frac{1}{2} \hat g^{\mu\nu}\hat R &=& \hat S^{\mu\nu},\\
 \lb{2.4bb}
 \hat \AE_{\mu} &=&0, \\
   \lb{2.4cc}
  \hat g_{\alpha\beta} \hat u^{\alpha}\hat u^{\beta} &=& -1,
 \eqn
 where $\hat R^{\mu\nu}$ and $\hat R$ are the Ricci tensor and scalar made of $\hat g_{\alpha\beta}$. 
  $\hat S^{\mu\nu}$ and $\hat \AE_{\mu}$ are given by Eq.(\ref{2.5}) simply by replacing $\left(g_{\mu\nu}, u^{\mu}, c_i\right)$  by
  $\left(\hat g_{\mu\nu}, \hat u^{\mu}, \hat c_i\right)$.

Therefore, {\it for any given vacuum solution of the  {Einstein-aether} field equations} $\left(g_{\mu\nu}, u^{\mu}, c_i, \lambda\right)$, 
{\it using the above field redefinitions, we can obtain a class of the vacuum solutions of the  {Einstein-aether field equations,} given by}  
$\left(\hat g_{\mu\nu}, \hat u^{\mu}, \hat c_i, \lambda\right)$  \footnote{It should be noted that this holds in general only for the vacuum case. 
In particular, when matter presence, the aether field will be directly coupled with matter through the metric redefinitions.}. 
Certainly, such obtained solutions may not always satisfy the  physical  and observational constraints found so far \cite{OMW18}. 
 
In this paper, we shall take advantage of such field redefinitions to simplify the corresponding  mathematic problems by assuming that 
the fields described by  $\left(g_{\mu\nu}, u^{\mu}, c_i, \lambda\right)$ are the physical ones, while the ones described by 
$\left(\hat g_{\mu\nu}, \hat u^{\mu}, \hat c_i,  \lambda\right)$ as the ``effective" ones, although both of the two metrics are the vacuum solutions of the 
Einstein-aether  field equations, and can be physical, provided that the constraints recently given in \cite{OMW18} are satisfied.

The gravitational sector described by $\left(\hat g_{\mu\nu}, \hat u^{\mu}, \hat c_i,  \lambda\right)$ has also three different propagation modes, 
with their speeds $\hat{c}_{A}$ given by Eq.(\ref{eq1.1}) with the replacement $c_i$ by $\hat{c}_i$. Each of these modes defines a horizon, which is now a null surface 
of the metric,
\bq
\lb{EHMb}
\hat g_{\alpha \beta}^{(A)} \equiv \hat g_{\alpha \beta} - \left(\hat c_{A}^2 - 1\right) \hat u_{\alpha}\hat u_{ \beta}, 
\eq
where $A = S, V, T$. It is interesting to note that  
\bq
\lb{eq2.7a}
\hat{c}^2_A = \frac{c^2_A}{\sigma}.
\eq
Thus, choosing $\sigma = c_{S}^2$, we have $\hat{c}_S =1$, and from Eq.(\ref{EHMb})  we find that
\bq
\lb{eq2.7}
\hat g^{(S)}_{\alpha\beta} = \hat g _{\alpha\beta},\; \left(\sigma = c_{S}^2\right), 
\eq
that is, {the S0H} of the metric $\hat g_{\alpha\beta}$ coincides with its MH. Moreover,    from Eqs.(\ref{EHM}) and (\ref{eq2.2}) we also find that 
\bq
\lb{eq2.7b}
g^{(S)}_{\alpha\beta} = \hat g _{\alpha\beta},\; \left(\sigma = c_{S}^2\right).
\eq
Therefore, {\it with the choice $\sigma = c_{S}^2$, the MH of $\hat g_{\alpha\beta}$ is also the S0H of the metric  $g_{\alpha\beta}$}. 

    Arnowitt-Deser-Misner (ADM)


\subsection{Hypersurface-Orthogonal Aether Fields}

When the aether field $u_{\mu}$ is hypersurface-orthogonal (HO), the Einstein-aether field equations depend only on   three combinations of the four coupling constants $c_i$'s.  
To see this clearly, let us first notice that, if the aether is HO, the twist $\omega^{\mu}$ vanishes \cite{Jacobson}, where $\omega^{\mu}$ is defined as 
$\omega^{\mu} \equiv \epsilon^{\mu\nu\alpha\beta}u_{\nu}D_{\alpha}u_{\beta}$. Since  
\bqn
\lb{eq2.9}
  \omega_{\mu}\omega^{\mu}   &=& \left(D_{\mu}u_{\nu}\right)\left(D^{\nu}u^{\mu}\right) - \left(D_{\mu}u_{\nu}\right)\left(D^{\mu}u^{\nu}\right) \nb\\
&&
- \left(u^{\mu}D_{\mu}u_{\alpha}\right)\left(u^{\nu}D_{\nu}u^{\alpha}\right),
\eqn
we can see that the addition of the term 
\bq
\lb{eq2.10}
\Delta {\cal{L}}_{\ae} \equiv  c_0\omega_{\mu}\omega^{\mu}, 
\eq
to ${\cal{L}}_{\ae}$ will not change the action, where $c_0$ is an arbitrary real constant. However,  this is equivalent to replacing $c_i$ by $\bar{c}_i$
in ${\cal{L}}_{\ae}$, where
\bqn
\lb{eq2.11}
&& \bar{c}_1 \equiv  c_1 + c_0, \quad \bar{c}_2 \equiv  c_2, \nb\\
&& \bar{c}_3 \equiv  c_3 - c_0, \quad \bar{c}_4 \equiv c_4 - c_0.
\eqn
Thus, by properly choosing $c_0$,  we can always eliminate one of the three parameters, $c_{1}, \; c_{3}$ and $c_{4}$, or one of their combinations. 
Therefore, in this case only three combinations of $c_i$'s appear in the field equations. Since
\bq
\lb{eq2.13}
\bar{c}_{13} = c_{13}, \quad \bar{c}_{14} = c_{14}, \quad \bar{c}_{2} = c_{2}, 
\eq
without loss of the generality, we can always choose these three combinations   as $c_{13}, \; c_{14}$ and $c_2$. 

To understand the above further, and also see the physical meaning of these combinations, following  Jacobson  \cite{Jacobson14}, we first
decompose $D_{\beta}u_{\alpha}$ into the form,
\bq
\lb{eq2.13aa}
D_{\beta}u_{\alpha}  = \frac{1}{3}\theta h_{\alpha\beta} + \sigma_{\alpha\beta} + \omega_{\alpha\beta} - a_\alpha u_\beta,
\eq
where $\theta$ denotes the expansion of the aether field, $h_{\alpha\beta}$ the spatial projection operator, $\sigma_{\alpha\beta}$
the shear, which is the symmetric trace-free part of the spatial projection of $D_{\beta}u_{\alpha}$, while $\omega_{\alpha\beta}$ 
denotes the antisymmetric part of the spatial projection of $D_{\beta}u_{\alpha}$, defined, respectively, by
\bqn
\lb{eq2.13bb}
 h_{\alpha\beta} &\equiv& g_{\alpha\beta} + u_{\alpha} u_{\beta}, \quad \theta \equiv D_{\lambda}u^{\lambda},  \nb\\
\sigma_{\alpha\beta} &\equiv & D_{(\beta}u_{\alpha)} + a_{(\alpha}u_{\beta)}  - \frac{1}{3}\theta h_{\alpha\beta},\nb\\
\omega_{\alpha\beta} &\equiv & D_{[\beta}u_{\alpha]} + a_{[\alpha}u_{\beta]},
\eqn
with  $(A, B) \equiv (AB + BA)/2$ and $[A, B] \equiv (AB - BA)/2$.  Recall that $a_{\mu}$ is the acceleration of the aether field,  
given by Eq.(\ref{2.6}).

In terms of these quantities,  Jacobson found 
\bqn
\lb{eq2.13cc}
\int d^{4}x \sqrt{-g}{\cal{L}}_{\ae} &=&  \int d^{4}x \sqrt{-g}\Bigg[R-\frac{1}{3}c_{\theta} \theta^2    \nb\\
&&  {+ c_a a^2 - c_\sigma\sigma^2- c_{\omega} \omega^2}\Bigg], ~~~~
\eqn
where
\bqn
\lb{eq2.13dd}
&& c_\theta \equiv c_{13}  + 3c_2, \quad
c_\sigma \equiv c_{13}, \nb\\
&& c_\omega \equiv c_1 - c_3, \quad 
c_a \equiv c_{14},
 \eqn
 and
 \bqn
 \lb{sigma2}
 \sigma^2 &=& - \frac{1}{3} \theta^2+(D_\mu u_\nu) (D^\mu u^\nu) + a^2.
 \eqn
Note that  in the above action,  there are no crossing terms of $(\theta,  \sigma_{\alpha\beta},  \omega_{\alpha\beta}, a_\alpha)$. 
This is because the four terms on the right-hand side of Eq.(\ref{eq2.13aa}) are orthogonal to each other, and  when forming  
quadratic combinations of these quantities, only their  ``squares" contribute  \cite{Jacobson14}.
 
 From Eq.(\ref{eq2.13dd}) we can see clearly that $c_{14}$ is related to the acceleration of the aether field, $c_{13}$ to its shear, 
while its expansion is related to both $c_2$ and $c_{13}$. More interesting, the coefficient of the twist is proportional to $c_1 - c_3$.
When $u_{\mu}$ is hypersurface-orthogonal, we have $\omega^2 =0$, so the last term in the above action vanishes
identically, and only the three free parameters $c_{\theta}, c_\sigma$ and $ c_a$  remain.

It is also interesting to note that the twist vanishes if and only if the four-velocity of the aether satisfies the conditions \cite{Wald94},
\bq
\lb{eq2.12}
u_{[\mu}D_{\nu}u_{\alpha]} = 0.
\eq
When the aether is HO, it can be shown that Eq.(\ref{eq2.12}) is satisfied. In addition, in the spherically symmetric case, Eq.(\ref{eq2.12}) holds identically.

 Moreover, it can be also shown    \cite{Wald94} that Eq.(\ref{eq2.12}) is the necessary and sufficient condition to write the four-velocity $u_{\mu}$ in terms the gradient 
of a timelike scalar field $\phi$,
\bq
\lb{eq2.13}
u_{\mu} = \frac{\phi_{,\mu}}{\sqrt{-\phi_{,\alpha}\phi^{,\alpha}}}.
\eq
 Substituting it into the action (\ref{eq2.13cc}), one obtains the action of the infrared limit of the healthy extension \cite{BPSa,BPSb} of the  Ho\v{r}ava theory
\cite{Horava}, which is
often referred to as {\it  the khronometric theory} \footnote{ {In \cite{Jacobson10,Jacobson14}, it was also referred to as T-theory.}}, where $\phi$ is called {\it the  khronon field}. 

 It should be noted that the khronometric theory and the HO $\ae$-theory are equivalent only in the action level. In particular, in addition to the  scalar mode, the khronometric theory has
also an  instantaneous mode \cite{BS11,LMWZ17}, a mode that propagates with an infinitely large speed. This is mainly due to the fact that the field equations of the khronometric theory are the four-order
differential equations of $\phi$. It is the presence of those high-order terms that lead to the existence of the instantaneous mode \footnote{In the Degenerate Higher-Order Scalar-Tensor (DHOST) theories,
this mode is also referred to as the ``shadowy" mode \cite{DLMNW18}.}. On the other hand, in $\ae$-theory, including the case with the HO  symmetry,
the field equations are of  {the second order} for both the metric $g_{\mu\nu}$ and the aether field $u_{\mu}$. As a result, this instantaneous mode is absent. For more details, we refer readers to
\cite{Wang17} and references therein.


\section{Spherically symmetric vacuum spacetimes}
 \renewcommand{\theequation}{3.\arabic{equation}} \setcounter{equation}{0}

\subsection{Field Equations for $g_{\mu\nu}$ and $ u^{\mu}$}

As shown in the last section, to be consistent with observations, we must assume $c_S \ge 1$. As a result, S0Hs must be inside MHs. 
Since now S0Hs define the boundaries of spherically symmetric BHs, in order to cover spacetimes both inside and outside the MHs, 
one way is to adopt the Eddington-Finkelstein (EF) coordinates,
\bqn
\lb{eq2.14}
d s^{2}&\equiv& g_{\mu\nu}dx^{\mu}dx^{\nu}\nb\\
&=& - F(r) d v^{2}+2 B(r) d v d r+r^{2} d \Omega^{2},
\eqn
where $d\Omega^2 \equiv d\theta^2 + \sin^2\theta d\phi^2$ and $x^\mu=(v, r, \theta, \phi)$, while the aether field  takes the general form,
\begin{equation}
\lb{eq2.15}
u^{\alpha} \partial_{\alpha}=A(r) \partial_{v}-\frac{1-F(r) A^{2}(r)}{2 B(r) A(r)} \partial_{r},
\end{equation}
which is respect to the spherical symmetry, and satisfies the constraint $u_{\alpha} u^{\alpha} = -1$. Therefore, in the current case, we have three unknown functions, $F(r), A(r)$ and $B(r)$.
 
Then, the vacuum field equations $E^{\mu\nu} \equiv G^{\mu\nu} - S^{\mu\nu} = 0$ and $\text{\AE}^\mu=0$ can be divided into two groups \cite{Eling2006-2,Enrico11}: 
one represents the  evolution equations, given by
\begin{equation}
\label{eq2.20}
E^{v v}=E^{\theta \theta}=\text{\AE}^{v}=0, 
\end{equation}
and the other represents the constraint equation, given by
\begin{equation}
\label{eq2.21}
C^{v}=0,
\end{equation}
where $C^{\alpha} \equiv E^{r \alpha}+u^{r}\text{\AE}^{\alpha}=0$, and   { $G^{\mu \nu} \left[\equiv R^{\mu\nu} - R g^{\mu\nu}/2\right]$ denotes  the  Einstein tensor}. Note that in Eq.(35) of \cite{Enrico11} two constraint equations $C^{v}= C^{r} = 0$
were considered. However, $C^r$ and $C^v$ are not independent. Instead, they are related to each other by the relation 
$C^r = (F/B) C^v$. Thus,  $C^v = 0$ implies $C^r = 0$, so there is only one independent constraint. 
On the other hand,  the three evolution  equations  can be cast in the forms \footnote{It should be noted that in \cite{Enrico11}
the second-order differential equation for $F$ [cf. Eq.(36) given there] also depends on $B$. But, since from the constraint $C^v = 0$, given by Eq.(\ref{eq2.23a}), one can express $B$
in terms of $A, F$ and their derivatives, as shown explicitly by Eq.(\ref{Bcv}), so there are no essential differences here, and it should only reflect the facts that different combinations of 
the field equations are used.},
\bqn
\lb{eq2.22a} 
F'' &=&  \mathcal{F} \left(A, A', F, F', r, c_i\right) \nb\\
&=&  \frac{1}{2r^2A^4{\cal{D}}}\Big(f_0+f_1F+f_2F^2+f_3F^3  \nb\\
&& ~~~~~~~~~~~~  +f_4F^4\Big),  \\ 
\lb{eq2.22b}
A'' &=&  \mathcal{A} \left(A, A', F, F', r, c_i\right) \nb\\
&=&  \frac{1}{2r^2A^2{\cal{D}}}\Big(a_0+a_1F+a_2F^2+a_3F^3\Big),  \\
\lb{eq2.22c}
\frac{B'}{B}&=&\mathcal{B} \left(A, A', F, F', r, c_i\right) \nb\\
&=& \frac{1}{2rA^2{\cal{D}}}\Big(b_0+b_1F+b_2F^2\Big), 
\eqn
where a prime stands for the derivative with respect to $r$, and 
\bq
\lb{eq2.23}
{\cal{D}}  \equiv d_{-}\left(J^2 + 1\right) +  {2} d_{+} J,
\eq
with $J \equiv FA^2$ and 
\bqn
\lb{eq2.23-a}
d_{\pm}&\equiv& (c_S^2 \pm 1)c_{14}(1-c_{13})(2+c_{13}+3c_2).
\eqn
The coefficients $f_n, \; a_n$ and $b_n$ are independent of $F(r)$ and $B(r)$ but depend on $F'(r)$, $A(r)$ and $A'(r)$, and are given explicitly by Eqs.(\ref{fns}), (\ref{ans}) and (\ref{bns}) in Appendix A.
The constraint equation (\ref{eq2.21}) now can be cast in the form,
\bqn
\lb{eq2.23a}
n_0+n_1F +n_2F^2 = 0,
\eqn
where $n_n$'s are given explicitly by Eq.(\ref{nns}) in Appendix A.

 Thus, we have three dynamical equations and one constraint for the three unknown functions, $F, A$ and $B$. As a result, the system seems over determined. However, a closer examination 
 shows that not all of them are  independent. For example, Eq.(\ref{eq2.22c}) can be obtained from Eqs.(\ref{eq2.22a}), (\ref{eq2.22b}), and (\ref{eq2.23a}). In fact, from Eq.(\ref{eq2.23a}),  we  find that   the function $B$ can be written in the form 
	\bqn
	\lb{Bcv}
	B(r)  &=& \pm \frac{1}{2 \sqrt{2} A^2} \Bigg\{2 A^2 \Big[ 4 J (1+2 c_2+ c_{13}) \nb\\
	&& ~~~-(2 c_2+c_{13}) \big(J+1\big)^2 \Big] \nb\\
	&& +4 r A \Big[2 A J'-4 J A'\nb\\
	&& ~~~+ c_2 \big(J-1\big) \big(J A'-A'-A J'\big)\Big] \nb\\
	&& +r^2 \Big[c_{14} \big(J A'+A'-A J'\big)^2 \nb\\
	&& ~~~-(c_2+c_{13}) \big(J A'-A'-A J'\big)^2 \Big]\Bigg\}^{1/2}.~~~~~~~~
	\eqn
Recall that $J=F A^2$. Note that there are two branches of solutions for $B(r)$ with opposite signs,  since Eq.(\ref{eq2.23a}) is a quadratic equation of $B$. However,  only
 the ``+" sign   will give us $B = 1$ at the spatial infinity, while  the ``-" sign  will yield $B(r\rightarrow \infty) = -1$. Therefore, in the rest of the paper, we shall choose the ``+" sign in Eq.(\ref{Bcv}).
Then, first taking the derivative of Eq.(\ref{Bcv}) with respective to $r$, and then combining the obtained result with Eqs.(\ref{eq2.22a}) and (\ref{eq2.22b}), one can obtain Eq.(\ref{eq2.22c}) \footnote{From this proof it can be seen that obtaining Eq.(\ref{eq2.22c}) from Eq.(\ref{Bcv}) the operation of taking the first-order  derivatives was involved. Therefore, in principle these two equations are equivalent modulated an integration constant.}.

To solve these equations, in this paper we shall adopt the following strategy:  choosing Eqs.(\ref{eq2.22a}), (\ref{eq2.22b}) and (\ref{Bcv}) as the three independent equations for the three unknown functions, $F$, $A$, and $B$. The advantage of this choice is that Eqs.(\ref{eq2.22a}) and (\ref{eq2.22b}) are independent of the function $B$. Therefore, we can first solve these two equations to find $F$ and $A$, and then obtain the function $B$ directly from Eq.(\ref{Bcv}). In this approach, we only need to solve two equations, which will significantly save the computation labor, although we do use Eq.(\ref{eq2.22c}) to monitor our numerical errors.

To solve Eqs.(\ref{eq2.22a}) and (\ref{eq2.22b}), we can consider  them as   the ``initial" value problem at a given ``moment", say, $r = r_0$ \cite{Eling2006-2,Enrico11}.
Since they are second-order differential equations,  the initial data will consist of the four initial values,
\bq
\lb{eq2.24a}
\Big\{A(r_0), A'(r_0), F(r_0), F'(r_0)\Big\}.
\eq
In principle, $r_0$ can be chosen as any given (finite) moment. However, in the following we shall show that the most convenient choice will be the locations 
of the S0Hs. It should be noted that  a S0H does not always exist for any given initial data. However, since in this paper we are mainly interested in the case in which a S0H
exists, so whenever we choose $r_0 = r_{S0H}$, it always means that we only consider the case in which  such a S0H is present.

To determine the location of the S0H for a given spherical solution of the metric (\ref{eq2.14}), let us first consider the out-pointing normal vector, $N_{\mu}$, of a hypersurface $r = $ Constant, say, $r_0$, 
which is given by $N_{\mu} \equiv \partial(r - r_0)/\partial x^{\mu} = \delta^r_{\mu}$.  Then, the metric and spin-0 horizons of $g_{\mu\nu}$ are given, respectively, by
\bqn
\lb{MH}
g_{\alpha \beta}N^\alpha N^\beta=0,\\
\lb{S0H}
g^{(S)}_{\alpha \beta}N^\alpha N^\beta=0,
\eqn
where $N^{\mu} \equiv g^{\mu\nu} N_{\nu}$, and $g^{(S)}_{\alpha \beta}$ is defined by Eq.(\ref{eq1.3}).  For the metric and aether given in the form of Eqs.(\ref{eq2.14}) and (\ref{eq2.15}), they become
\bqn
\lb{MH2}
&&   {F(r_{MH})}  = 0,\\
\lb{S0H2}
&&  \left(c_S^2-1\right) \left(J(r_{S0H})^2 + 1\right) +2 \left(c_S^2+1\right) J(r_{S0H}) =0, \nb\\
\eqn 
where  $r = r_{MH}$ and $r = r_{S0H}$ are the locations of the metric and spin-0 horizons, respectively. Note that Eqs.(\ref{MH2}) and (\ref{S0H2}) may have multiple roots, say,  $r_{MH}^i$ and $r_{S0H}^j$. In these cases, 
the  location of the metric (spin-0) horizon  is always taken to be the largest root of  $r_{MH}^i$ ($r_{S0H}^j$).

 Depending on the value of $c_S$, the solutions of Eq.(\ref{S0H}) are given, respectively, by
\bq
\lb{eq2.24b}
J(r_{S0H}^\pm)=\frac{1\mp c_S}{1\pm c_S} \equiv J^{\pm} , \quad c_S\neq1,
\eq
  and 
\bq
\lb{eq2.24c}
J(r_{S0H})=0, \quad c_S=1.
\eq 

It is interesting to note that on  {S0Hs}, we   have 
\bq
\lb{eq2.24d}
{\cal{D}}(r_{S0H}) = 0,
\eq
as it can be seen from Eqs.(\ref{eq2.23}), (\ref{eq2.23-a}) and (\ref{S0H2}). 

\begin{figure}[htb]
 	\includegraphics[width=\columnwidth]{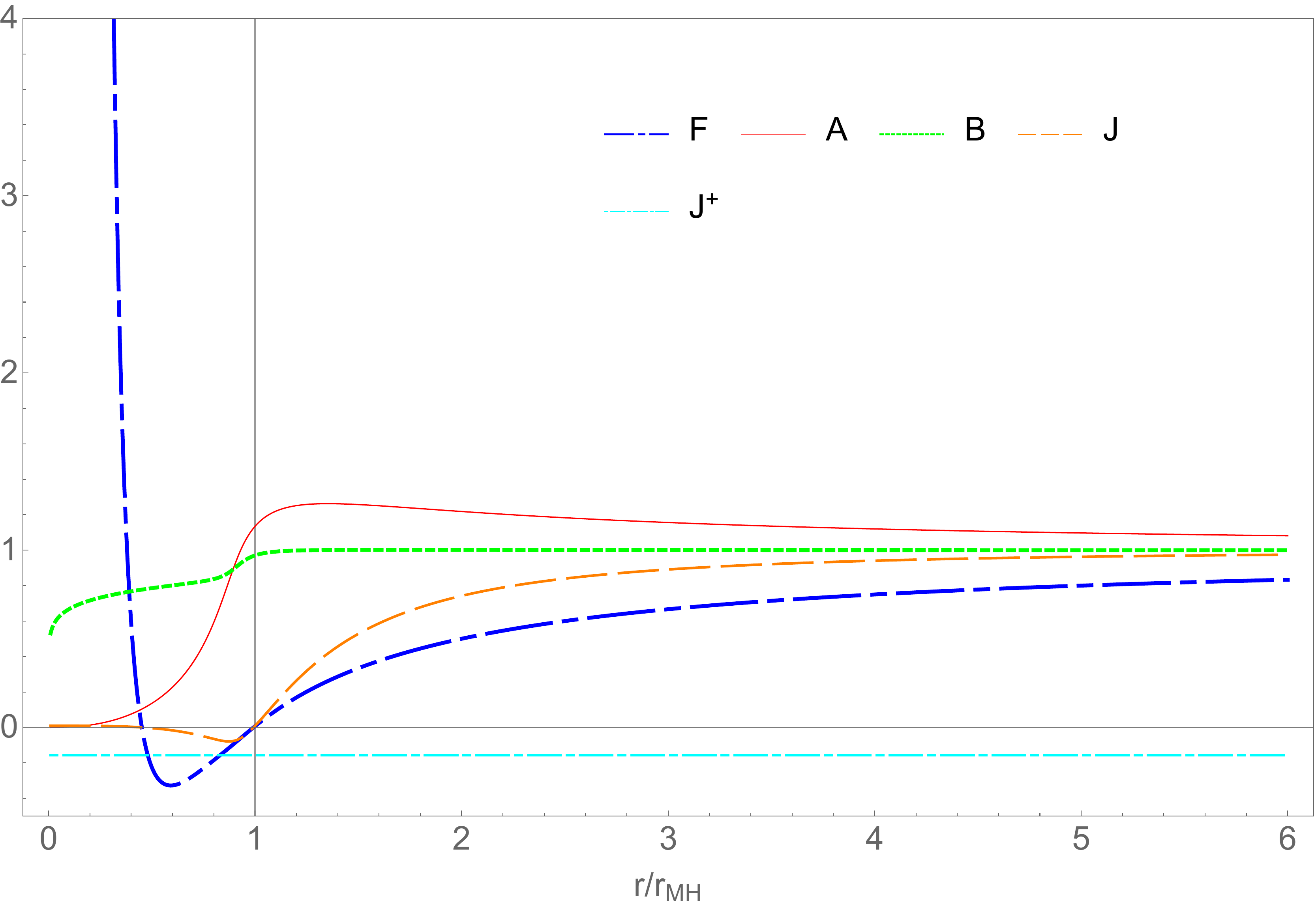} 
 	\caption{ { The solution for  $c_1=0.051$, $c_2=0.116$, $c_3=-c_1$ and $c_4=0$, first considered numerically in \cite{Eling2006-2}. There are an outer and inner MHs, 
	at which $F$ vanishes. But,  $J$ does not cross the constant line of {$J^+$},
	 so that a S0H is absent. This graph is the same as the one given in \cite{Eling2006-2} (up to the numerical errors).}} 
 	\label{plot18}
 \end{figure}

 As  mentioned above, for some choices of $c_i$, Eq.(\ref{S0H}) does not always admit a solution, hence a S0H does not exist in this case. A particular choice was considered in
 \cite{Eling2006-2}, in which we have  $c_1=0.051$, $c_2=0.116$, $c_3=-c_1$ and $c_4=0$.  For this choice, we find that $c_S \simeq 1.37404$, $J^+ \simeq -0.157556$
and  $J^- \simeq -6.34696$. As shown in Fig. \ref{plot18}, the function $J(r)$ is always greater than
 $J^{\pm}$, so no S0H is formed, as first noticed in \cite{Eling2006-2}.   
  Up to the numerical errors, Fig. \ref{plot18} is the same as that given in \cite{Eling2006-2}, which provides another way to check our general  expressions of the field equations given above.

 In addition, we also find that the two exact solutions obtained in \cite{Per12} satisfy these equations identically, as it is expected.

\subsection{Exact Solutions with $c_{14}=c_{13}=0$}
  
 From  Eqs.(\ref{2.8a}) - (\ref{c1234bb}) we can see that the choice  $c_{14}=c_{13}=0$ satisfies these constraints, provided that $c_2$ satisfies the condition \footnote{When $c_{14} = c_{13} = 0$,
 the speeds of the spin-0 and spin-1 modes can be infinitely large, as it can be seen from Eq.(\ref{eq1.1}). Then,  cautions must be taken, including the calculations of the PPN parameters \cite{FJ06}.}, 
 \bq
 \lb{eq2.24e}
 0 \lesssim c_2 \lesssim  0.095.
 \eq
 
Then,   we find that Eqs.(\ref{eq2.22a})-(\ref{eq2.22b}) now reduce to
 \bqn
 \lb{eqF}
 F^{\prime \prime} &=&  -\frac{2}{r} F'+\frac{c_2 \hat{\cal{F}}(r)}{4 r^2 A^4}, \\
 \lb{eqA}
 A^{\prime \prime} &=& \frac{2}{r^2 (A+A^3 F)} \Big[r^2 (A')^2-r A A' - A^2 \nb\\
 && -r A^3 A'(F+r F')+A^4 F\Big] \nb\\
 &&   -\frac{c_2  \hat{\cal{F}}(r)}{4 r^2 (A+A^3 F)},
 \eqn 
 where 
 \bqn
 \lb{eqD}
\hat {\cal{F}}(r) \equiv  \left[r A'-2 A+r A^2 A' F+A^3 \left(2 F+r F'\right)\right]^2.~~~~
 \eqn
Combining Eqs.(\ref{eqF}) and (\ref{eqA}), we find the following equation, 
 \bqn
 \lb{eqW}
 W''+{W'}^2+\frac{2}{r} W'-\frac{2}{r^2}=0,
 \eqn 
 where
 \bqn
 \lb{eqW1}
 W &\equiv& \ln\left(\frac{1-FA^2}{A}\right).
 \eqn 
Eq.(\ref{eqW}) has the general solution, 
  \bqn
 \lb{eqW2}
 W =\ln w_2 +\ln\left(\frac{1+w_1 r^3}{r^2}\right),
 \eqn 
 where $w_1$ and $w_2$ are two integration constants. Then, the combination of Eqs.(\ref{eqW1}) and (\ref{eqW2}) yields,
\bqn
\lb{eqF2}
F(r) &=& \frac{1}{A^2} - \frac{w_2}{A} \left(\frac{1}{r^2}+w_1 r\right).
\eqn
Substituting Eq.(\ref{eqF2}) into Eq.(\ref{eqF}), we find
 \bqn
\lb{eqF3}
F'' &=& -\frac{2}{r} F'+ F_0, 
\eqn
where $F_0 \equiv 9c_2  w_1^2w_2^2/4$. Integrating Eq.(\ref{eqF3}), we find
 \bqn
\lb{eqF4}
F(r) &=& F_2\left(1 - \frac{2m}{r}\right) + \frac{F_0}{6}r^2,
\eqn
where $m$ and $F_2$ are two other  integration constants. On the other hand,   from Eq.(\ref{eqF2}), we find that 
 \bqn
\lb{eqA2}
A(r) &=& - \frac{w_2}{2F} \Bigg[\left(\frac{1}{r^2} + w_1 r\right) \nb\\
&&   \left. \pm   \sqrt{\frac{4F}{w_2^2} +\left(\frac{1}{r^2} + w_1 r\right)^2}\right].
\eqn
Substituting the above expressions for $A$ and $F$ into the constraint (\ref{Bcv}), we find that
\bqn
\lb{eqA3}
B  &=& \sqrt{F_2}.
\eqn

Note that the above solution is asymptotically flat only when   $w_1=0$, for which we have
\bqn
\lb{eqA4}
F(r) &=& F_2\left(1 - \frac{2m}{r}\right), \quad B(r)  = \sqrt{F_2},\nb\\
A(r) &=& - \frac{w_2}{2F} \left(\frac{1}{r^2}  \pm   \sqrt{\frac{4F}{w_2^2} + \frac{1}{r^4}}\right).
\eqn
Using the gauge residual $v ' = C_0 v + C_1$ of the metric (\ref{eq2.14}), without loss of the generality, we can always set $F_2 = 1$, so the corresponding metric takes the 
precise form of the Schwarzschild solution,  
\bqn
\lb{eqA5}
ds^2 = - \left(1 - \frac{2m}{r}\right)dv^2 + 2 dv dr + r^2d\Omega^2,  
\eqn
while the aether field is given by
\bqn
\lb{eqA6}
A(r) &=& -\frac{w_2 \pm \sqrt{w_2^2 +4 r^3(r-2 m)}}{2r (r-2 m)}.~~~~~
\eqn
 It is remarkable to note that now the aether field has no contribution to the spacetime geometry, although it does feel the gravitational field, 
as it can be seen from Eq.(\ref{eqA6}).

It should be also noted that Eqs.(\ref{eqA5}) and (\ref{eqA6}) were a particular case of the solutions first found in \cite{Per12} for the case $c_{14}=0$ by further setting $c_{13} = 0$. But,
the general solutions given by Eqs.(\ref{eqF4}) - (\ref{eqA3}) are new, as far as we know.

\subsection{Field Equations for $\tilde g_{\mu\nu}$ and $ \tilde u^{\mu}$}

Note that, instead of solving the three independent equations directly for $A$, $B$ and $F$, we shall first solve the corresponding
 three equations for $\tilde A$, $\tilde B$ and $\tilde F$,  by taking the advantage 
 of the field redefinitions introduced in the last section, and then obtain the functions $A$, $B$ and $F$ by the inverse transformations 
 of Eqs.(\ref{FBtilde}) and (\ref{Atilde}) to be given below. This will considerably simplify mathematically the problem of solving such complicated equations. 
 
 To this goal, let us first note that,
with the filed redefinitions (\ref{eq2.2}), the line element  corresponding to $\hat{g}_{\mu\nu}$ in the  coordinates ($v, r, \theta, \phi$),  
 takes the form, 
\bqn
\lb{eq2.16b}
d\hat{s}^{2}&\equiv& \hat g_{\mu\nu}dx^{\mu}dx^{\nu}\nb\\
&=& -\left[F+ \frac{(\sigma -1) \left( A^2  F+1\right)^2}{4 A^2} \right] d v^{2} \nb\\
&&  +2\left[B+ \frac{1}{2} (\sigma -1) B \left( A^2  F+1\right)\right]d v d r \nb\\
&& -(\sigma -1) A^2 B^2  dr^2 +r^{2} d \Omega^{2}.
\eqn
To bring the above expression into the standard EF form, we first make the coordinate transformation,
\bq
\lb{eq2.17}
  \tilde v = C_0 v - C(r), 
\eq
 where   $C_0$ is an arbitrary real constant, and   $C(r)$ is a function of $r$. Then, choosing $C(r)$ so that
  \bqn
\lb{C0Cr}
\frac{dC(r)}{dr} &=& \frac{2 C_0 A^2 B \left(\sqrt{\sigma }-1\right)}{J \left(\sqrt{\sigma }-1\right)+(\sqrt{\sigma }+1)},
\eqn
we find that  in the coordinates ${\tilde x}^\mu =(\tilde v,  r,  \theta, \phi)$  the line element (\ref{eq2.16b}) takes the form,
\bqn
\lb{eq2.19}
d\hat{s}^{2}&\equiv& \hat g_{\mu\nu}dx^{\mu}dx^{\nu} = \tilde g_{\mu\nu} d\tilde x^{\mu}d\tilde x^{\nu}\nb\\
&=& -\tilde F(r)  d \tilde v^{2}    +2\tilde B(r) d \tilde v d r + r^{2} d \Omega^{2},
\eqn
where
 \bqn
\lb{FBtilde}
{\tilde {F}} &=&  \frac{J^2 (\sigma -1)+2 J (\sigma +1)+(\sigma -1)}{4 C_0^2 A^2}, \nb\\
{\tilde {B}} &=& \frac{\sqrt{\sigma} B}{C_0}.
\eqn

 On the other hand, in terms of the coordinates $\tilde{x}^{\mu}$, the aether four-velocity is given by 
 \bqn
\lb{eq2.20b}
  \hat u^{\alpha} \frac{\partial}{\partial {x}^{\alpha}}  =  \tilde u^{\alpha} \frac{\partial}{\partial \tilde{x}^{\alpha}}   
  =   \tilde A(r) \partial_{\tilde v}-\frac{1-\tilde F(r) \tilde A^{2}(r)}{2 \tilde B(r) \tilde A(r)} \partial_{r}, ~~~~~~
\eqn
where
\bq
\lb{Atilde}
{\tilde {A}} = \frac{2 C_0 A}{J \left(\sqrt{\sigma }-1\right)+(\sqrt{\sigma }+1)},
\eq
which satisfies the constraint  $\tilde u^{\alpha} \tilde u^{\beta} \tilde g_{\alpha\beta}= -1$, with
 \bqn
 \lb{eq2.18}
  \tilde g_{\mu\nu}  \equiv  \frac{\partial x^{\alpha}}{\partial \tilde x^{\mu}}  \frac{\partial x^{\beta}}{\partial \tilde x^{\nu}} \hat g_{\alpha \beta},\quad
  \tilde u_{\mu}  \equiv  \frac{\partial x^{\alpha}}{\partial \tilde x^{\mu}}    \hat u_{\alpha}.
\eqn

It should be noted that the metric (\ref{eq2.19}) still has the gauge residual,  
\bq
\lb{eq2.22}
 \tilde{\tilde v} = C_1 \tilde v + C_2,
 \eq
 where $C_1$ and $C_2$ are two arbitrary constants, which will keep the   {line element} in the  same form, after the rescaling, 
  \bq
 \lb{eq2.23b} 
 \tilde{\tilde F} =  \frac{\tilde F}{C_{1}^{2}}, \quad \tilde{\tilde B} =  \frac{\tilde B}{C_{1}}.
 \eq
 Later we shall use this gauge freedom to fix one of the initial conditions.

In the rest of this paper, we always refer $\left(\tilde g_{\mu\nu}, \tilde u^{\alpha}\right)$ as the field obtained by the field  redefinitions. The latter is related  
to $\left(\hat g_{\mu\nu}, \hat u^{\alpha}\right)$  via the inverse coordinate transformations of Eq.(\ref{eq2.18}).
Then, the Einstein-aether field equations for $\left(\tilde g_{\mu\nu}, \tilde u^{\alpha}\right)$ will take the same forms as those given by Eqs.(\ref{2.4aa}) - (\ref{2.4cc}), but now in 
terms of  $\left(\tilde g_{\mu\nu}, \tilde u^{\alpha}, \tilde{c}_i\right)$  in the coordinates $\tilde{x}^{\mu}$, where $\tilde c_i \equiv \hat c_i$.

On the other hand, since the metric   (\ref{eq2.19}) for $\tilde g_{\mu\nu}$ takes the same form as the metric (\ref{eq2.14})  for $g_{\mu\nu}$, and so does the aether field   {(\ref{eq2.20b})} for
  {$\tilde u^{\mu}$} as the one (\ref{eq2.15}) for  {$u^{\mu}$,} it is not difficult to see that the field equations for $\tilde{F}(r), \; \tilde{A}(r)$ and $\tilde{B}(r)$ will be given  precisely
 by Eqs.(\ref{eq2.22a}) - (\ref{eq2.23a}), if we simply make the following replacement,
  \bq
\lb{eq2.25}
\left(F, A, B, c_i\right) \rightarrow \left(\tilde F, \tilde A, \tilde B, \tilde c_i\right).
\eq
 As a result, we have
\bqn
\lb{eq2.22aB}
\tilde F'' &=&  \tilde{\mathcal{F}} \left(\tilde A, \tilde A', \tilde F, \tilde F', r, \tilde c_i\right) \nb\\
&=&  \frac{1}{2r^2 \tilde A^4\tilde{\cal{D}}}\Big[\tilde f_0+\tilde f_1\tilde F+\tilde f_2\tilde F^2+\tilde f_3\tilde F^3  \nb\\
&& ~~~~~~~~~~~~  +\tilde f_4\tilde F^4\Big],  \\ 
\lb{eq2.22bB}
\tilde A'' &=&  \tilde{\mathcal{A}} \left(\tilde A, \tilde A', \tilde F, \tilde F', r, \tilde c_i\right) \nb\\
&=&  \frac{1}{2r^2\tilde A^2\tilde {\cal{D}}}\Big[\tilde a_0+\tilde a_1\tilde F+\tilde a_2\tilde F^2+\tilde a_3\tilde F^3\Big],  \\
\lb{eq2.22cB}
\frac{\tilde B'}{\tilde B} &=& \tilde{\mathcal{B}} \left(\tilde A, \tilde A', \tilde F, \tilde F', r, \tilde c_i\right) \nb\\
&=& \frac{1}{2r\tilde A^2\tilde {\cal{D}}}\Big[\tilde b_0+\tilde b_1\tilde F+\tilde b_2\tilde F^2\Big], 
\eqn
and
\bq
\lb{eq2.23aB}
 {\tilde C}^{\tilde v}  \equiv \tilde n_0+\tilde n_1 \tilde F 
+\tilde n_2\tilde F^2 = 0,
\eq
where   
\bqn
\lb{eq2.25aaa}
\tilde{\cal{D}}(r)  &\equiv& \tilde d_{-}\left(\tilde J^2(r) + 1\right) + {2} \tilde d_{+} \tilde J(r), \nb\\
 \tilde J(r) &\equiv& \tilde F(r) \tilde A^2(r), \nb\\
\tilde d_{\pm} &\equiv& (\tilde c_S^2 \pm 1)\tilde c_{14}(1- \tilde c_{13})(2+\tilde c_{13}+3\tilde c_2). 
\eqn
The coefficients $ \tilde f_n, \;  \tilde a_n, \; \tilde b_n $ and $ \tilde n_n$ are given by $f_n, \;  a_n,\; b_n $ and $n_n$  after the replacement 
(\ref{eq2.25}) is carried out.

Then,  the metric and spin-0 horizons for $\tilde g_{\mu\nu}$  are given, respectively, by 
\bqn
\lb{eq2.25a}
&& \tilde g_{\alpha\beta} \tilde N^{\alpha} \tilde N^{\beta} = 0, \\
\lb{eq2.25b}
&& \tilde g^{(S)}_{\alpha\beta} \tilde N^{\alpha} \tilde N^{\beta} = 0,
\eqn
 where $\tilde N_{\alpha} = (\partial x^{\mu}/\partial \tilde x^{\alpha}) N_{\mu}
=  {\delta^r_{\tilde \alpha}=\delta^r_{ \alpha}}$ and 
\bq
\lb{ew2.25c}
\tilde g^{(S)}_{\alpha\beta} \equiv \tilde g_{\alpha\beta} - \left(\tilde{c}_S^2 -1 \right) \tilde u_{\alpha} \tilde u_{\beta}.
\eq
In terms of $\tilde F$ and $ \tilde A$, Eqs.(\ref{eq2.25a})  {and (\ref{eq2.25b})} becomes, 
\bqn
\lb{MHt}
&&  \tilde F( \tilde r_{MH})  = 0,\\
\lb{S0Ht}
&&  \left( \tilde c_S^2-1\right) \left( \tilde J( \tilde r_{S0H})^2 + 1\right) +2 \left( \tilde c_S^2+1\right)  \tilde J( \tilde r_{S0H}) =0, \nb\\
\eqn 
where $r =  \tilde r_{MH}$ and $r =  \tilde r_{S0H}$ are respectively the locations of the metric and spin-0 horizons for the metric $\tilde g_{\mu\nu}$. Similarly, at $r =  \tilde r_{S0H}$ we  have
\bq
\lb{eq2.26}
\tilde{\cal{D}}(\tilde r_{S0H}) = 0.
\eq

Comparing the field equations given in this subsection with the corresponding ones given in the last subsection, we  see that we can get one set from the other simply by the replacement
(\ref{eq2.25}).

In addition, in terms of $\hat g_{\alpha\beta}$ and $N_{\alpha}$, Eqs.(\ref{eq2.25a}) and (\ref{eq2.25b}) reduce, respectively, to
\bqn
\lb{eq2.25aa}
&& \hat g_{\alpha\beta}  N^{\alpha}  N^{\beta} = 0, \\
\lb{eq2.25bb}
&& \hat g^{(S)}_{\alpha\beta}  N^{\alpha} N^{\beta} = 0.
\eqn
Since $\tilde r = r$, we find that
\bq
\lb{eq2.25cc}
\tilde r_{MH} = \hat r_{MH}, \quad \tilde r_{S0H} = \hat r_{S0H},
\eq
where $\tilde r_{MH}$ and $\tilde r_{S0H}$ ($\hat r_{MH},  \hat r_{S0H}$)
are the locations of the metric and spin-0 horizons of the metric $\tilde g_{\alpha\beta}$ ($\hat g_{\alpha\beta}$). The above analysis shows that {\it  these horizons determined
by $\tilde g_{\alpha\beta}$ are precisely equal to those determined by $\hat g_{\alpha\beta}$}.

\subsection{ $\sigma = c_S^2$}  

To solve Eqs.(\ref{eq2.22aB}) - (\ref{eq2.23aB}), we take the advantage of the choice $\sigma = c_S^2$, so that the speed of the spin-0 mode of the metric $\hat g_{\mu\nu}$ becomes 
unity, i.e., $\hat c_S = 1$. Since $\tilde c_i = \hat c_i$, we also have $\tilde c_S  = \hat c_S  = 1$. Then, from Eq.(\ref{eq1.1}) we find that this leads to, 
\bq
\lb{eq2.27}
\tilde c_2=\frac{2\tilde c_{14} -2\tilde c_{13}- \tilde c_{13}^2\tilde c_{14}}{2-4\tilde c_{14}+3 \tilde c_{13}\tilde c_{14}}.
\eq
For such a choice, from Eq.(\ref{eq2.25aaa}) we find that $\tilde d_{-} = 0$, and
\bq
\lb{eq2.28}
\tilde{\cal{D}}(r)   =  {2} \tilde d_{+} \tilde J(r) = {2} \tilde d_{+}  \tilde A^2(r) \tilde F(r).
\eq
 Then, Eq.(\ref{eq2.26}) yields $\tilde{F}(\tilde r_{S0H}) = 0$, since $\tilde{A} \not= 0$, which also represents the location  of the MH, defined by Eq.(\ref{MHt}). Therefore, for the choice $\sigma = c_S^2$
the  MH coincides with the  S0H for the effective metric $\tilde{g}_{\mu\nu}$, that is, 
\bq
\lb{eq2.29}
 \tilde r_{MH} = \tilde r_{S0H}, \; \left(\sigma = c_S^2\right).
\eq
As shown below, this will significantly simplify our computational labor. In particular, if we choose this surface as our initial moment, it will reduce the phase space of initial data from 4 dimensions
to one dimension only. 

For $\tilde c_S=1$, Eqs.(\ref{eq2.22aB})-(\ref{eq2.23aB}) reduce to, 
\bqn
\lb{eq2.30a}
&& \tilde F'' =  \frac{1}{ {4} \tilde d_+ r^2\tilde A^6}\left(\frac{\tilde f_0}{\tilde F} + \tilde f_1 + \tilde f_2 \tilde F + \tilde f_3 \tilde F^2 \right.\nb\\
&&~~~~~~~~~~~~~~~~~~~~~~~~ \left.+\tilde f_4 \tilde F^3\right), \\
\lb{eq2.30b}
&& \tilde A'' = \frac{1}{ {4}\tilde d_+ r^2 \tilde A^4}\left( \frac{\tilde a_0}{\tilde F} + \tilde a_1 + \tilde a_2 \tilde F +  \tilde a_3 \tilde F^2\right),~~~\\
\lb{eq2.30c}
&& \frac{\tilde B'}{\tilde B}= \frac{1}{ {4}\tilde d_+ r\tilde A^4}  \left( \frac{\tilde b_0}{\tilde F} + \tilde b_1 + \tilde b_2 \tilde F\right),\\
\lb{eq2.30d}
 &&  \tilde n_0+\tilde n_1 \tilde F  +\tilde n_2\tilde F^2 = 0.
\eqn

As shown previously, among these four equations, only three of them  are  independent, and our strategy in this paper is to take Eqs.(\ref{eq2.30a}), (\ref{eq2.30b}) and (\ref{eq2.30d}) as the three independent
equations. The advantage of this approach is that Eqs.(\ref{eq2.30a}), (\ref{eq2.30b}) are independent of $\tilde{B}(r)$, and Eq.(\ref{eq2.30d}) is a quadratic polynomial  of $\tilde{B}(r)$. So, we can 
solve Eqs.(\ref{eq2.30a}), (\ref{eq2.30b}) as the initial value problem first to find $\tilde{F}(r)$ and $\tilde{A}(r)$, and then insert them into  Eq.(\ref{eq2.30d}) to obtain directly  $\tilde{B}(r)$, as explicitly given by 
Eq.(\ref{Bcv}), after taking  the replacement (\ref{eq2.25}) and the choice of $\tilde c_2$ of Eq.(\ref{eq2.27}) into account.
 
From  Eqs.(\ref{eq2.30a}) and (\ref{eq2.30b}) we can see that they become singular at $r = \tilde r_{S0H}$ (Recall $\tilde{F}( \tilde r_{S0H}) = 0$), unless 
$\tilde  f_0(\tilde r_{S0H}) =  \tilde a_0(\tilde r_{S0H}) =0$. 
As can be seen from the expressions of $ f_0(r), \;  a_0( r)$ given in Appendix A,   $\tilde f_0(\tilde r_{S0H}) = \tilde  a_0(\tilde r_{S0H}) = 0$ imply $\tilde b_0(\tilde r_{S0H}) = 0$. Therefore, 
to have the field equations regular across the S0H, we must require 
$\tilde b_0(\tilde  r_{S0H}) =0$. It is interesting that this is also the condition for Eq.(\ref{eq2.30c}) to be non-singular across the S0H. 
In addition, using the gauge residual (\ref{eq2.22}), we shall set $\tilde B_H  = 1$, so Eq.(\ref{eq2.30d}) [which can be written in the form of Eq.(\ref{Bcv}), after the replacement (\ref{eq2.25})] will provide a constraint among the initial values of $\tilde F'_H$, $\tilde A_H$ and $\tilde A'_H$,
where $\tilde F'_{H} \equiv \tilde F'(\tilde r_{S0H})$ and so on. 
  In summary, on the S0H we have the following
\bqn
\lb{eq2.32a}
&& \tilde F_H = 0, \\
\lb{eq2.32c}
&& \tilde b_0\left(\tilde A_H, \tilde A'_H, \tilde F'_H, \tilde r_{S0H}\right) = 0, \\   
\lb{eq2.32d}
&&\tilde B_H = 1.
\eqn
From the expression for $\tilde  b_0$ given in Appendix A, we can see that Eq.(\ref{eq2.32c}) is quadratic in $\tilde A'_H$, and solving  it on the S0H, in general 
we obtain two solutions,
\bq
\lb{eq2.33}
\tilde {A'}^{\pm}_H =\tilde {A'}^{\pm}_H\left( \tilde A_H, \tilde F'_H, \tilde r_{S0H}\right).  
\eq
Then, inserting it, together with Eqs.(\ref{eq2.32a}) and (\ref{eq2.32d}), into Eq.(\ref{eq2.23a}),  we get  
\bq
\lb{eq2.34}
\tilde n^{\pm}_0\left(\tilde A_H, \tilde F'_H, \tilde r_{S0H}\right) = 0,
\eq
where the ``$\pm$" signs correspond to the choices of $\tilde {A'}_H = \tilde {A'}^{\pm}_H$.  
In general, Eq.(\ref{eq2.34})  is a fourth-order polynomial of $\tilde F'_H$,
so it normally has four roots,  denoted as
\bq
\lb{eq2.35a}
\tilde {F'}^{(\pm,n)}_H=\tilde {F'}^{(\pm,n)}_H\left(\tilde  A_H, \tilde r_{S0H}\right),
\eq
where $n=1, 2, 3, 4$.
For each given $\tilde {F'}^{(\pm,n)}_H$, substituting it into Eq.(\ref{eq2.33}) we find a corresponding $\tilde {A'}^{(\pm,n)}_H$, given by
\bq
\lb{eq2.35b}
\tilde {A'}^{(\pm,n)}_H=\tilde {A'}^{(\pm,n)}_H\left( \tilde A_H, \tilde r_{S0H}\right).
\eq
Thus, once $\tilde A_H$ and $\tilde r_{S0H}$ are given, the quantities $\tilde {F'}^{(\pm,n)}_H$ and $\tilde {A'}^{(\pm,n)}_H$ are uniquely determined from Eqs.(\ref{eq2.35a}) and (\ref{eq2.35b}). For each set of { ($\tilde A_H, \tilde r_{S0H}$),} in general there are eight sets of $\left(\tilde A'_H, \tilde F'_H\right)$.
 
 If we choose $r = \tilde r_{S0H}$ as the initial moment, such obtained $\left(\tilde A'_H, \tilde F'_H\right)$, together with $\tilde F_H = 0$, and a proper choice of $\tilde A_H$, can be considered as the initial conditions for the differential equations (\ref{eq2.30a}) and (\ref{eq2.30b}).
 
 However,  it is unclear which one(s) of these eight sets of initial conditions will lead to asymptotically flat solutions, except that the one with $\tilde F'_H < 0$, which can be discarded immediately, as it would lead to
 $\tilde F = 0$ at some radius $r > \tilde r_{S0H}$, which is inconsistent with our assumption that $r = \tilde r_{S0H}$ is the location of the S0H  \cite{Enrico11}. So,  in general what one  needs to do is to try all the possibilities.

 Therefore, {\it if we choose $r =\tilde  r_{S0H}$ as the initial moment,  the four-dimensional phase space of the initial conditions,
 $\left(\tilde F_H, \tilde F'_H,\tilde  A_H, \tilde A'_H \right)$,  reduces to one-dimensional, spanned by $\tilde A_H$ only}.
 
 In the following, we shall show further that $\tilde r_{S0H}$ can be chosen arbitrarily.  In fact, introducing the dimensionless quantity,
 $\xi \equiv \tilde r_{S0H}/r$, we find that Eqs. (\ref{eq2.30a}) - (\ref{eq2.30c}) and (\ref{eq2.23aB}) can be written in the forms, 
 \bqn
\lb{eq2.36a}
&& \frac{d^2 \tilde F(\xi)}{d\xi^2} =  {\cal{G}}_1\left(\xi, \tilde c_i\right),\\
\lb{eq2.36b}
&& \frac{d^2 \tilde A(\xi)}{d\xi^2} =  {\cal{G}}_2\left(\xi, \tilde c_i\right),\\
\lb{eq2.36c}
&& \frac{1}{\tilde B(\xi)}\frac{d\tilde B(\xi)}{d\xi} =  {\cal{G}}_3\left(\xi, \tilde c_i\right),\\
\lb{eq2.36d}
&& C^{\tilde v} \left(\tilde A(\xi), \tilde A'(\xi), \tilde F(\xi), \tilde F'(\xi), \tilde B(\xi), \xi, \tilde c_i\right)= 0,~~~~~ 
\eqn
where ${\cal{G}}_i$'s are all independent of $\tilde r_{S0H}$,  $C^{\tilde v} \equiv r_{S0H}^2 \tilde C^{\tilde v}$, and  the primes  in the last equation stand for the derivatives respect to $\xi$. 
Therefore, Eqs.(\ref{eq2.36a})-(\ref{eq2.36d}), or equivalently,   Eqs.(\ref{eq2.22aB})-(\ref{eq2.23aB}), are  scaling-invariant and independent of $\tilde r_{S0H}$. 
Thus, without  loss of the generality, we can always set
\bq
\lb{eq2.37}
\tilde r_{S0H} = 1,
\eq
which does not affect Eqs.(\ref{eq2.36a}) - (\ref{eq2.36d}), and also explains  the reason why in  \cite{Eling2006-2,Enrico11} the authors set $\tilde r_{S0H} = 1$ directly.  
At the same time, it should be noted that once $\tilde r_{S0H}=1$ is taken, it implies that the unit of length is fixed. 
For instance,  if we have a BH with $\tilde r_{S0H}=1$ km, then setting $\tilde r_{S0H}=1$ means the unit of length is in km.

Once $\tilde A_H$ is chosen, we can integrate    Eqs.(\ref{eq2.36a}) and (\ref{eq2.36b}) in both directions to find $ \tilde F(\xi)$ and $ \tilde A(\xi)$,
 one is toward the center, $\xi =\tilde r_{S0H}/r =  \infty$,  in which we have $\xi \in [1, \infty)$, 
and the other is toward infinity, $\xi = \tilde r_{S0H}/r = 0$, in which we have $\xi \in (0, 1]$. Then, from
 Eq.(\ref{Bcv}) we   can find $\tilde B(\xi)$ uniquely,  after the replacement of Eq.(\ref{eq2.25}).  Again, to have a proper asymptotical behavior of
$\tilde B(r)$, the ``+" sign will be chosen.

At the spatial infinity $\xi = \tilde r_{S0H}/r \rightarrow 0$,   we  require that the spacetime be asymptotically flat, that is \cite{Eling2006-2,Enrico11} \footnote{Note that 
in \cite{Eling2006-2,Enrico11} a factor $1/2$ is missing in front of $A_2$ in  the expression of $A(x)$.},
\bqn
\label{eq2.38}
\tilde F(\xi)&=& 1+\tilde F_{1} \xi +\frac{1}{48} \tilde c_{14} \tilde F_{1}^{3} \xi^{3}+\cdots,  \nb\\
\tilde A(\xi)&=& 1-\frac{1}{2} \tilde F_{1} \xi +\frac{1}{2} \tilde A_{2} \xi^{2} -\Bigg(\frac{1}{96} \tilde c_{14} \tilde F_{1}^{3}\nb\\
&& -\frac{1}{16} \tilde F_{1}^{3}+ { \frac{1}{2} \tilde F_{1} \tilde A_{2}}\Bigg) \xi^{3}+\cdots,  \nb\\
\tilde B(\xi)&=& 1+\frac{1}{16} \tilde c_{14} \tilde F_{1}^{2} \xi^{2}-\frac{1}{12} \tilde  c_{14} \tilde F_{1}^{3} \xi^{3}+\cdots, 
\eqn
where   $\tilde F_{1}\equiv \tilde  F^{\prime}(\xi=0)$ and $\tilde A_{2}\equiv \tilde A^{\prime \prime}(\xi=0)$.
 
 It should be noted that   the Minkowski spacetime   is given by
\bqn
\label{Finfty}
\tilde F =  \tilde F_{M}, ~~\quad \tilde A = \frac{1}{\sqrt{\tilde F_{M}}}, ~~\quad \tilde B = \sqrt{\tilde F_{M}},~~~~~~~
\eqn
 where $\tilde F_{M}$ is a positive otherwise arbitrary constant. 
Therefore, in the asymptotical expansions of Eq.(\ref{eq2.38}), we had set $\tilde F_{M}=1$ at the zeroth order of $\xi$. However, the  initial conditions imposed at $r = \tilde r_{S0H}$ given above usually leads to $\tilde F_{M} \not= 1$, even for spacetimes that are asymptotically flat. Therefore, we need first to use the gauge residual  (\ref{eq2.22})  to bring $\tilde F(\xi=0)=
\tilde A(\xi=0) = \tilde{B}(\xi=0) = 1$, before using  Eq.(\ref{eq2.38})  to calculate the constants $\tilde A_2$ and $\tilde F_1$.

From the above analysis we can see that {\it finding spherically symmetric solutions of the $\ae$-theory now reduces to finding the initial condition $\tilde A_H$ that leads to the asymptotical behavior 
(\ref{eq2.38}), for a given set of $c_i$'s}.

Before proceeding to the next section, we would like to recall that when $\sigma = c_S^2$, we have $g^{(S)}_{\alpha\beta} = \hat g_{\alpha\beta}$, as shown by Eq.(\ref{eq2.7b}). That is, the S0H for the
metric $g_{\alpha\beta}$ now coincides with the MH of $\hat g_{\alpha\beta}$. With this same very choice, $\sigma = c_S^2$, the MH for $\hat g_{\alpha\beta}$ also coincides with its 
S0H. Thus, we have
\bqn
\label{eq2.39}
r_{S0H} = \hat r_{S0H}  =  \hat r_{MH} =  \tilde r_{S0H}  =  \tilde r_{MH} \equiv r_H,\; \left(\sigma = c_S^2\right). \nb\\
\eqn
It must be noted that $ r_{H}$ defined in the last step denotes the location of the S0H of $g_{\alpha\beta}$, which is usually different from its MH,  defined by
\bqn
\label{eq2.40}
\left. g_{\alpha\beta} N^{\alpha} N^{\beta} \right|_{r = r_{MH}} = 0,  
\eqn
since in general we have $c_S \not= 1$, so $g^{(S)}_{\alpha\beta} \equiv g_{\alpha\beta} - \left(c_S^2 -1\right) u_{\alpha} u_{\beta} 
\not = g_{\alpha\beta}$.  As a result, we have $r_{MH} \not= r_{S0H}$ for $c_S \not= 1$. 

However, it is worth emphasizing again that,  for the choice $\sigma = c_S^2$
we have $\tilde c_S = \hat c_S = 1$, so the  metric and spin-0 horizons of both $\hat g_{\alpha\beta}$ and $\tilde g_{\alpha\beta}$ all coincide, and
 are given by the same $r_H$, as explicitly shown by Eq.(\ref{eq2.39}). More importantly, it is also the location of the S0Hs of the metric $g_{\alpha\beta}$.


\section{Numerical Setup and Results}
 \renewcommand{\theequation}{4.\arabic{equation}} \setcounter{equation}{0}

\subsection{General Steps}  

It is difficult to find  analytical solutions to Eqs.(\ref{eq2.36a})-(\ref{eq2.36d}). Thus, in this paper we are going to solve them numerically, using the shooting method, with the asymptotical conditions (\ref{eq2.38}).  
  In particular, our strategy is the following:
  
 (i) Choose a set of {physical} $c_i$'s satisfying the constraints (\ref{c1234a})-(\ref{c1234bb}), 
and then calculate the corresponding $\tilde{c}_i$'s with $\sigma = {c}_S^2$.

 (ii) Assume that for such chosen $c_i$'s the corresponding solution possesses  a S0H located at $r = r_H$, and then follow the analysis given in the last section to impose the conditions 
$\tilde F_H=0$ and $\tilde B_H=1$.

(iii) Choose   a test value for $\tilde A_H$, and then solve Eq.(\ref{eq2.32c}) for $\tilde A'_H$ in terms of $\tilde F'_H$ and $ \tilde A_H$, i.e., $\tilde A'_H=\tilde A'_H(\tilde F'_H, \tilde A_H)$.

(iv) Substitute $\tilde A'_H$ into Eq.(\ref{eq2.34}) to obtain a quartic equation for $\tilde F'_H$ and then solve it to find $\tilde F'_H$.

(v) With the initial conditions $\{\tilde F_H, \tilde A_H, \tilde F'_H, \tilde A'_H\}$, {integrate Eqs.(\ref{eq2.36a}) and (\ref{eq2.36b})} from $\xi=1$ to $\xi=0$. 

However, since the field equations are singular at $\xi = 1$,   
we will actually integrate these equations 
from $\xi=1-\epsilon$ to $\xi \simeq 0$, where $\epsilon$ is a very small quantity. To obtain the values of $\digamma(\xi)$ at $\xi=1-\epsilon$, we first Taylor expand them in the form,
\bq
\lb{eq4.1}
 {\digamma (1-\epsilon) =  \sum_{k=0}^2  \frac{\digamma^{(k)}|_{\xi=1}}{k!} (-1)^k \epsilon^k+ {\cal{O}}\left(\epsilon^3\right), }
\eq
{where $\digamma  \equiv \left\{\tilde A, \tilde A', \tilde F, \tilde F' \right\}$} and $\digamma^{(k)} \equiv d^k\digamma/d\xi^k$. For each $\digamma$, 
we shall expand it to  {the second order of $\epsilon$, so the errors are of the order $\epsilon^3$}. 
Thus, if we choose $\epsilon = 10^{-14}$, the errors in the initial conditions $\digamma (1-\epsilon)$ is of the order   {$10^{-42}$}. 
 For $\digamma = \tilde A, \; \tilde F$, we already obtained
 $\digamma(1)$  and $\digamma'(1)$ from the initial conditions. In these cases, to get  $\tilde A''(1)$ and $\tilde F''(1)$, we use the field equations (\ref{eq2.36a}) and (\ref{eq2.36b}) and L'Hospital's rule.
 On the other hand, for $\digamma=\tilde A'$, expanding  it to  {the second order} of $\epsilon$, we have 
 \bq
\lb{eq4.1}
\tilde A'(1-\epsilon) =  \tilde A'(1) - \tilde A''(1) \epsilon + \frac{1}{2}\tilde A^{(3)}(1) \epsilon^2 + {\cal{O}}\left(\epsilon^3\right), 
\eq
{where $\tilde A^{(3)}(1) \equiv \left. d^3\tilde A(\xi)/d\xi^3\right|_{\xi =1}$ can be obtained by first taking the derivative of Eq.(\ref{eq2.36b}) and then taking the limit $\xi \rightarrow 1$, as now we have already known $\tilde A(1), \tilde A'(1), \tilde A''(1), \tilde F(1), \tilde F'(1)$ and $\tilde F''(1)$.} Similarly, for $\digamma=\tilde F'$, from Eq.(\ref{eq2.36a}) we can find $\tilde F^{(3)}(1)$.
 
(vi) Repeat (iii)-(v) until a numerical solution matched to Eq.(\ref{eq2.38}) is obtained, by choosing different  values of $\tilde A_H$ with a bisectional search. Clearly, once such a value of $\tilde A_H$ is found,
it means that we obtain numerically an asymptotically flat solution of the Einstein-aether field equations outside the S0H. Note that, to guarantee that Eq.(\ref{eq2.38}) is satisfied, the normalization of $\{\tilde F, \tilde A, \tilde B\}$ need to be done according to Eq.(\ref{Finfty}), by using the remaining gauge residual  of Eq.(\ref{eq2.22}).

(vii) To obtain the solution in the internal region $\xi \in (1, \infty)$, we simply {integrate Eqs.(\ref{eq2.36a}) and (\ref{eq2.36b})} from $\xi=1$ to  {$\xi \to \infty$} with the same value
of  $\tilde A_H$ found in the last step.  As in the region $\xi \in (0, 1)$, we can't really set the ``initial'' conditions precisely at 
$\xi = 1$. Instead,  we will integrate them from $\xi=1+\epsilon$ to $\xi=\xi_{\infty}\gg 1$. The initial values at $\xi = 1+\epsilon$ can be obtained by following what we did in 
Step (v), that is, Taylor expand $\digamma(\xi)$  {at $\xi=1+\epsilon$}, and then use the field equations to get all the quantities up to the  {third-order} of $\epsilon$.

(viii) Matching the results obtained from steps (vi) and (vii) together,  we finally obtain a solution of { {$\{\tilde F(\xi), \tilde A(\xi)\}$}} on the whole spacetime $\xi \in (0, \infty)$  (or $r\in (0, \infty)$).

(ix) Once $\tilde{F}$ and $\tilde{A}$ are known, from Eq.(\ref{Bcv}), we can calculate $\tilde{B}$, so that an asymptotically flat black hole solution for $\{\tilde{A}, \tilde{B}, \tilde{F} \}$ is finally obtained over the whole space $r \in (0, \infty)$.

Before proceeding to the next subsection to consider the physically allowed region of the parameter space of $c_i$'s, let us first reproduce the results presented in Table I of \cite{Enrico11}, in order to check our numerical code, although all these choices  have been ruled out currently by observations \cite{OMW18}. To see this explicitly, let us first note that the parameters chosen in \cite{Eling2006-2,Enrico11} correspond to
\bqn
\lb{eq4.1a}
\hat{c}_2 = -\frac{\hat{c}_1^3}{3 \hat{c}_1^2 - 4 \hat{c}_1 + 2}, \quad
  		\hat{c}_3 = 0 = \hat{c}_4 = 0,
\eqn
so that now only $\hat{c}_1$ is a free parameter. With this choice of $\hat c_i$'s, the corresponding $c_i$'s can be obtained from Eqs.(\ref{eq2.8}) with $\sigma=c_S^2$, which are given by, 
\bqn
\lb{eqc14}
c_{14}&=&\hat c_1,\nb\\
c_2&=&\frac{-2c_{13}+2\hat{c}_1+2c_{13}\hat{c}_1-2\hat{c}_1^2-c_{13}\hat{c}_1^2}{2-4\hat{c}_1+3\hat{c}_1^2},
\eqn
where $c_{13}$ is arbitrary. This implies that Eqs.(\ref{eq2.8}) are degenerate for the choices of  Eqs.(\ref{eq4.1a}). It can be seen from Eq.(\ref{eqc14}), in all the cases considered in \cite{Enrico11}, we have  $c_{14} > 2.5 \times10^{-5}$. Hence all the cases considered in  \cite{Eling2006-2,Enrico11} do not satisfy the current constraints  \cite{OMW18}.

With the above in mind, we reproduce all the cases considered in  \cite{Eling2006-2,Enrico11}, including the ones {with $\tilde c_1 > 0.8$.} Our results are presented in Table \ref{table1}, where
\bqn
\lb{eq4.1b}
{\tilde \gamma}_{ff} &\equiv& \tilde u^{\alpha} u^{\text{obs}}_{\alpha}, \\
\label{rg}
\tilde r_g &\equiv& -{r_{H}}\times \lim_{\xi \to 0}\frac{d \tilde F(\xi)}{d \xi} = 2 G_{\ae} M_{\text{ADM}},
\eqn
where $u^{\text{obs}}_{\alpha}$ is the tangent (unit) vector to a radial free-fall trajectory that starts at rest at spatial infinity, and 
$ M_{\text{ADM}}$ denotes the Komar mass, which is equal to the Arnowitt-Deser-Misner (ADM) mass in the spherically symmetric case for the metric $\tilde{g}_{\alpha\beta}$ \cite{Per12}. 

From Table \ref{table1} we can see that our results are exactly the same as those given in  \cite{Enrico11} up to the same accuracy. But, due to the improved accuracy of our numerical code,
for each of the physical quantity, we provided two more digits.

  \begin{table}[]
  	\caption{ The cases considered in  \cite{Eling2006-2, Enrico11} for various $\hat{c}_1$ with the choice of the parameters 
  		$\hat{c}_2$, $\hat{c}_3$ and $\hat{c}_4$ given by Eq.(\ref{eq4.1a}). Note that for each physical quantity, we have added two more digits,
		 due to the improved accuracy of our numerical code.}   
  	\label{table1}  
  	\begin{tabular}{| c c c c c c c |}
  		\hline 
  		$\hat c_1$    &\hspace{23pt}      & $\tilde r_g/r_H$      &\hspace{23pt} & $\tilde F'_H \tilde A_H^2$                  &\hspace{23pt} & ${\tilde \gamma}_{ff}$                 \\\hline
  		0.1           &\hspace{23pt}      & 0.98948936 &\hspace{23pt} & 2.0961175                     &\hspace{23pt} & 1.6028048                     \\[-8pt]
  		0.2           &\hspace{23pt}      & 0.97802140 &\hspace{23pt} & 2.0716798                     &\hspace{23pt} & 1.5769479                     \\[-8pt]
  		0.3           &\hspace{23pt}      & 0.96522924 &\hspace{23pt} & 2.0391972                     &\hspace{23pt} & 1.5476848                     \\[-8pt]
  		0.4           &\hspace{23pt}      & 0.95054650 &\hspace{23pt} & 1.9965155                     &\hspace{23pt} & 1.5140905                     \\[-8pt]
  		0.5           &\hspace{23pt}      & 0.93304411 &\hspace{23pt} & 1.9405578                     &\hspace{23pt} & 1.4748439                     \\[-8pt]
  		0.6           &\hspace{23pt}      & 0.91106847 &\hspace{23pt} & 1.8666845                     &\hspace{23pt} & 1.4279611                     \\[-8pt]
  		0.7           &\hspace{23pt}      & 0.88131278 &\hspace{23pt} & 1.7673168                     &\hspace{23pt} & 1.3702427                     \\[-8pt]
  		0.8           &\hspace{23pt}      & 0.83583029 &\hspace{23pt} & 1.6283356                     &\hspace{23pt} & 1.2959142                     \\[-8pt]
  		0.9           &\hspace{23pt}      & 0.74751927 &\hspace{23pt} & 1.4155736                     &\hspace{23pt} & 1.1921231                     \\[-8pt]
  		0.91          &\hspace{23pt}      & 0.73301185 &\hspace{23pt} & 1.3870211                     &\hspace{23pt} & 1.1790400                     \\[-8pt]
  		0.92          &\hspace{23pt}      & 0.71650458 &\hspace{23pt} & 1.3563710                     &\hspace{23pt} & 1.1652344                     \\[-8pt]
  		0.93          &\hspace{23pt}      & 0.69745439 &\hspace{23pt} & 1.3232418                     &\hspace{23pt} & 1.1506047                     \\[-8pt]
  		0.94          &\hspace{23pt}      & 0.67507450 &\hspace{23pt} & 1.2871125                     &\hspace{23pt} & 1.1350208                     \\[-8pt]
  		0.95          &\hspace{23pt}      & 0.64816499 &\hspace{23pt} & 1.2472379                     &\hspace{23pt} & 1.1183101                     \\[-8pt]
  		0.96          &\hspace{23pt}      & 0.61476429 &\hspace{23pt} & 1.2024805                     &\hspace{23pt} & 1.1002331                     \\[-8pt]
  		0.97          &\hspace{23pt}      & 0.57133058 &\hspace{23pt} & 1.1509356                     &\hspace{23pt} & 1.0804355                     \\[-8pt]
  		0.98          &\hspace{23pt}      & 0.51038168 &\hspace{23pt} & 1.0889067                     &\hspace{23pt} & 1.0583387                     \\[-8pt]
  		0.99          &\hspace{23pt}      & 0.41063001 &\hspace{23pt} & 1.0068873                     &\hspace{23pt} & 1.0328120                     \\\hline                    
  	\end{tabular}
  \end{table}

   \begin{figure}[htb]
   	\includegraphics[width=\columnwidth]{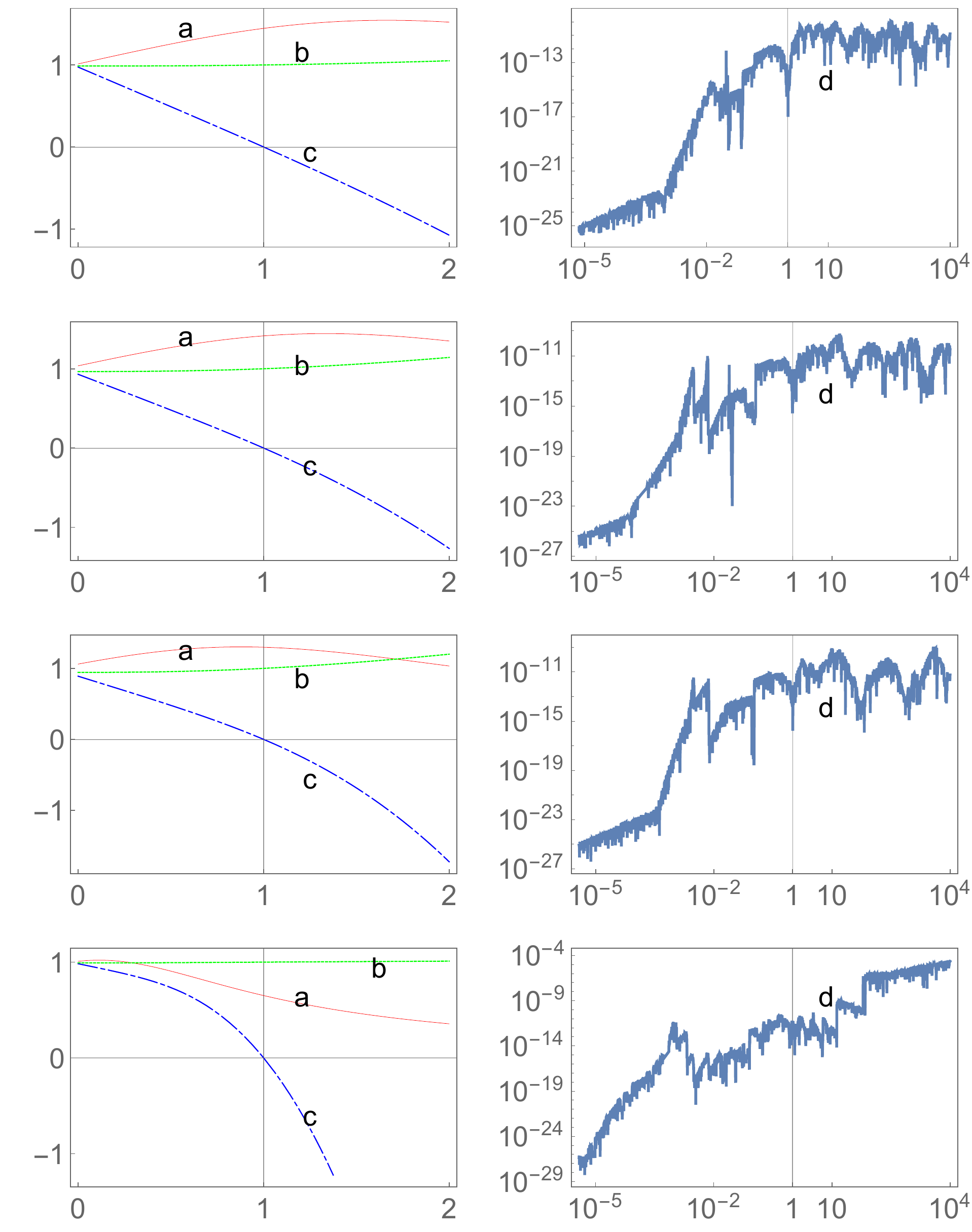} 
   	\caption{{In the above graphs, we use a, b, c and d to represent $\tilde A$, $\tilde B$, $\tilde F$ and $\tilde{\cal{C}}$. In each row,  $\hat{c}_1$ is chosen, respectively,
	 as $\hat{c}_1= 0.1, 0.3, 0.6, 0.99$, as  listed in Table \ref{table1}. The horizontal axis is $r_{H}/r$.}}
   	\label{FABEnrico}
   \end{figure}

 Additionally, in Fig. \ref{FABEnrico} we plotted the functions $\tilde F$, $\tilde B$, $\tilde A$ and $\tilde{\cal{C}}$ for four representative cases listed in Table \ref{table1} ($\hat c_1=0.1, 0.3, 0.6, 0.99$). Here, the quantity $\tilde {\cal{C}}$ is defined as
 \bq
 \lb{scC}
\tilde {\cal{C}} \equiv \left|\frac{d\ln\tilde{B}}{d\xi} - {\cal{G}}_3\right|,
 \eq
 which vanishes identically for the solutions of the field equations, as it can be seen from Eq.(\ref{eq2.36c}). In the rest of this paper, we shall use it to check the accuracy of our numerical code. 
 
 From Fig. \ref{FABEnrico}, we note that the properties of $\{\tilde F, \tilde A, \tilde B\}$ depend on the choice of $\hat{c}_1$. The quantity $\tilde{\cal{C}}$ is approximately zero within the whole integration range, which means
 that  our numerical  solutions are quite reliable.

  
  \subsection{Physically Viable Solutions with  S0Hs}

 With the above verification of our numerical code, we turn to the physically viable solutions of the Einstein-aether  field equations, in which {a S0H always exists}. Since $c_{13}$ is very small, 
 without loss of the generality, in this subsection we only consider the   cases with $c_{13}=0$.

 \begin{figure}[htb]
	\includegraphics[width=\columnwidth]{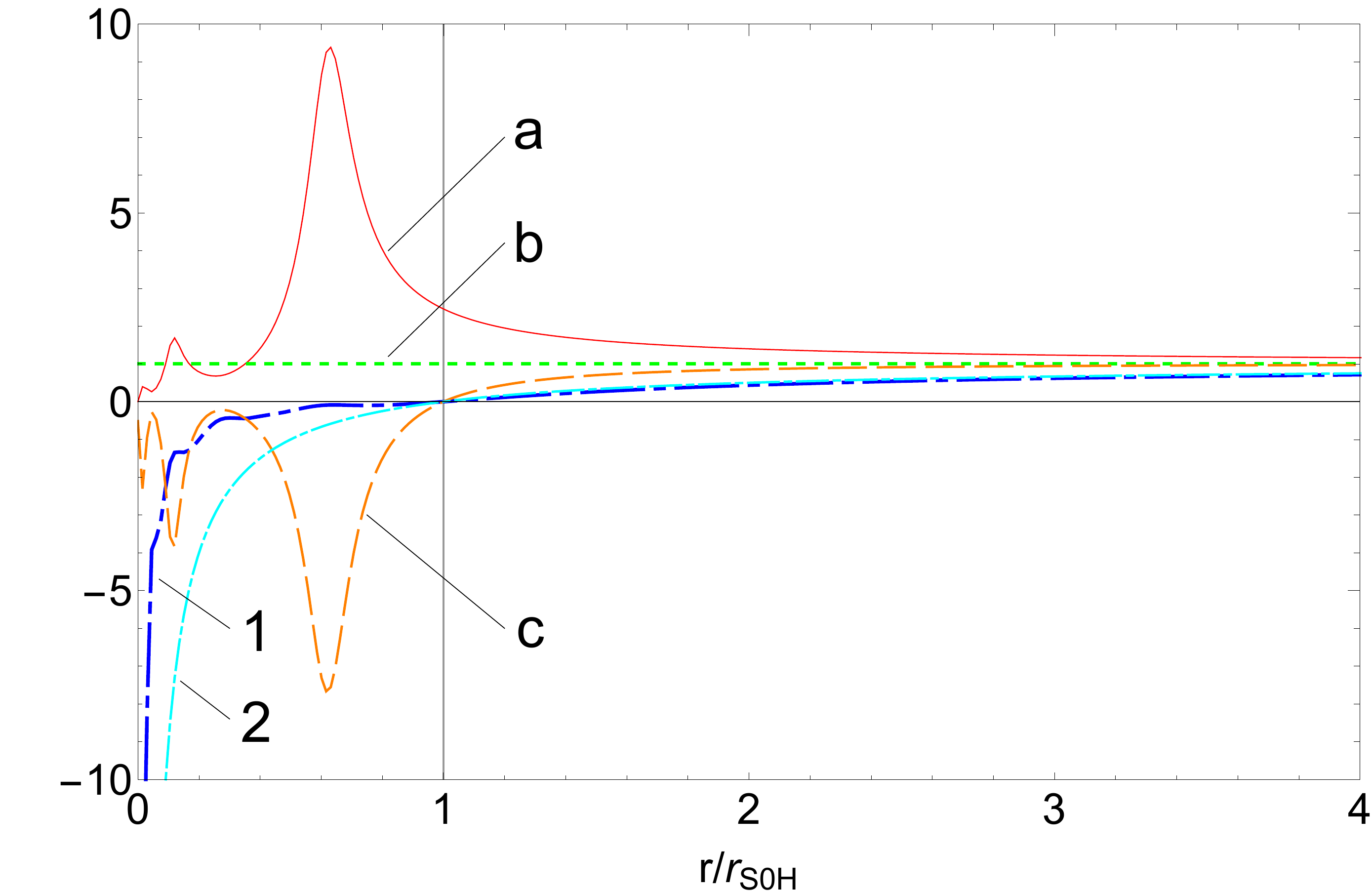} 
	\caption{ The solution for  $c_{14}=2\times 10^{-7}$, $c_2=9\times 10^{-7}$, and $c_3=-c_1$. { Here, $\tilde A$, $\tilde B$, $\tilde J$,  $\tilde F$ and $\tilde F^{GR}$  
	are represented by the red line (labeled by a), green line (labeled by b), orange line (labeled by c), blue line (labeled by 1) and cyan line (labeled by 2) respectively.}}
	\label{FABtilde3}
\end{figure}
 
 As the first example,  let us  consider   the case $c_{14}=2\times 10^{-7}$, $c_2=9\times 10^{-7}$, and $c_3=-c_1$, which satisfies the constraints (\ref{c1234}).
  Fig. \ref{FABtilde3} shows the functions ${\tilde F, \tilde A, \tilde B}$, in which  we also plot $\tilde J \equiv \tilde F \tilde A^2$ and the GR limit of $\tilde F$, {denoted by $\tilde F^{GR}$ with $\tilde F^{GR} \equiv 1-r_{H}/r$. }
  
     In plotting Fig. \ref{FABtilde3}, we chose   $\epsilon = 10^{-14}$. 
  With the shooting method,  $\tilde A_H$ is determined to be  $\tilde A_H \simeq 2.4558992$ \footnote{During the numerical calculations, we find that the asymptotical behavior (\ref{eq2.38}) of
  	the metric coefficients at $\xi \equiv r_H/r \simeq 0$ sensitively depends on the value of $\tilde A_H$. 
  	To make our results reliable, among all the steps in our codes, the precision is chosen to be not less than 37.}.
  In our calculations, we stop repeating the bisection search  for $\tilde A_H$, when
  the value $\tilde A_H$ giving an asymptotically flat solution is
  determined to within $10^{-23}$.
    Technically, these accuracies could be further  improved. However, for our current purposes, they are already sufficient. 
 
\begin{figure}
  	\begin{tabular}{cc}
  		\includegraphics[width=37mm]{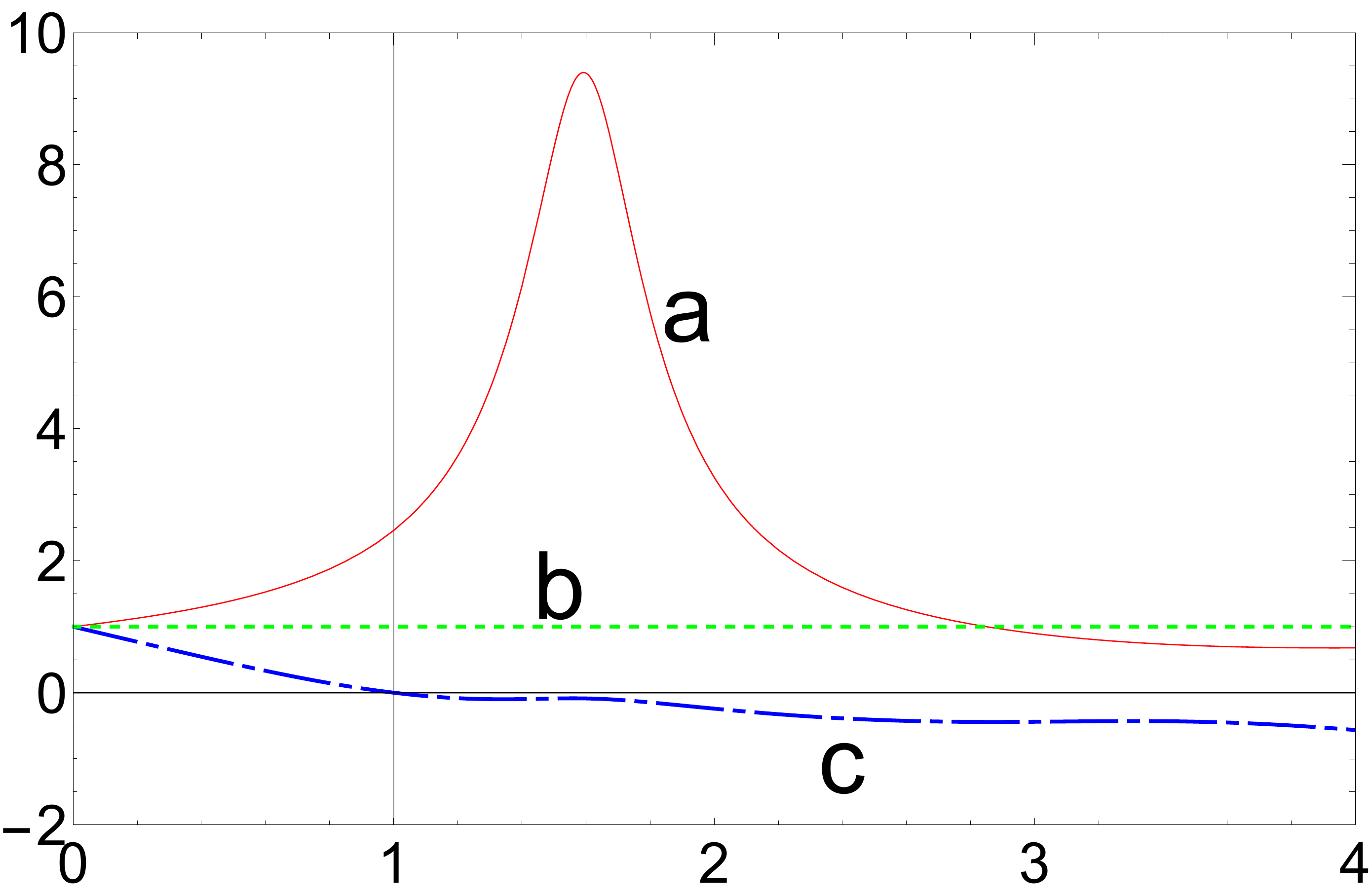} &   \includegraphics[width=37mm]{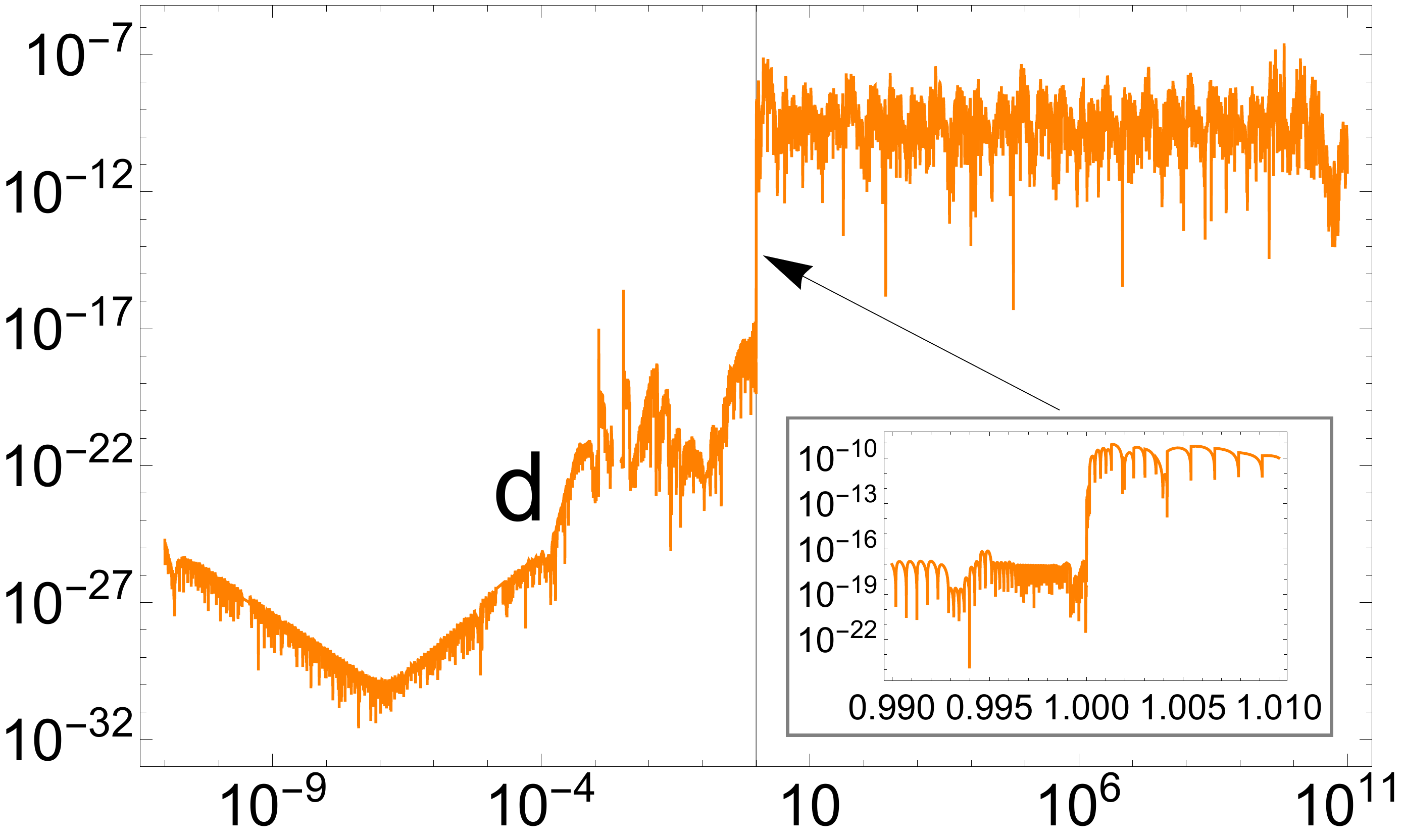} \\
  		(a)  & (b)  \\[6pt]
  		\includegraphics[width=37mm]{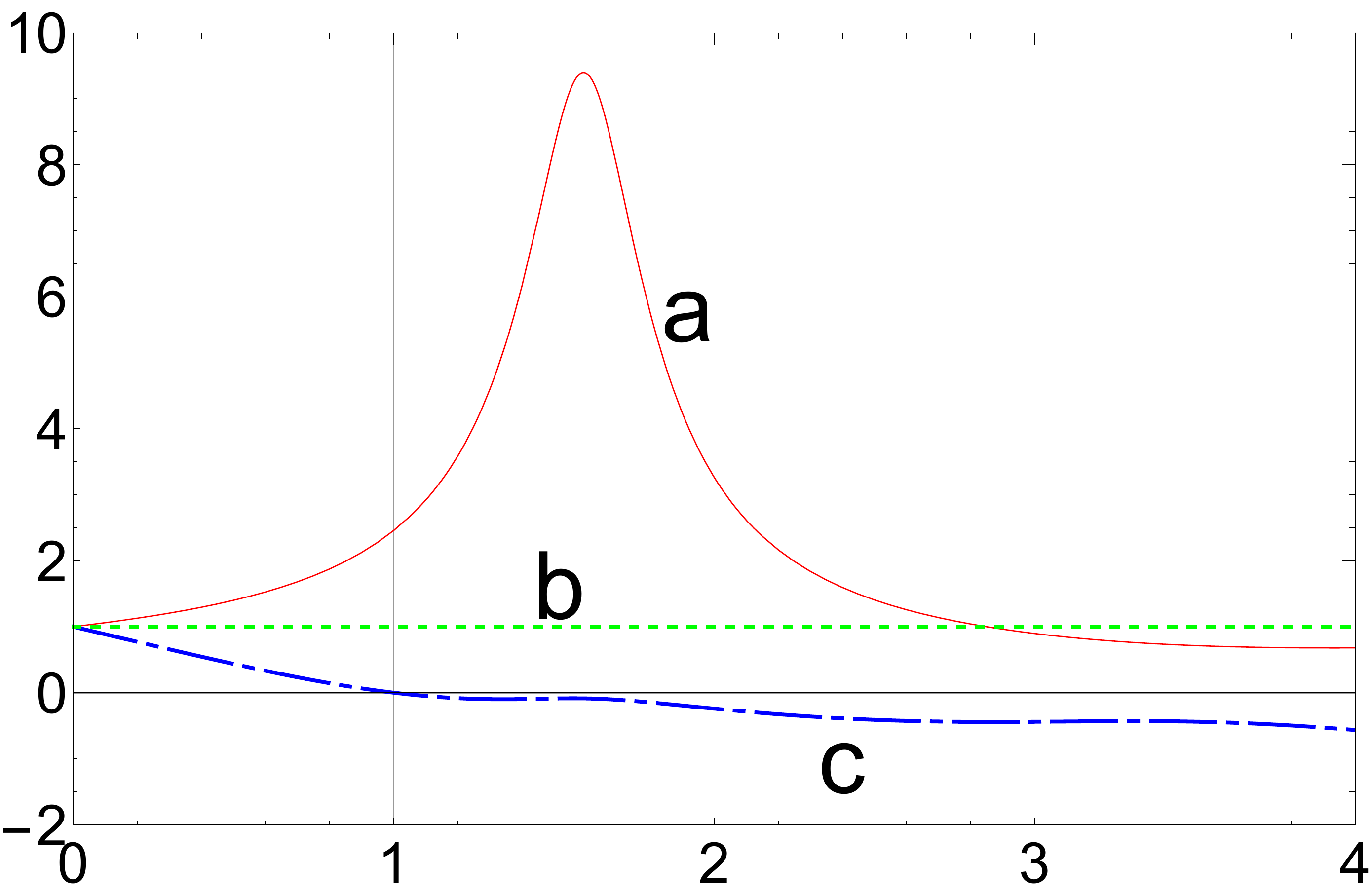} &   \includegraphics[width=37mm]{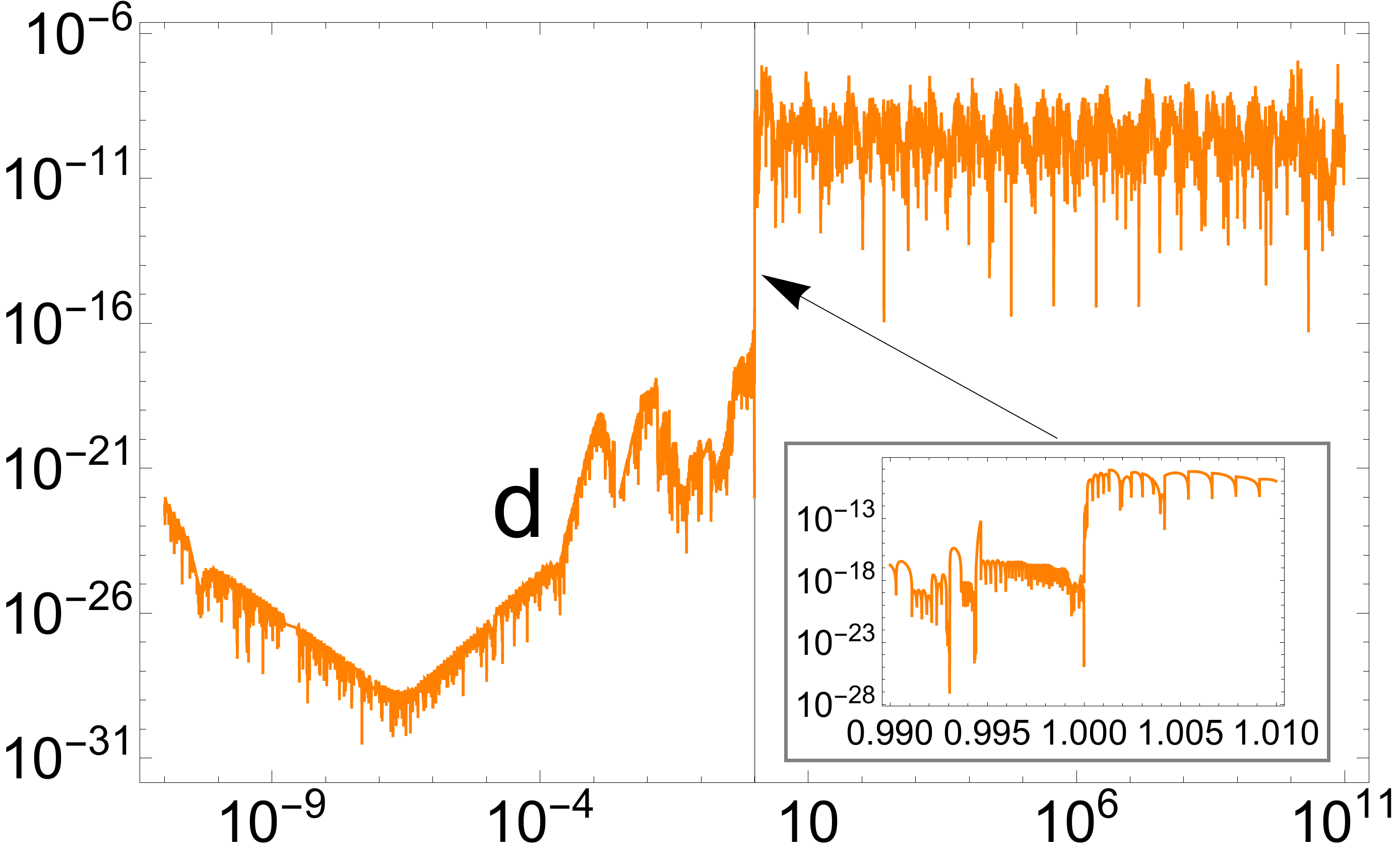} \\
  		(c) & (d) \\[6pt]
  		\includegraphics[width=37mm]{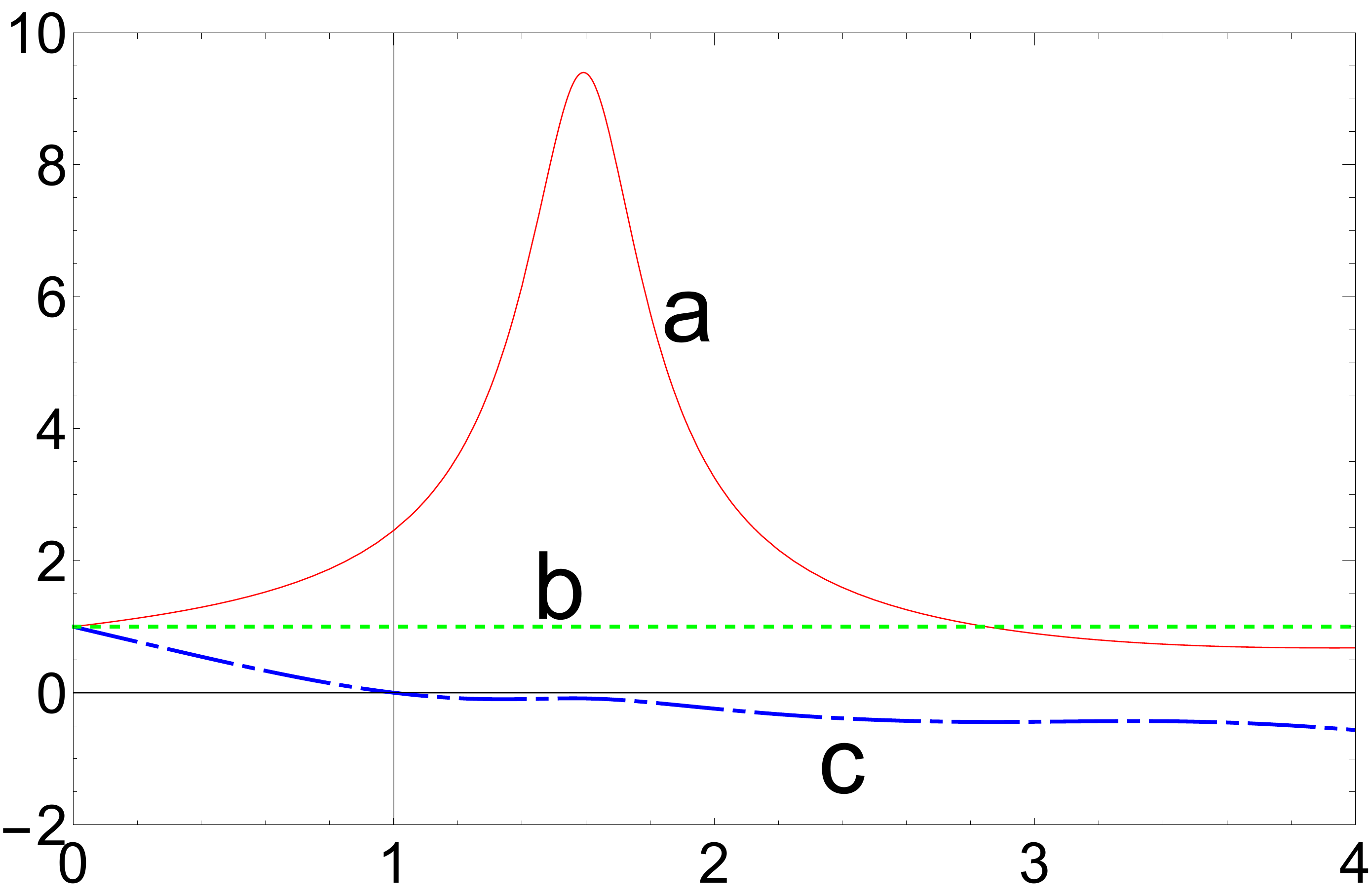} &   \includegraphics[width=37mm]{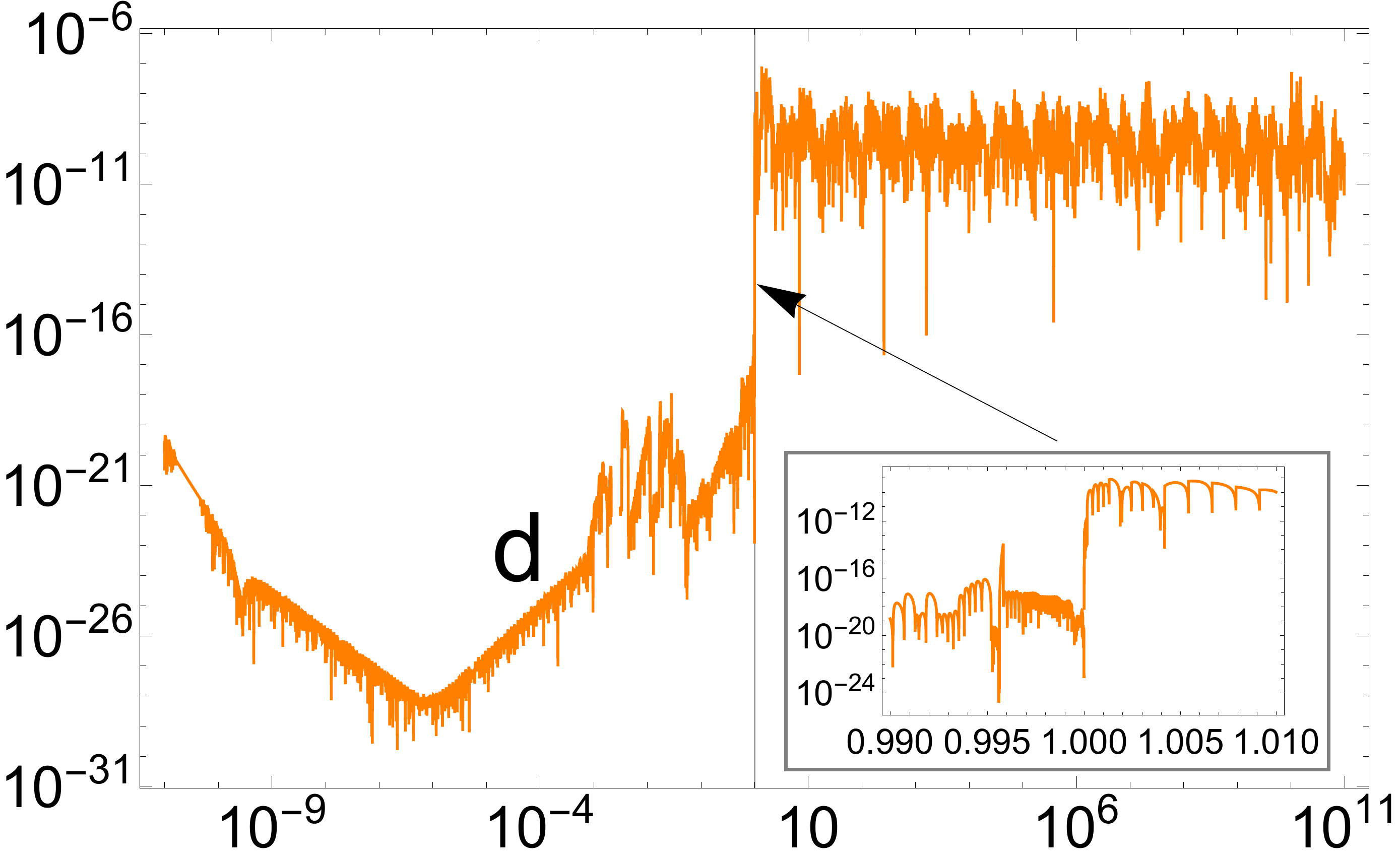} \\
  		(e)  & (f)  \\[6pt]
  	\end{tabular}
  	\caption{$\tilde A$, $\tilde B$ and $\tilde F$ for different combinations of $\{c_2, c_{14}\}$ listed in Table \ref{table2} and their corresponding $\tilde{\cal{C}}$'s. Here the horizontal axis is $r_{H}/r$. {$\tilde A$, $\tilde B$, $\tilde F$ and $\tilde{\cal{C}}$ are represented by the red solid line (labeled by a), green dotted line (labeled by b), blue dash-dotted line (labeled by c) and orange solid line (labeled by d) respectively.} To be specific, (a) and (b) are for the case  $\{9 \times 10^{-7}, 2 \times 10^{-7}\}$, (c) and (d) are for the case  $\{9 \times 10^{-8}, 2 \times 10^{-8}\}$, (e) and (f) are for the case  $\{9 \times 10^{-9}, 2 \times 10^{-9}\}$. Note that the small graphs inserted in (b), (d) and (f) show the amplifications of $\tilde{\cal{C}}$'s near $r=r_{H}$.}
  	\label{FABtilde}
  \end{figure}

As we have already mentioned, theoretically Eq. (\ref{eq2.36c}) will be automatically satisfied once Eqs. (\ref{eq2.36a}), (\ref{eq2.36b}) and (\ref{eq2.36d})  hold. However, due to numerical errors,
in practice,  it can never be zero numerically. Thus, to monitor our numerical errors, we always plot out the quantity $\tilde {\cal{C}}$ defined  by Eq.(\ref{scC}), from which we can see clearly the numerical errors 
in our calculations. So, in the right-hand panels of Fig. \ref{FABtilde}, we plot out the curves of  $\tilde{\cal{C}}$, denoted by  $d$,  in each case. 

Clearly, outside the S0H, $\tilde{\cal{C}} \lesssim 10^{-17}$, while inside the S0H we have $\tilde{\cal{C}} \lesssim 10^{-10}$. Thus,
 the solutions inside the horizon are not as accurate as the ones given outside of the horizon.
 However, since in this paper we are mainly concerned with the spacetime outside of the S0H, we shall not consider  further improvements of our numerical code inside the horizon.
 The other quantities,  such as $c_S^2$ and $\tilde r_g$,  are all given by the first row of Table \ref{table2}.
 
\begin{table}
	\caption{$c_S^2$, ${\tilde A}_H$ and $\tilde r_g/{r_{H}}$  calculated from different $\{c_2, c_{14}\}$ with $c_{13}=0$ {and a fixed ratio of $c_2/c_{14}$.}}  
	\label{table2}   
	\begin{tabular}{|c|c|c|c|c|} 
		\hline  
		$c_2$ & $c_{14}$ & $c_S^2$ & ${\tilde A}_H$ &  $\tilde r_g/{ r_{H}}$ \\  
		\hline
		$9 \times 10^{-7}$ & 	$2 \times 10^{-7}$ & $ 4.4999935 $ & 2.4558992 & 1.1450729 \\  
		\hline
		$9 \times 10^{-8}$ & 	$2 \times 10^{-8}$ & $ 4.4999994 $ & 2.4559003 & 1.1450730    \\  
		\hline
		$9 \times 10^{-9}$ & 	$2 \times 10^{-9}$ & $ 4.4999999 $ & 2.4559004 & 1.1450730  \\  
		\hline
	\end{tabular}
\end{table}

\begin{table}
	\caption{$c_S^2$, ${\tilde A}_H$ and $\tilde r_g/{r_{H}}$  calculated from different $\{c_2, c_{14}\}$ with $c_{13}=0$ {and changing $c_2/c_{14}$}.}  
	\label{table3}   
	\begin{tabular}{|c|c|c|c|c|} 
		\hline  
		$c_2$ & $c_{14}$ & $c_S^2$ & ${\tilde A}_H$ &  $\tilde r_g/{ r_{H}}$ \\  
		\hline
		$2.01 \times 10^{-5}$ & 	$2 \times 10^{-5}$ & $ 1.0049596 $ &  1.4562430  &   1.0005850  \\  
		\hline
		$7 \times 10^{-7}$ & 	$5 \times 10^{-7}$ & $ 1.3999982 $ &   1.6196457   &  1.0381205  \\  
		\hline
		$9 \times 10^{-7}$ & 	$2 \times 10^{-8}$ & $ 44.999939 $ & 6.4676346 &  1.2629671   \\  
		\hline
		$9 \times 10^{-5}$ & 	$2 \times 10^{-7}$ & $ 449.93921 $ & 19.053220  & 1.3091657  \\  
		\hline
	\end{tabular}
\end{table}

 Following the same steps, we also consider other cases, and some of them are presented in Tables \ref{table2}-\ref{table3}. In particular, in Table \ref{table2}, we fix the ratio of $c_2/c_{14}$ to be 9/2. In addition, the values of $\{c_2, c_{14}\}$ are chosen so that they satisfy the constraints of Eq.(\ref{c1234}). In Table \ref{table3}, the ratio $c_2/c_{14}$ is changing and the values of $\{c_2, c_{14}\}$ are chosen so that they are spreading over the whole 
 viable range of $c_{14}$, given by Eqs.(\ref{c1234a})-(\ref{c1234bb}).  
  
   From these tables we can see that quantities like $\tilde A_H$ and $\tilde r_g$ are sensitive only to the ratio of $c_2/c_{14}$, instead of their individual values. 
 This is understandable, as for $c_{13} = 0$ and $c_{14} \lesssim 2.5\times 10^{-5}$, Eq.(\ref{eq1.1}) shows that $c_S \simeq c_{S}(c_2/c_{14})$. Therefore, 
 the same ratio of  $c_2/c_{14}$ implies the same velocity of the spin-0 graviton. Since S0H is defined by the speed of this massless particle, it is quite reasonable
  to expect that the related quantities are sensitive only to the value of $c_S$. 
 
  \begin{figure}
  	\begin{tabular}{cc}
  		\includegraphics[width=37mm]{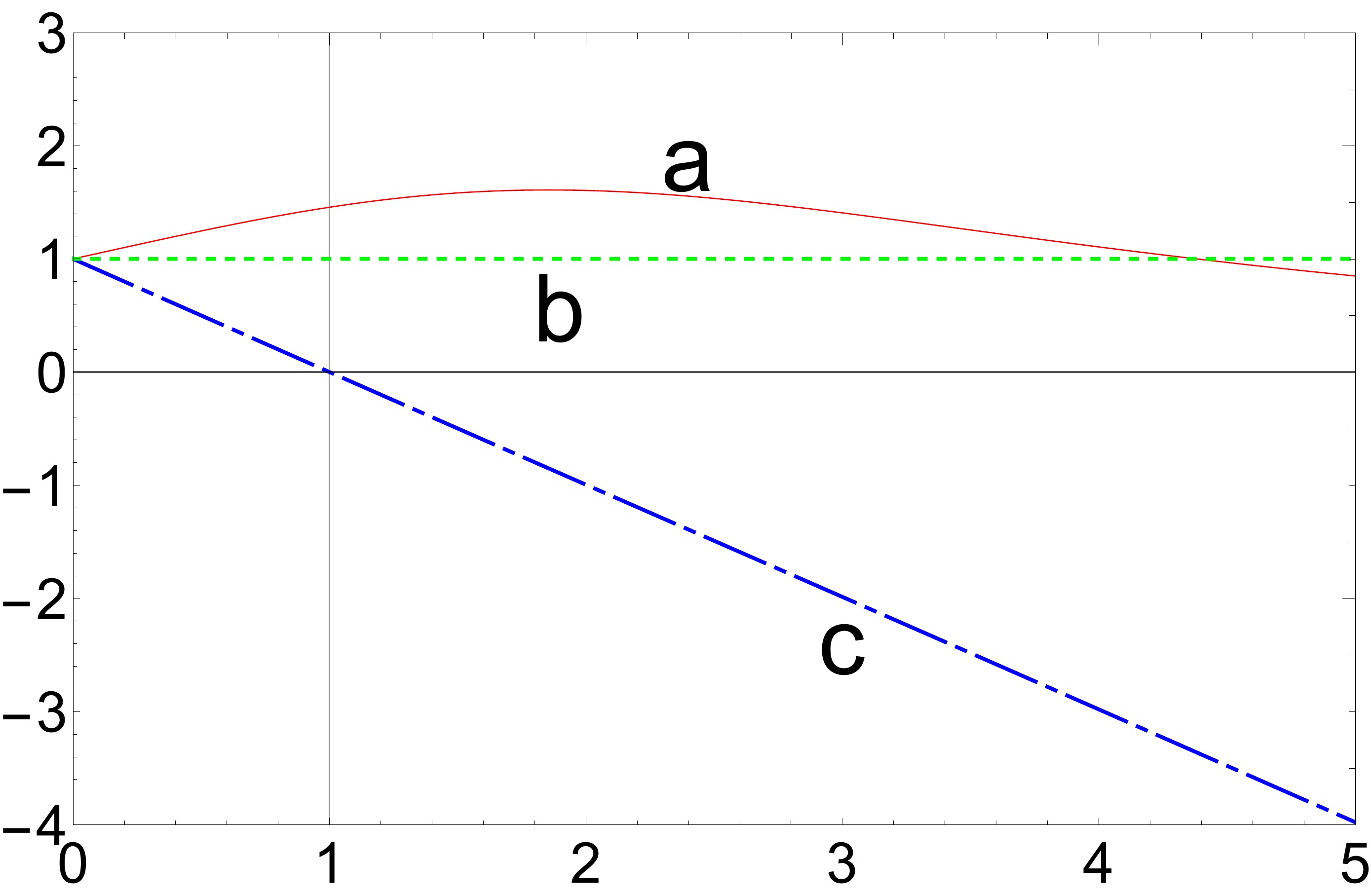} &   \includegraphics[width=37mm]{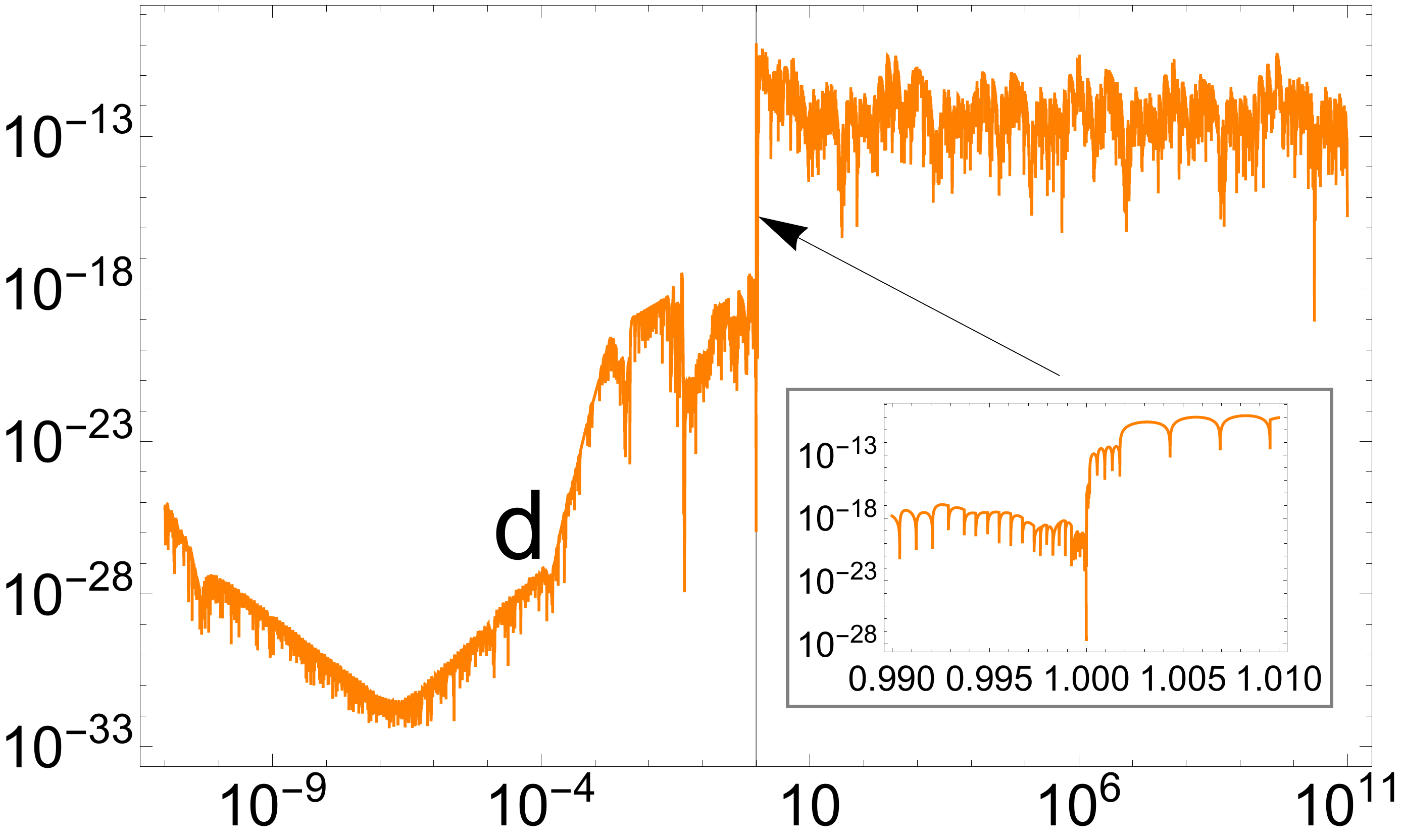} \\
  		(a) & (b)  \\[6pt]
  		\includegraphics[width=37mm]{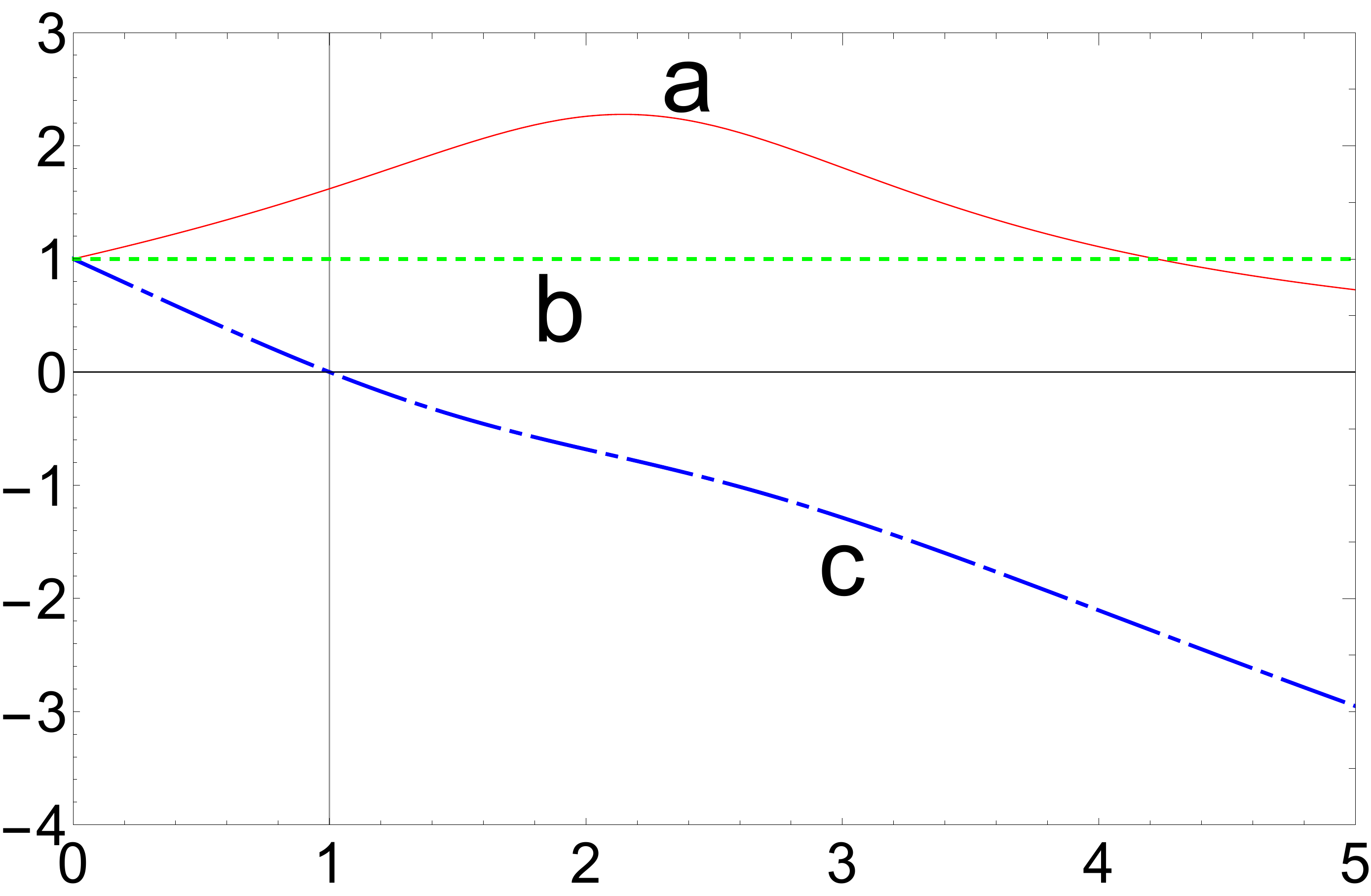} &   \includegraphics[width=37mm]{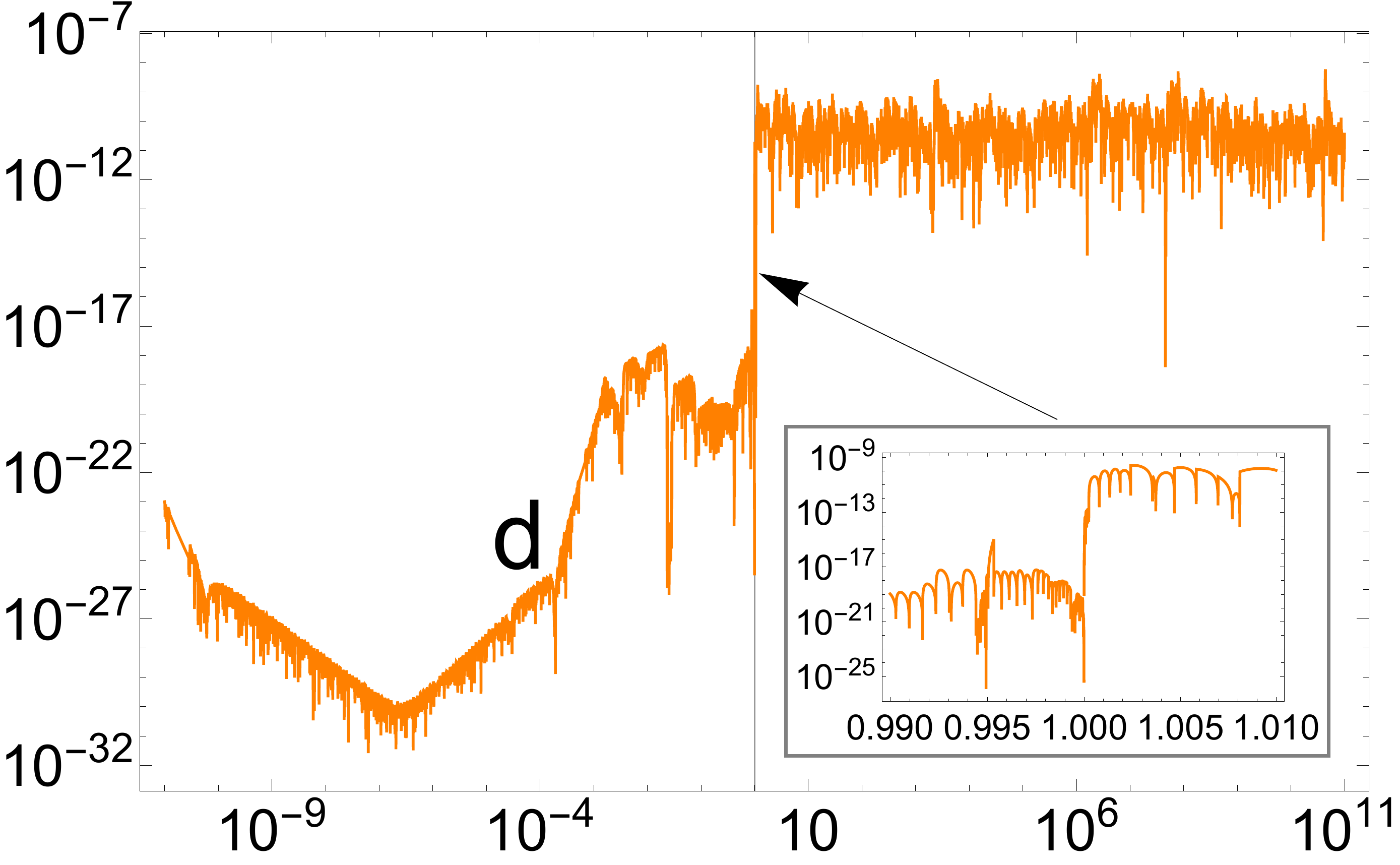} \\
  		(c)  & (d)  \\[6pt]
  		\includegraphics[width=37mm]{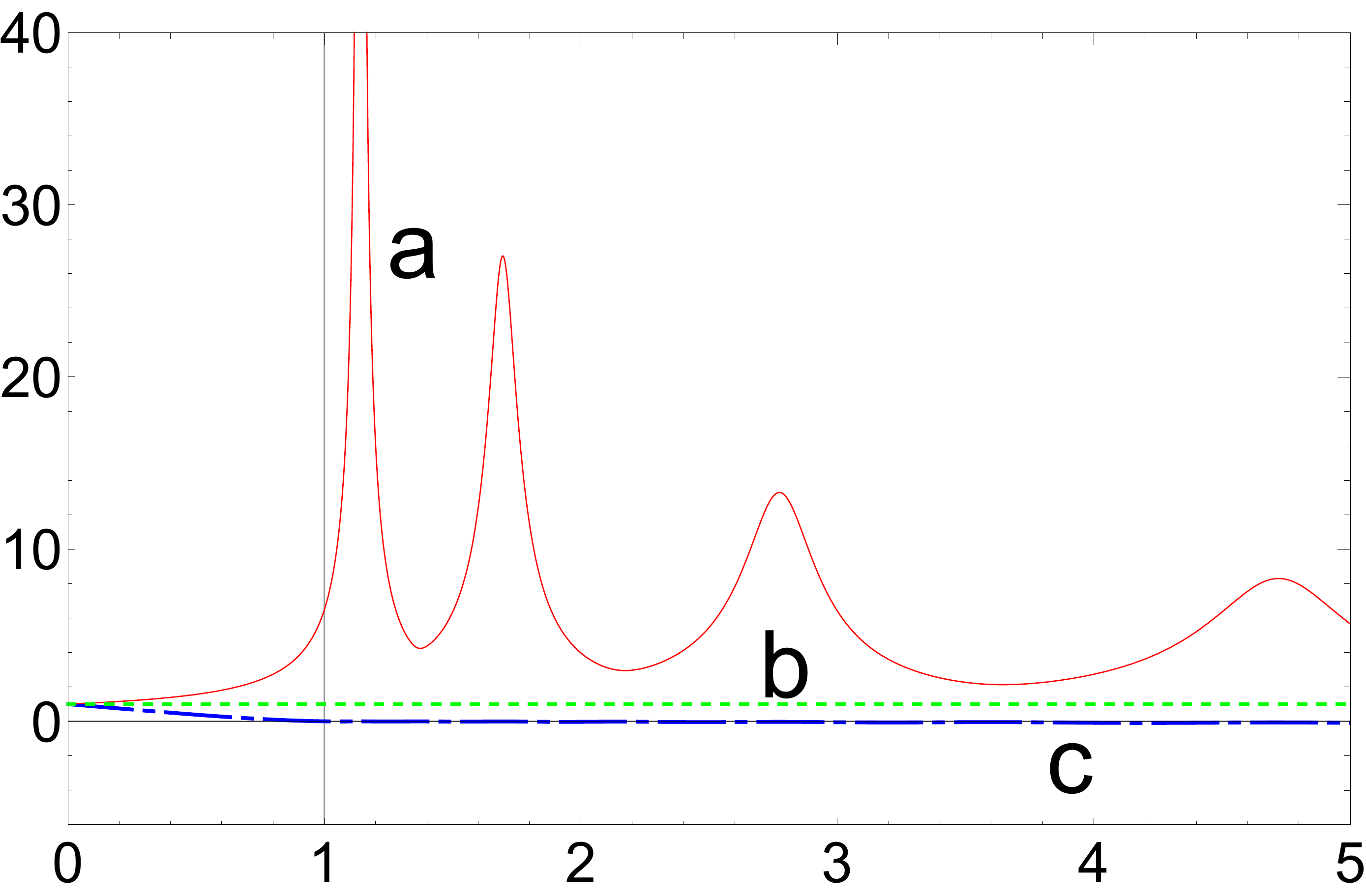} &   \includegraphics[width=37mm]{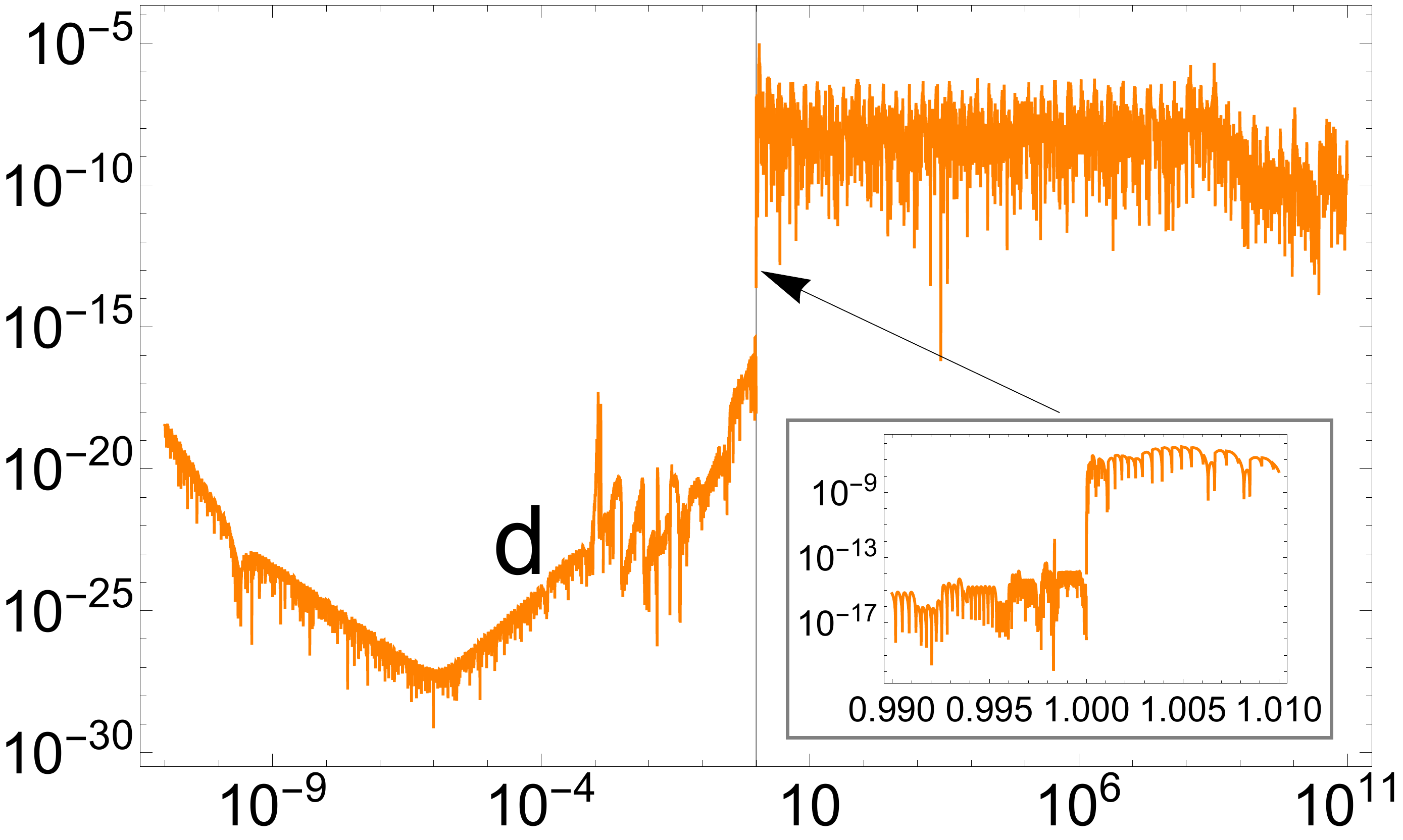} \\
  		(e)  & (f) \\[6pt]
  		\includegraphics[width=37mm]{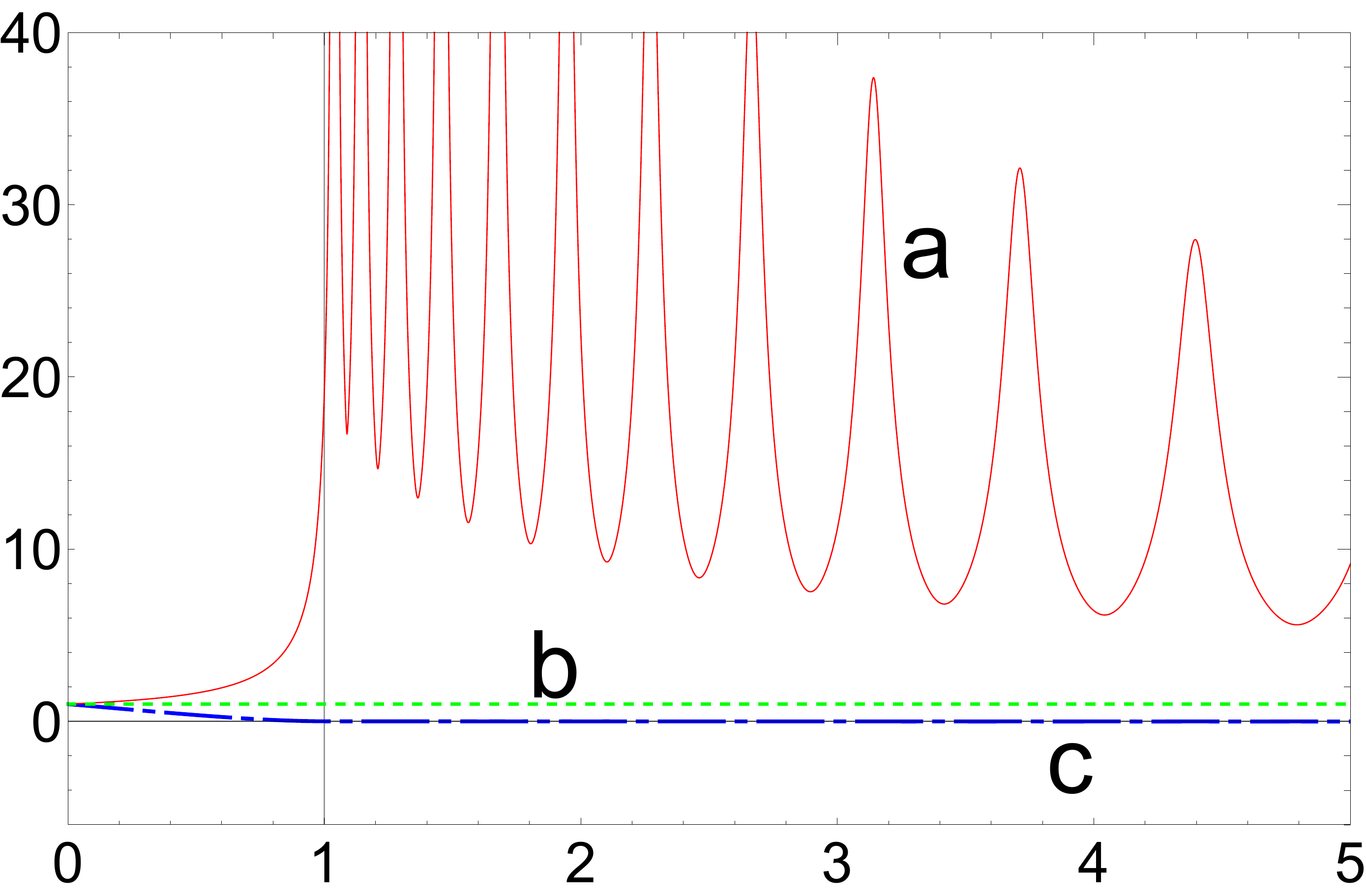} &   \includegraphics[width=37mm]{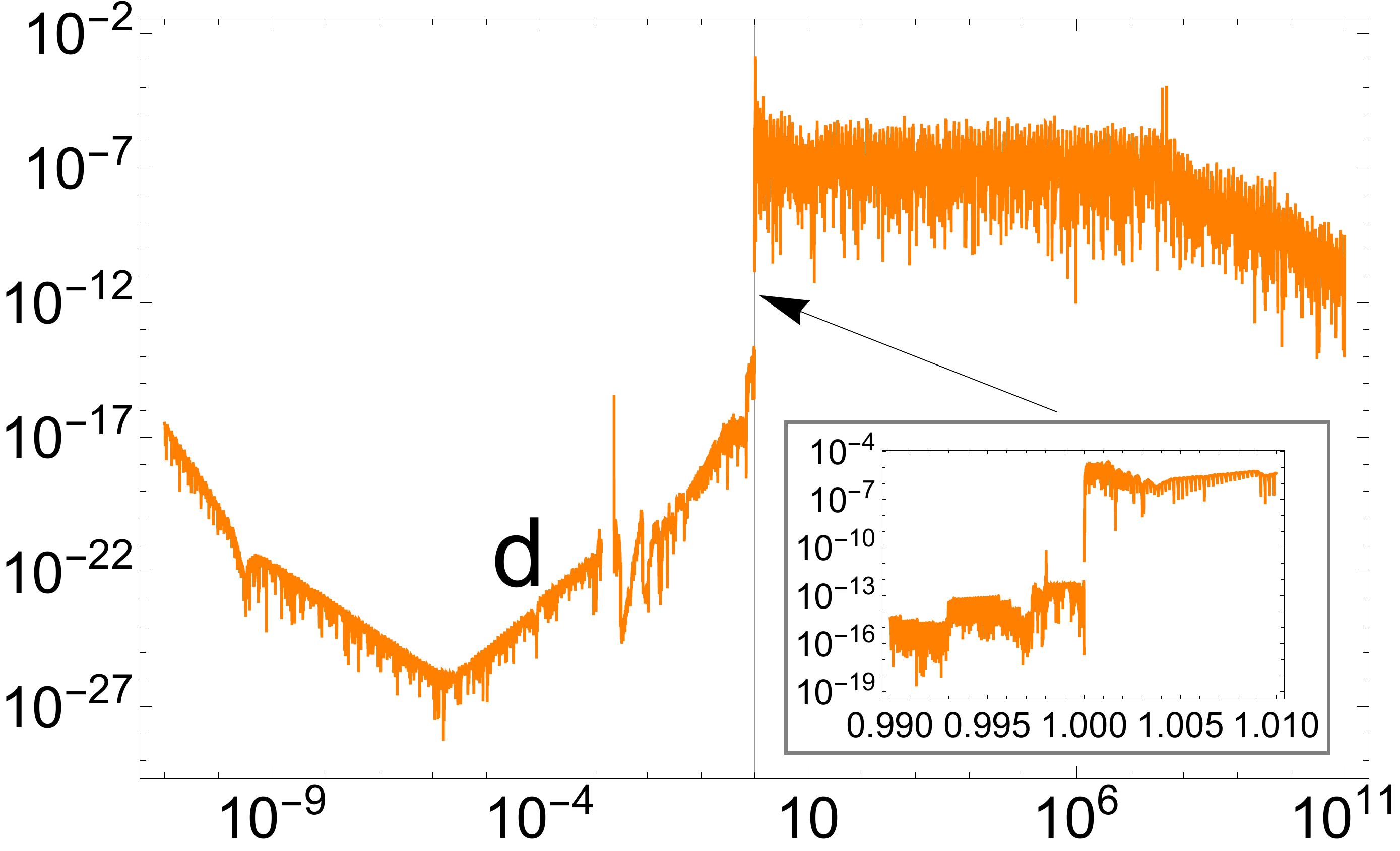} \\
  		(g)  & (h) \\[6pt]
  	\end{tabular}
  	\caption{$\tilde A$, $\tilde B$ and $\tilde F$ for different combinations of $\{c_2, c_{14}\}$ listed in Table \ref{table3} and their corresponding $\tilde{\cal{C}}$'s. Here the horizontal axis is $r_{H}/r$. {$\tilde A$, $\tilde B$, $\tilde F$ and $\tilde{\cal{C}}$ are represented by the red solid line (labeled by a), green dotted line (labeled by b), blue dash-dotted line (labeled by c) and orange solid line (labeled by d) respectively.} To be specific, (a) and (b) are for the case   $\{2.01 \times 10^{-5}, 2 \times 10^{-5}\}$, (c) and (d) are for the case  $\{7 \times 10^{-7}, 5 \times 10^{-7}\}$, (e) and (f) are for the case  $\{9 \times 10^{-7}, 2 \times 10^{-8}\}$, (g) and (h) are for the case  $\{9 \times 10^{-5}, 2 \times 10^{-7}\}$.  Note that the small graphs inserted in (b), (d), (f) and (h) show the amplifications of $\tilde{\cal{C}}$ near  $r=r_{H}$.}
  	\label{FABtilde2}
  \end{figure}
 
The resulting $\tilde F$, $\tilde A$, $\tilde B$ and $\tilde{\cal{C}}$ for the cases listed in Tables \ref{table2} and \ref{table3} are plotted in Figs. \ref{FABtilde} and \ref{FABtilde2}, respectively.


\section{Physical Solutions ($g_{\alpha\beta},\;u^{\mu}$)}
 \renewcommand{\theequation}{5.\arabic{equation}} \setcounter{equation}{0}

The above steps reveal how we find the solutions of the effective metric  $\tilde g_{\mu \nu}$ and aether field $\tilde u^\mu$. To find the corresponding physical  
quantities $g_{\mu \nu}$ and $u^\mu$, we shall follow two steps: 
 (a) Reverse Eqs.(\ref{FBtilde}) and (\ref{Atilde}) to find a set of the physical quantities $\{ F(\xi),  A(\xi),  B(\xi)\}$ (Note that we have $\xi=\tilde r_{S0H}/r=r_{S0H}/r$).
 (b) Apply the rescaling $ v \rightarrow C_0 v$  to make the set of $\{ F(\xi),  A(\xi),  B(\xi)\}$ take the standard form at spatial infinity $r = \infty$. 
 
 To these purposes,  let us first note that, near the spatial infinity, Eqs.(\ref{eq2.38}), (\ref{FBtilde}) and (\ref{Atilde}) lead to
\bqn
\label{normal}
F(\xi) &=& \frac{C_0^2}{\sigma } \left(1+{F}_1 \xi +\frac{1}{48} {c}_{14} {F}_1^3 \xi ^3 \right){+{\cal O}\left(\xi ^4\right)}, \nb\\ 
B(\xi) &=& \frac{C_0}{\sqrt{\sigma }} \left(1+\frac{1}{16} {c}_{14}  {F}_1^2 \xi ^2- \frac{1}{12}  {c}_{14}  {F}_1^3 \xi ^3 \right)\nb\\
&& +{\cal O}\left(\xi ^4\right), \nb\\
A(\xi)&=&  \frac{\sqrt{\sigma }}{C_0} \left[ 1- \frac {1}{2} {F}_1\xi+ \frac{1}{2} A_2 \xi ^2 \right. \nb\\
&& \left. -\left( \frac{1}{2} {A}_2 {F}_1 -\frac{1}{16} {F}_1^3+\frac{1}{96} {c}_{14}  {F}_1^3\right) \xi ^3 \right]\nb\\
&& +{\cal O}\left(\xi ^4\right),
\eqn
where 
\bqn
\label{normal2}
F_1 &=& \tilde F_1, \quad c_{14}=\tilde c_{14}, \nb\\
A_2 &=& \sqrt {\sigma} \tilde A_2-\frac{3}{4} ( \sqrt {\sigma}-1) \tilde F_1^2.
\eqn
The above expressions show clearly that {\it the spacetimes described by $\left(g_{\mu\nu}, u^{\mu}\right)$  are  asymptotically flat, provided that the effective fields  $\left(\tilde g_{\mu\nu}, \tilde u^{\mu}\right)$ are}. 
In particular, setting   $C_0 = \sqrt{\sigma}$, a condition that will be assumed in the rest of this section, the functions $F, \; A$ and $B$ will take their standard  asymptotically-flat forms. 

It is remarkable to note that the asymptotical behavior of the functions $F, \; A$ and $B$ depends only on $c_{14}$ up to the third-order of $\xi$, but   $c_2$  will show up starting from the four-order 
of $\xi^4$.

\begin{figure*}[htb]
	\includegraphics[width=\columnwidth]{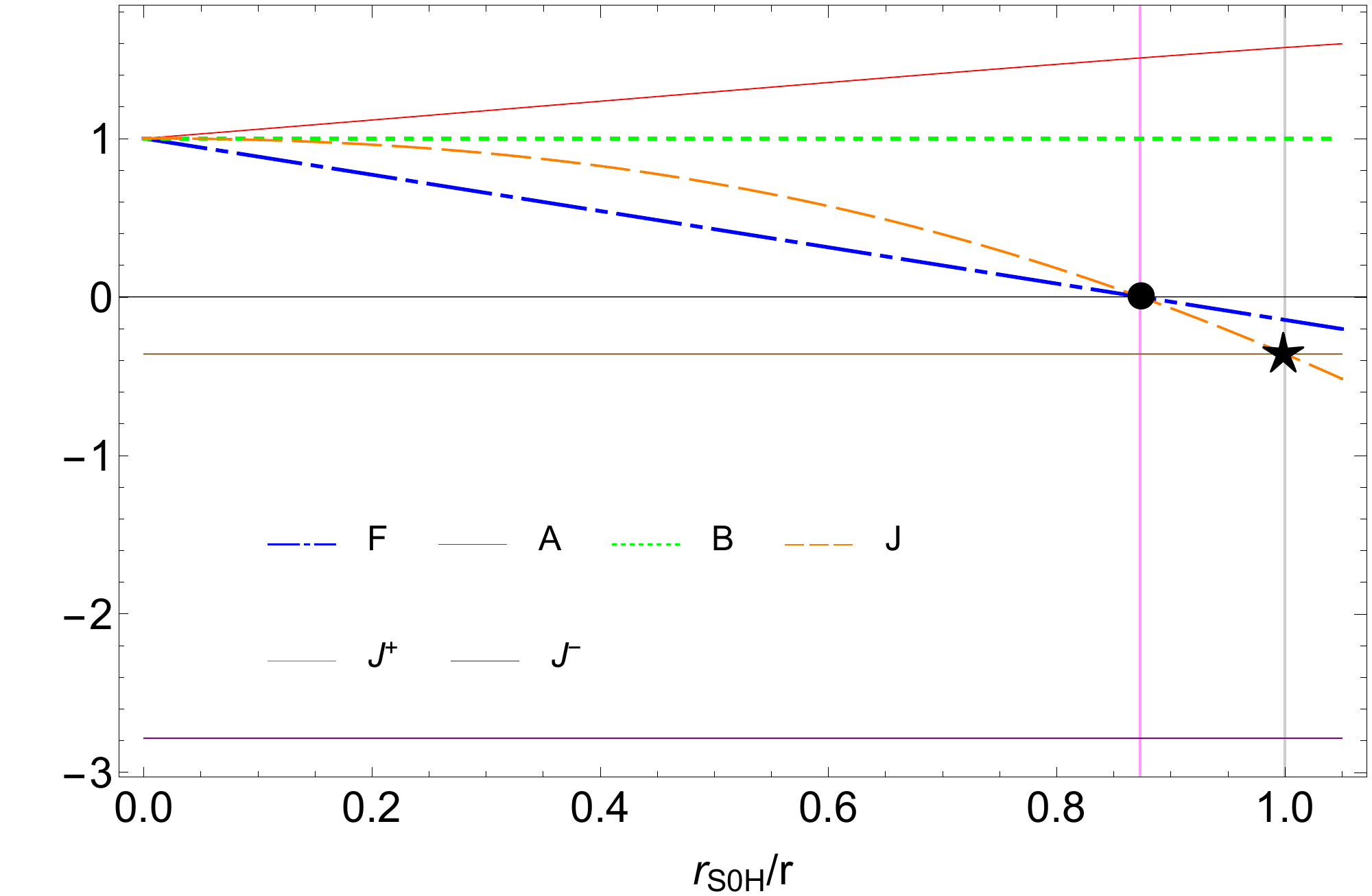} 
	\includegraphics[width=\columnwidth]{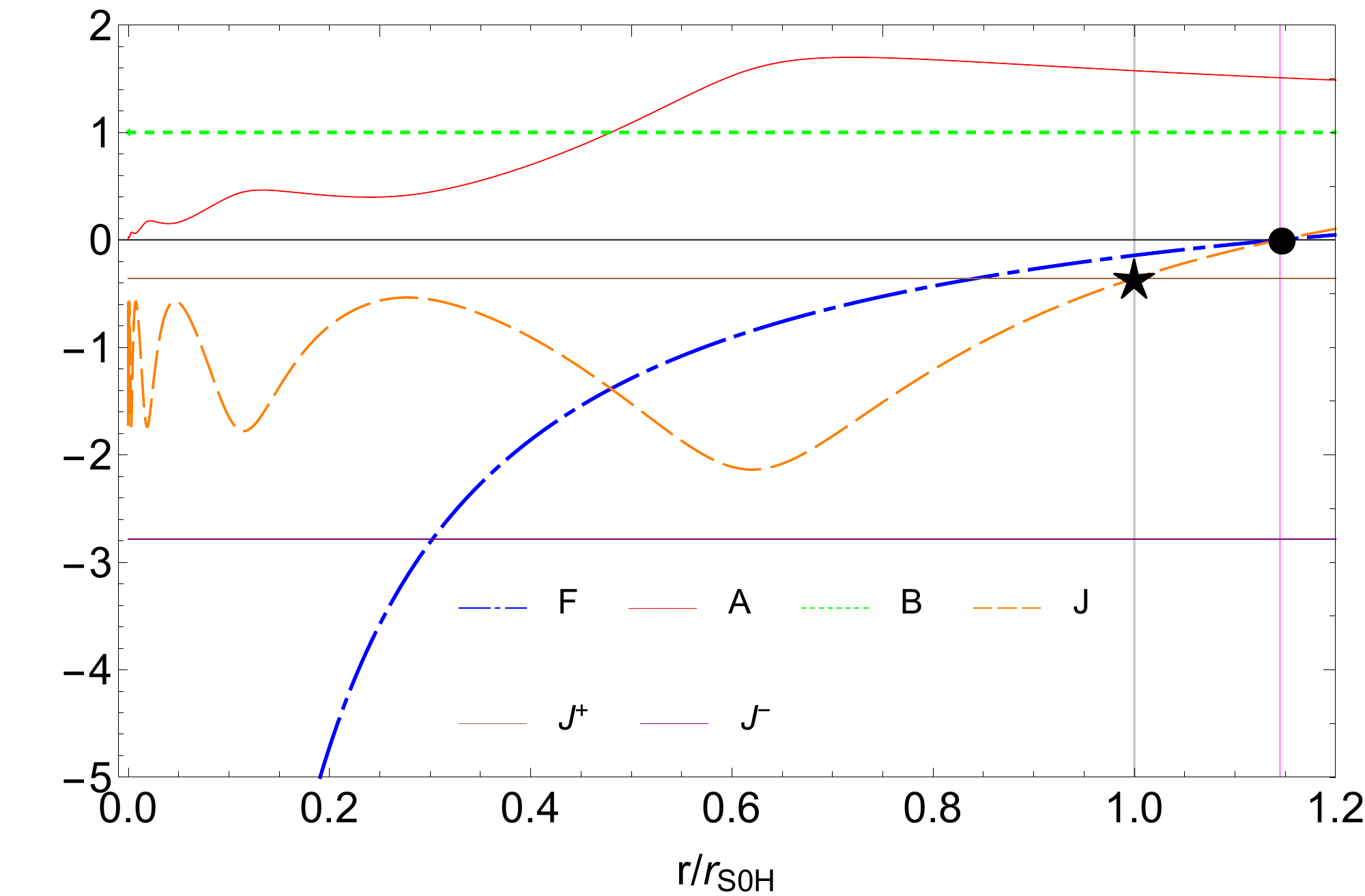} 
	\caption{The evolutions of the physical quantities $ F$, $ A$, $ B$ and $ J$ for the case $c_{13}=0$,  {$c_2=9 \times 10^{-7}$ and $c_{14}=2 \times 10^{-7}$.} Here, $A$, $B$, $J$ and $F$ are represented by the red solid line, green dotted line, orange dashed line, and blue dash-dotted line, respectively. The positions of $r=r_{MH}$ and $r=r_{S0H}$ are marked by a small full solid circle and a pentagram,  respectively. Note that we have $r_{MH}>r_{S0H}$. The values $J^+$ and $J^-$ are given respectively by the brown and purple solid lines with $J^+ > J^-$.
	The left panel shows the main behaviors of the functions outside the S0H in the  range $r_{S0H}/r \in (0, 1.105)$, while the right panel shows their main behaviors inside the S0H in  the  range $r/r_{S0H} \in (0, 1.2)$.}
	\label{FAB}
\end{figure*}

\subsection{Metric and Spin-0 Horizons}

Again, we take the case of $c_{14}=2\times 10^{-7}$, $c_2=9\times 10^{-7}$, and $c_3=-c_1$ as the first example. The results for the normalized $F$, $A$, $B$ and $J$ in this case are plotted in Fig. \ref{FAB}. To see the whole picture of these functions on $r\in(0, \infty)$, they are plotted as functions of $r/r_{H}$ inside the horizon, while  outside the horizon they are plotted as functions of $(r/r_H)^{-1}$. This explains why in the left-hand panel of Fig. \ref{FAB}, the MH ($r = r_{MH}$) stays in the left-hand side of the S0H, while in the right-hand panel, they just reverse the order. In this figure, we didn't plot the GR limits for $B$ and $F$ since they are almost overlapped with their counterparts. 
From the analysis of this case,  we find the following:  

(a) The values of $F$ and $B$ are almost equal to their GR limits all the time.  This is true even when $r$ is approaching the center $r=0$, at which  a spacetime curvature singularity is expected to be located. 

(b) Inside the S0H,  the oscillations   of $A$ and $J$ become visible, which was also noted  in \cite{Eling2006-2} \footnote{In \cite{Eling2006-2}, the author just considered the oscillational behavior of $\tilde A$. 
The physical quantities $F$, $A$,  and $B$ were not considered.}. Such oscillations continue, and become more violent as the curvature singularity at the center is approaching.

        \begin{figure*}
  	\begin{tabular}{cc}
  		\includegraphics[width=83mm]{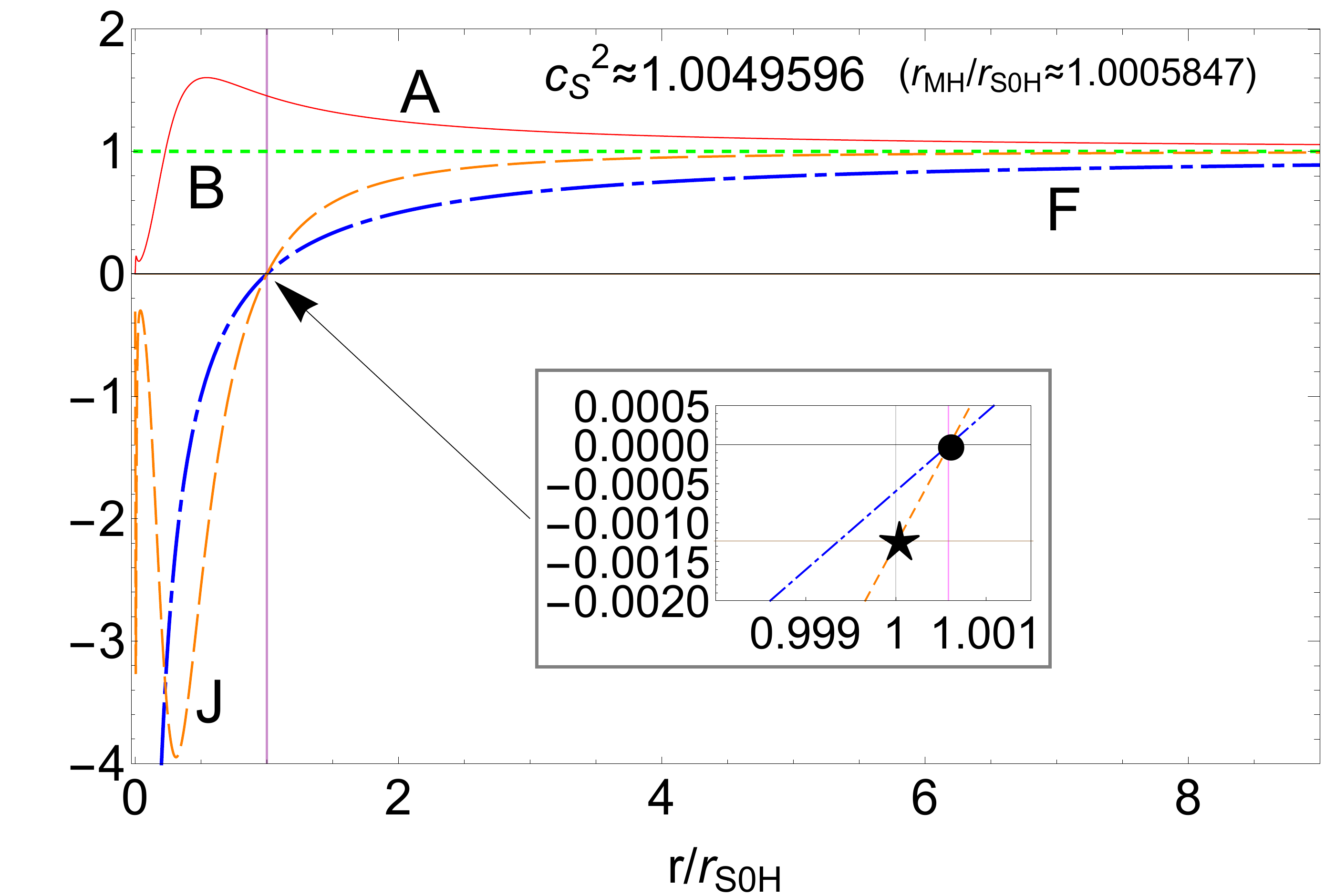} &   \includegraphics[width=83mm]{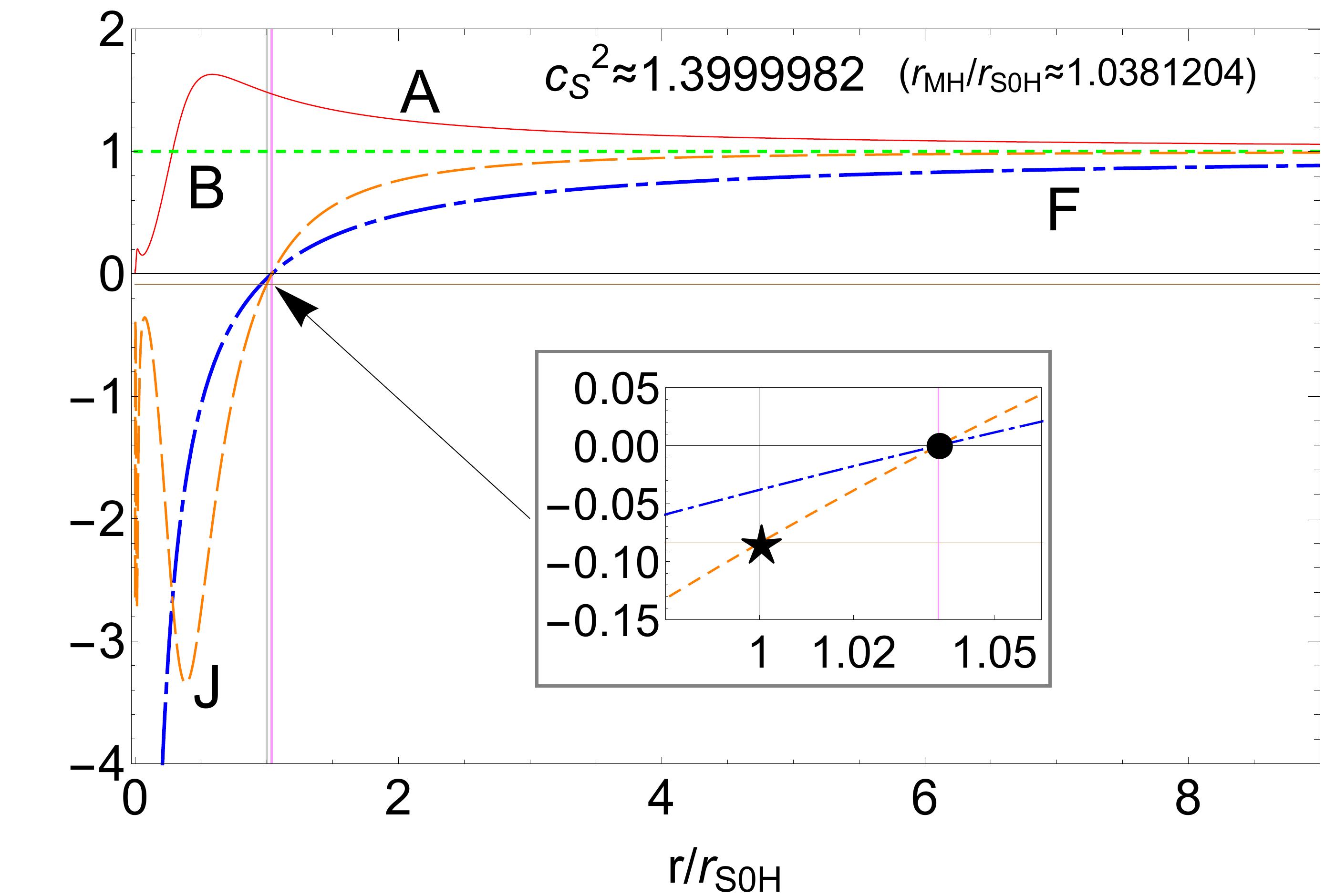} \\
  		(a) & (b)  \\[6pt]
  		\includegraphics[width=83mm]{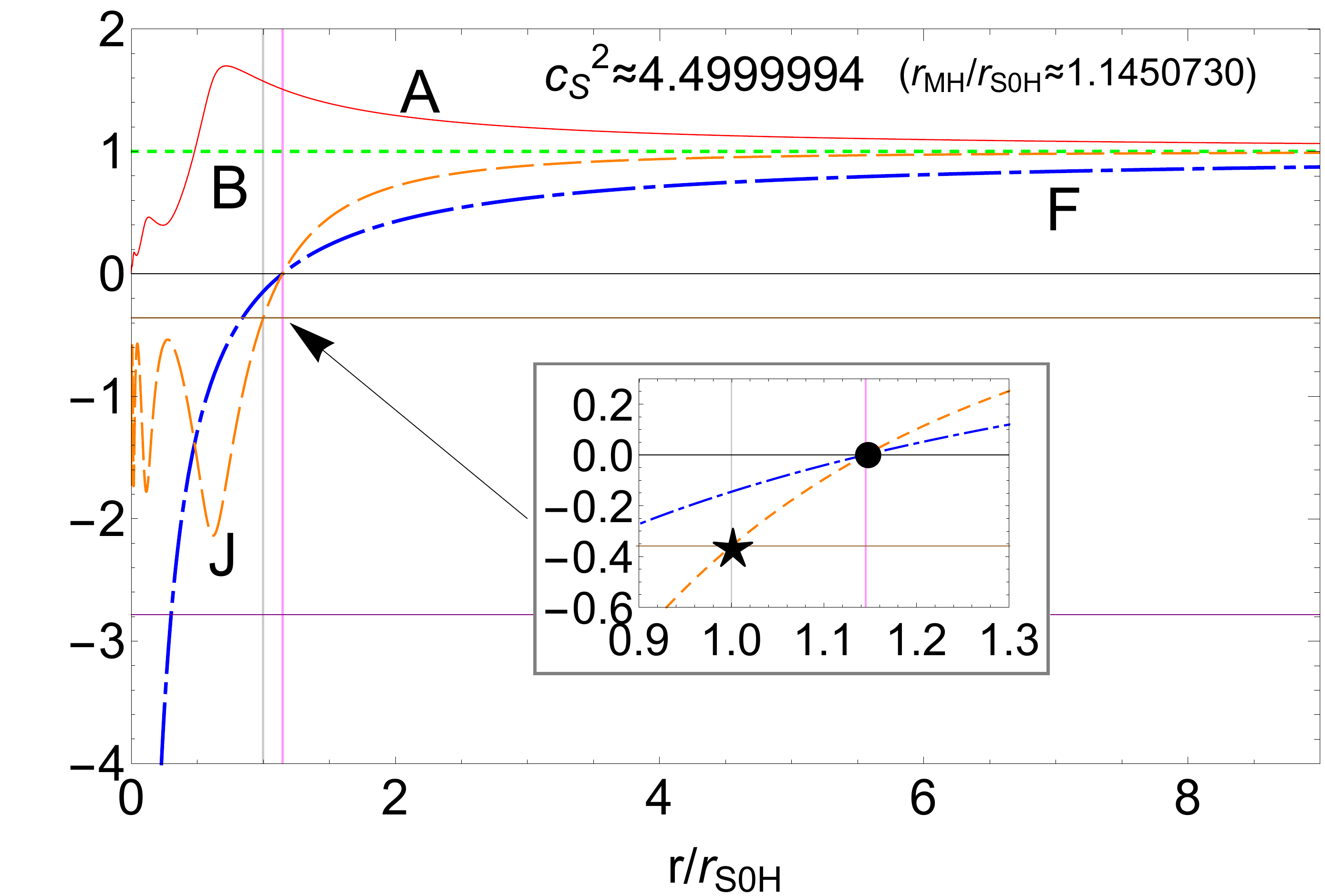} &   \includegraphics[width=83mm]{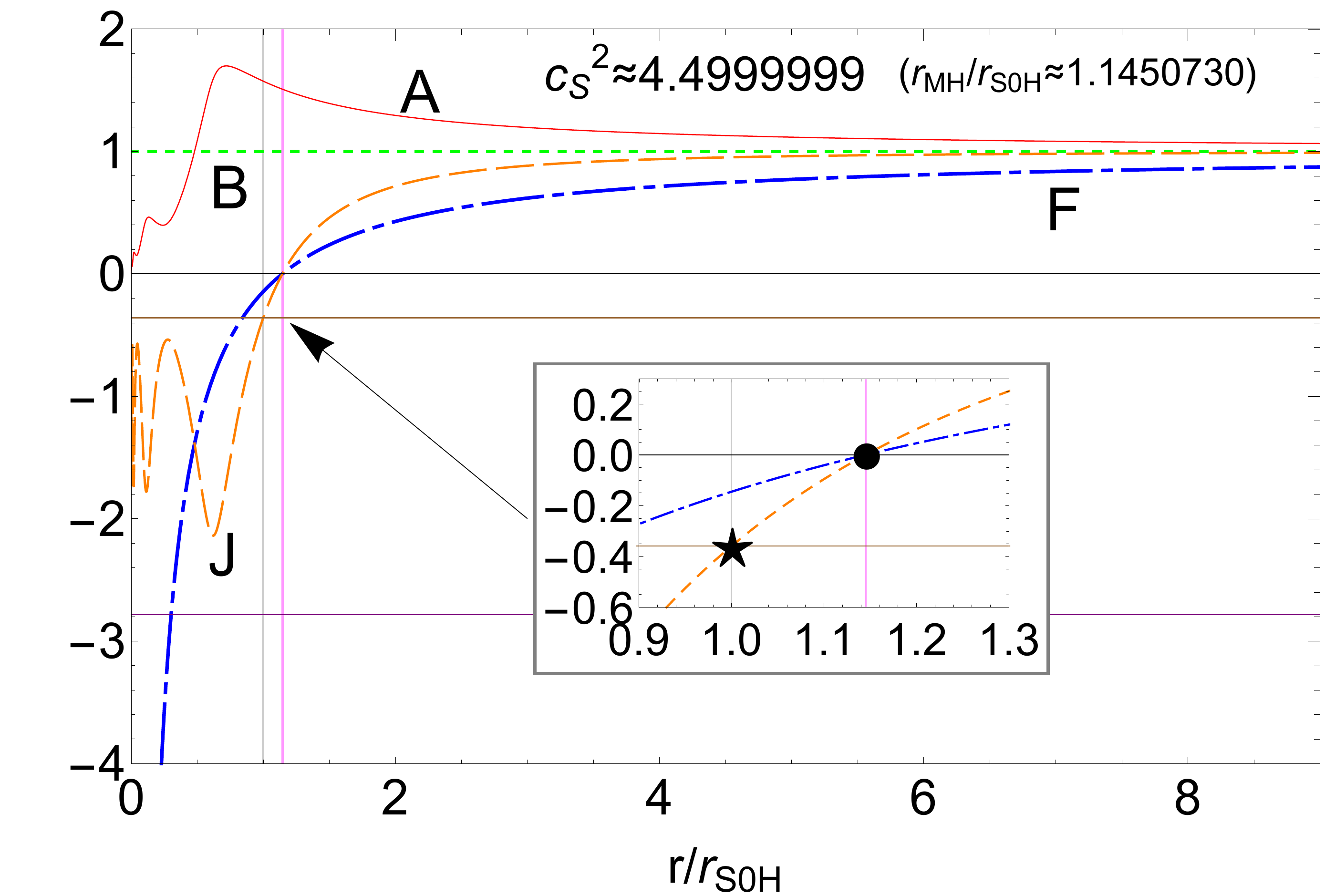} \\
  		(c)  & (d)  \\[6pt]
  		\includegraphics[width=83mm]{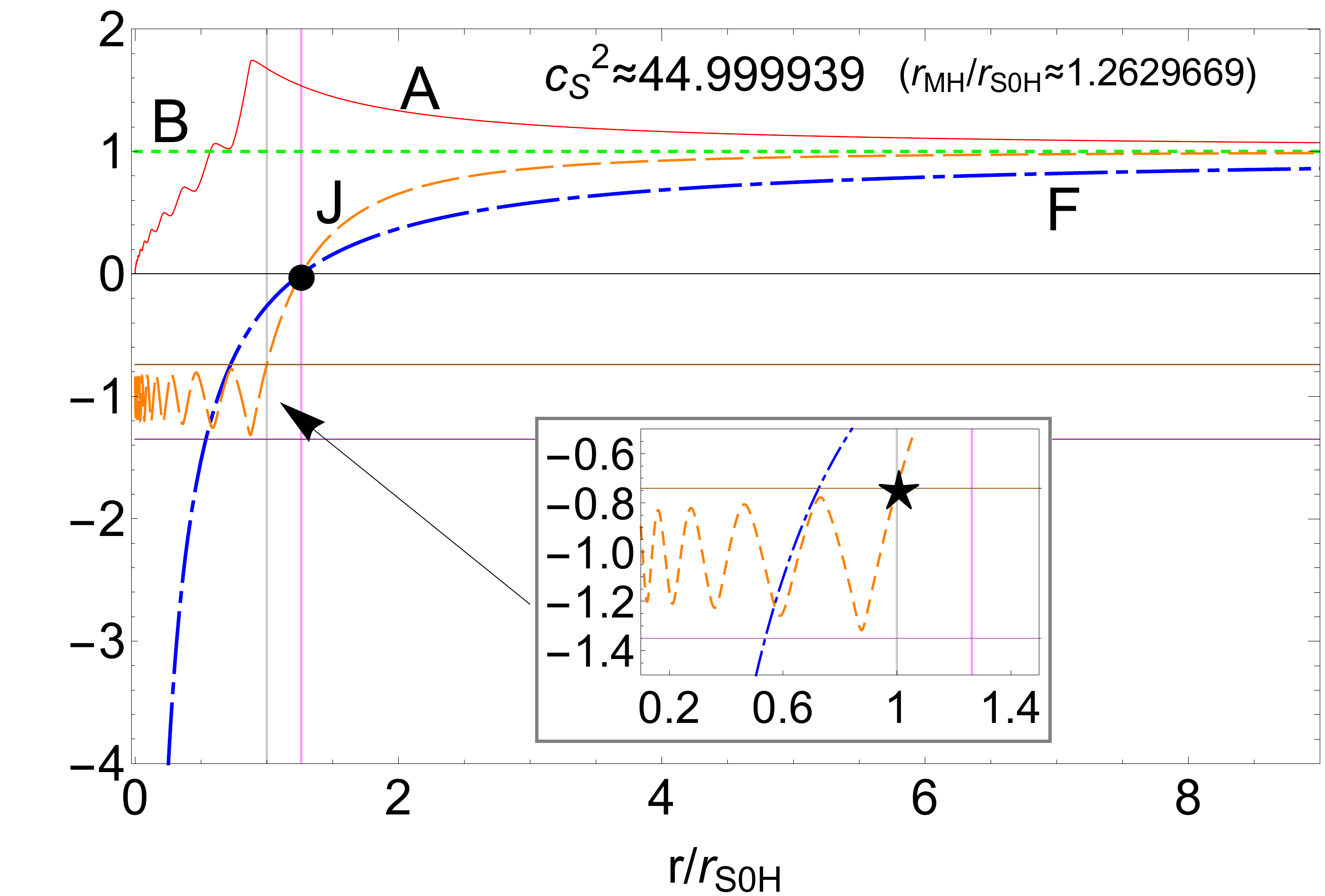} &   \includegraphics[width=83mm]{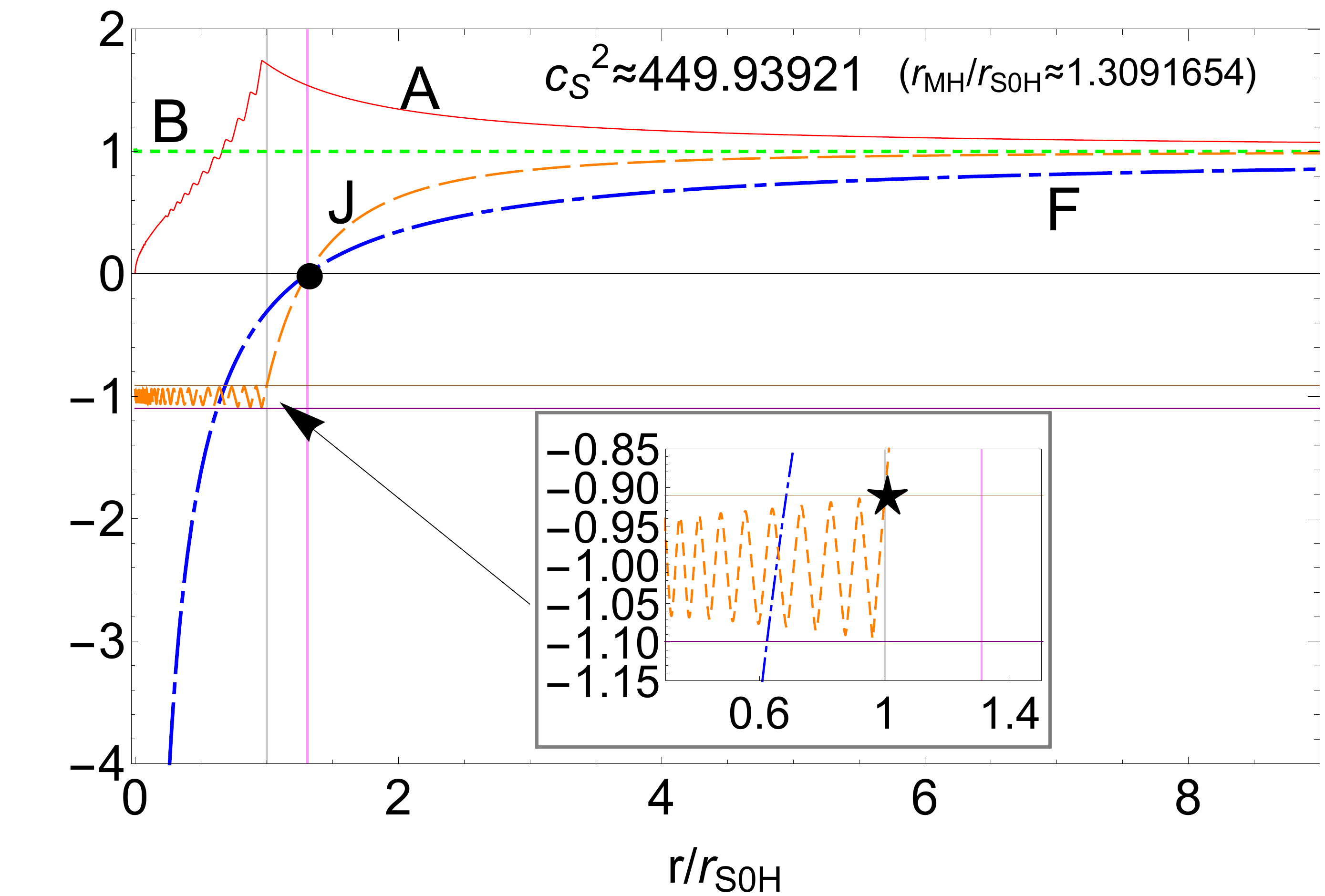} \\
  		(e) & (f) \\[6pt]
  	\end{tabular}
  	\caption{Solutions for different combinations of $\{c_2, c_{14}\}$ listed in Tables \ref{table2}-\ref{table3}. Here, $A$, $B$, $J$ and $F$ are represented by the red solid line, green dotted line, orange dashed line, and blue dash-dotted line, respectively. These figures are ordered according to the magnitude of $c_S^2$.  In each of the figure, the values $J^+$ and $J^-$ are given respectively by the brown and purple solid lines
	with $J^+ > J^-$, while the positions of $r=r_{MH}$ and $r=r_{S0H}$ are marked by a small full solid circle and a pentagram,  respectively. Additionally, the value of $r_{MH}/r_{S0H}$ is also given in each case. }
  	\label{FAB2}
  \end{figure*}

  The functions of $\{F, A, B, J\}$ for the other cases listed  in Tables \ref{table2}-\ref{table3} are plotted in Fig. \ref{FAB2}. In this figure, the plots are ordered according to the magnitude of $c_S^2$. Besides, some amplified figures are inserted in (a)-(d) near the region around the point of $F=0$. Similarly, in (e)-(f), some amplified figures are inserted near the region around the point of $J=J^+$.  The position of $r=r_{MH}$, at which we have $F(r_{MH})=0$, is marked by a  full solid circle, while  the position of $r=r_{S0H}$, at which we have $J(r_{S0H})=J^+$  [cf., Eq. (\ref{eq2.24b})], is marked by a  pentagram, and  in all these cases we always have $r_{MH}>r_{S0H}$. The values of $J^+$ and $J^-$ are given by the brown and purple solid lines, respectively. Note we always have $J^+ > J^-$ for $c_S > 1$. By using these two lines, we can easily find that there is only one $r_{S0H}$ in each case, i.e.,  $r_{S0H}^+$ in  Eq. (\ref{eq2.24b}).

From the studies of these representative cases, we find the following: 
   (i) As we have already mentioned, in all these cases the functions $B$ and $F$ are very  {close to}   their GR limits. 
   (ii) Changing $c_S^2$ won't influence the maximum of $A$  much. In contrast, the maximum of $|J|$ inside the S0H is sensitive to $c_S^2$. 
   (iii) The oscillation of $A(r)$  gets more violent as $c_S^2$ is increasing.
   (iv) The value of $|r_{MH}-r_{S0H}|$ is getting bigger as $c_S^2$ deviating from 1.
   (v) In all these cases,  we have only one $r_{S0H}$, i.e., only one intersection between $J(r)$ and $J^{\pm}$, in each case. 
     (vi) Just like what we saw in Tables \ref{table2}-\ref{table3},   in the cases with the same  $c_S$ (but different values of $c_{14}$ and $c_2$), the corresponding functions $\{F, A, B, J\}$ are  quite similar.

 From Tables \ref{table2}-\ref{table3} and Fig. \ref{FAB2}, we would like also to note that the value of $r_{MH}$ is always  {close to}  the corresponding $\tilde r_g$.  To understand this,  let us consider Eq. (\ref{normal}), from
 which  we find that
  \bqn
  \label{Fphy}
  F(\xi) &=& 1+{F}_1 \xi +\frac{1}{48} {c}_{14} {F}_1^3 \xi ^3 +{\cal O}\left(\xi ^4, c_{14}, c_2\right),~~~
  \eqn
  after normalization. Recall $\xi \equiv r_{H}/r$ and  $r_H \equiv r_{S0H}$. Then, from Eqs.(\ref{rg}),  (\ref{eq2.38}), (\ref{normal2}) and (\ref{Fphy}), we  also find that
  \bqn
  \label{rg2}
  \frac{\tilde r_g}{r_{S0H}} = -\tilde F_1=-F_1.
  \eqn
 On the other hand, from Eq. (\ref{MH2}), we have
  \bqn
  \label{MH3}
  \left.F(\xi)\right|_{r=r_{MH}} &=& 1+{F}_1 \frac{r_{S0H}}{r_{MH}} +\frac{1}{48} {c}_{14} {F}_1^3 \left( \frac{r_{S0H}}{r_{MH}}\right) ^3 \nb\\
  && ~~+{\cal O}\left(\xi ^4, c_{14}, c_2\right) \nb\\
  &=& 0,
  \eqn
  from which we obtain,  
  \bqn
  \label{MH4}
  \frac{r_{MH}}{r_{S0H}} &=& -{F}_1 - \frac{1}{48} {c}_{14} {F}_1^3 \left( \frac{r_{S0H}}{r_{MH}}\right) ^2 +{\cal O}\left( \frac{r_{S0H}}{r_{MH}}\right) ^3 \nb\\
  &=&  \frac{\tilde r_g}{r_{S0H}} + \frac{1}{48} {c}_{14}  \left( \frac{ \tilde r_g}{r_{S0H}}\right) ^3 \left( \frac{ r_{S0H}}{r_{MH}}\right) ^2 \nb\\
  && +{\cal O}\left(\xi ^3, c_{14}, c_2\right),~~~~
  \eqn
 where Eq.(\ref{rg2}) was used. For the expansion of $F$ to be finite, we must assume
    \bqn
  \label{MH5}
 {\cal O}\left(\xi ^3, c_{14}, c_2\right) \lesssim {\cal{O}} \left[ \frac{1}{48} {c}_{14}  \left( \frac{ \tilde r_g}{r_{S0H}}\right) ^3 \left( \frac{ r_{S0H}}{r_{MH}}\right) ^2\right].~~~~
  \eqn
  At the same time, recall that we have $c_{14} \lesssim 2.5\times 10^{-5}$ and $r_{S0H} \leqslant r_{MH}$. Besides,  we also have $\tilde r_g/r_{S0H} \simeq {\cal{O}}(1)$. Thus, from Eq. (\ref{MH4}) we
    find 
  \bqn
  \label{rgandrMH}
  \left|\frac{r_{MH}}{r_{S0H}}-\frac{\tilde r_g}{r_{S0H}}\right| \lesssim {\cal{O}} (c_{14}).
  \eqn 
  This result reveals why the values of $r_{MH}/r_{S0H}$ and $\tilde r_{g}/r_{S0H}$ are very  {close to}  each other, although  not necessarily  the same exactly.

\begin{widetext}

    \begin{figure}
	\begin{tabular}{cc}
		\includegraphics[width=83mm]{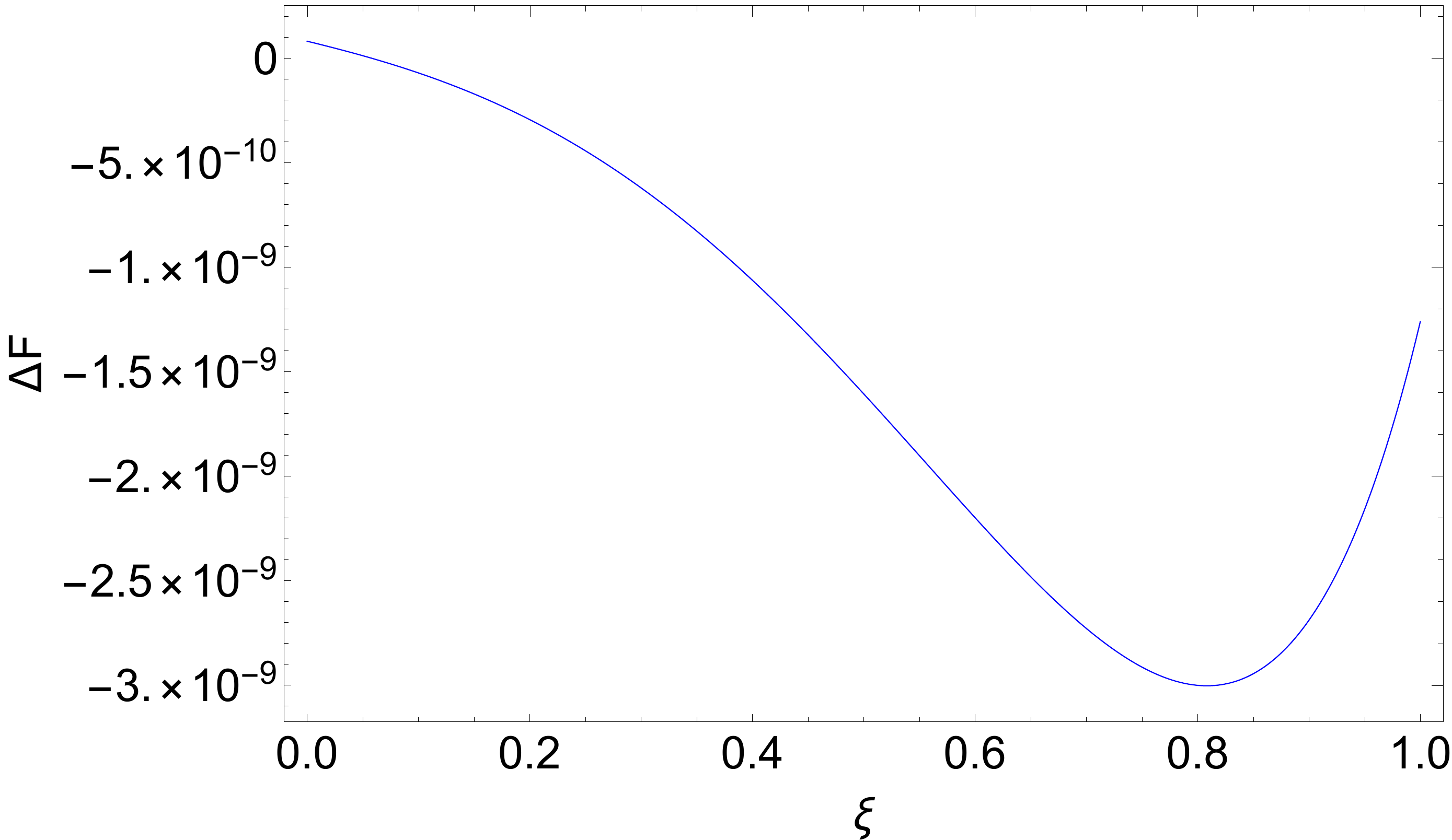} &   \includegraphics[width=83mm]{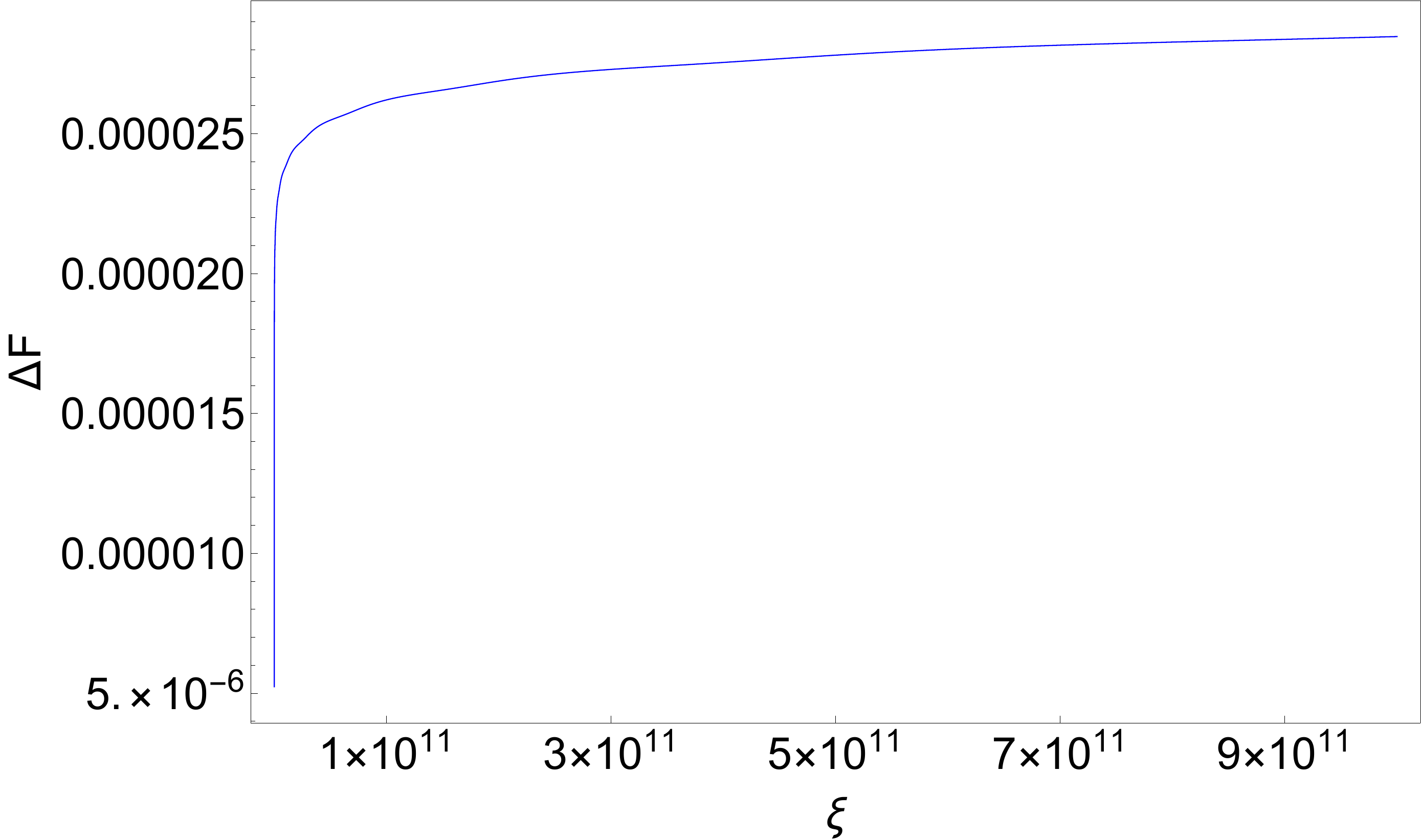} \\
		(a)  & (b)  \\[6pt]
		\includegraphics[width=83mm]{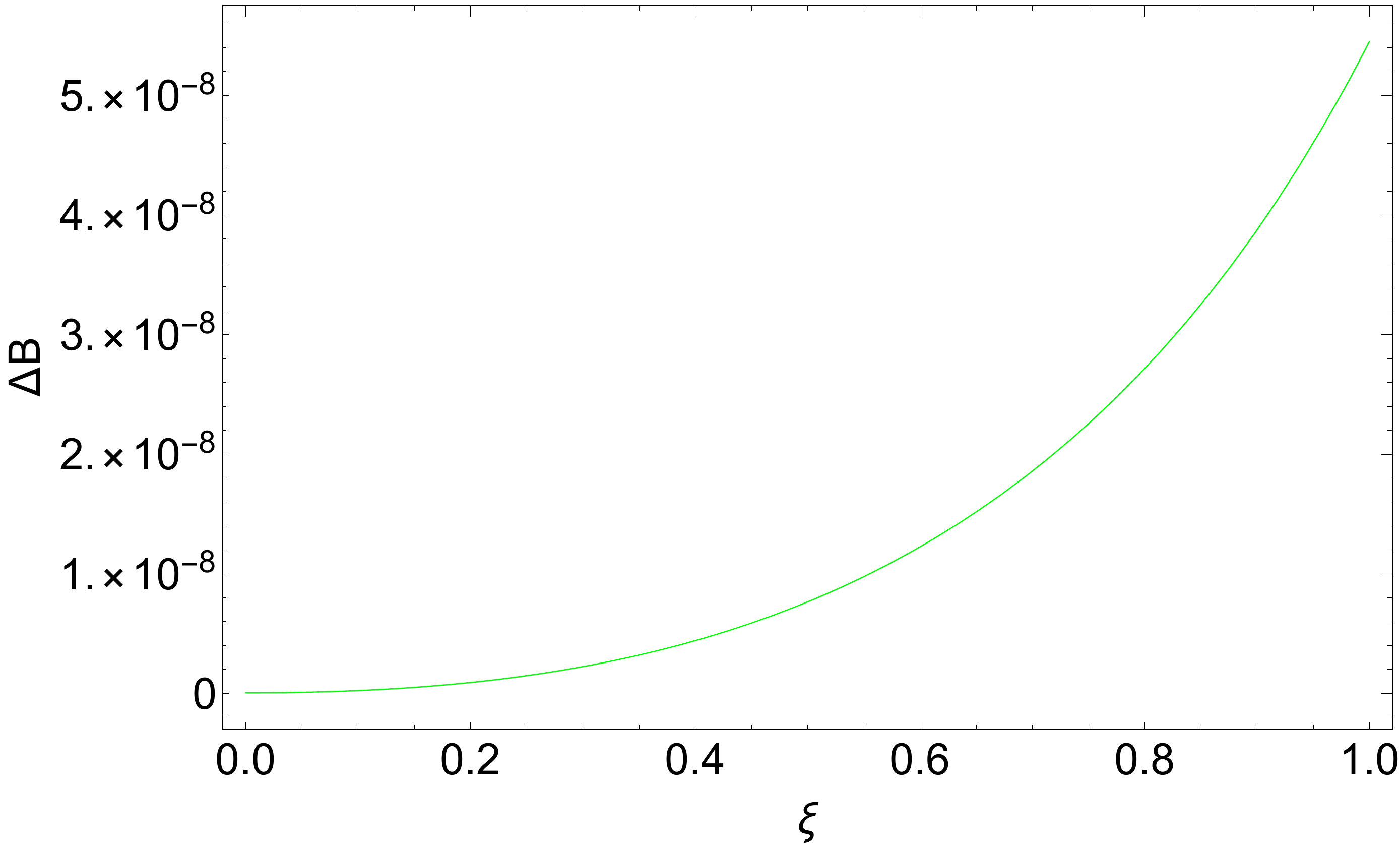} &   \includegraphics[width=83mm]{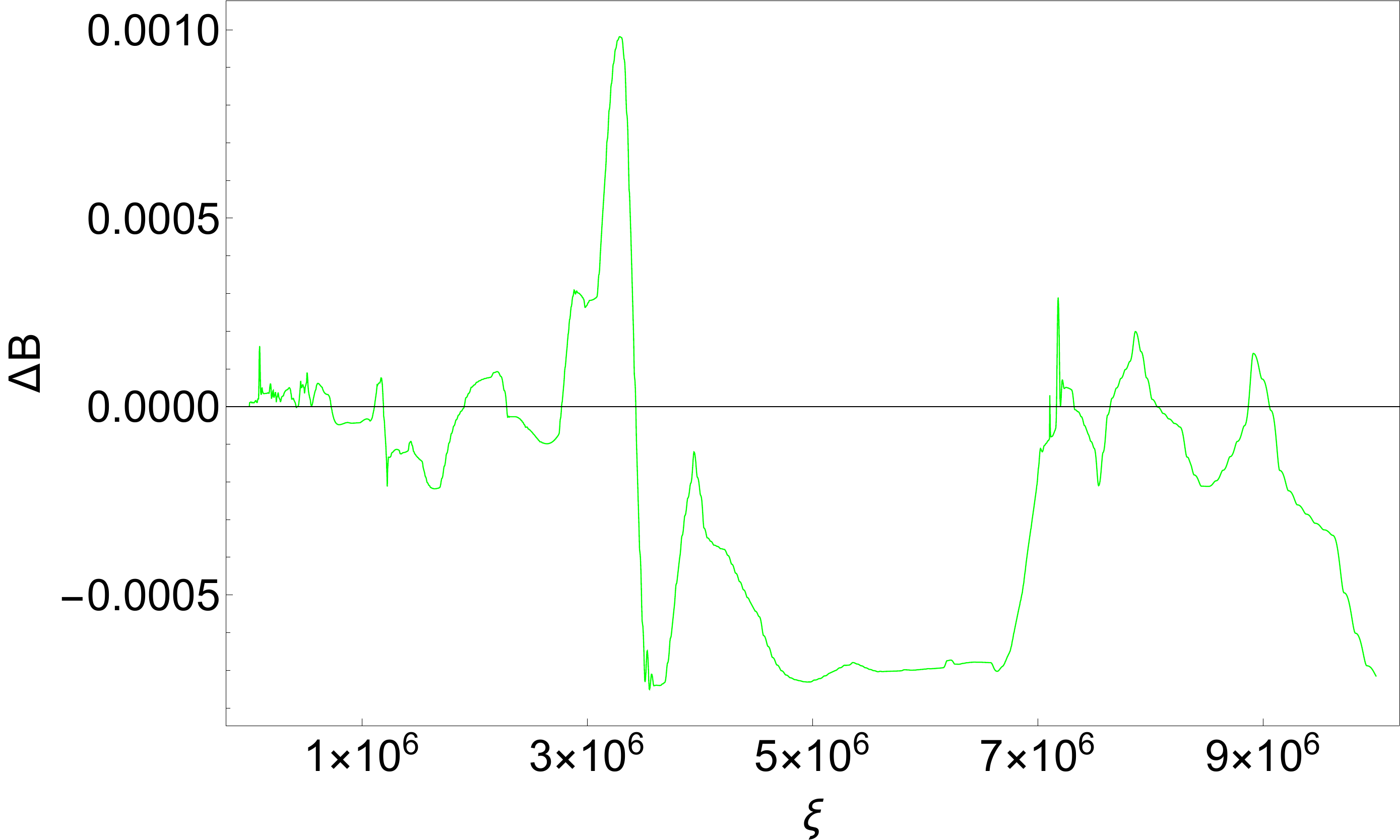} \\
		(c) & (d) \\[6pt]
	\end{tabular}
	\caption{$\Delta F$ and $\Delta B$ for $c_2=9 \times 10^{-7}$, $c_{14}=2 \times 10^{-7}$ and $c_{13}=0$. The panels (a) and (c) show the region outside the S0H,
	 while the panels (b) and (d) show the region inside the S0H. }
	\label{DeltaFB}
\end{figure}

\end{widetext}

  Finally, let us take a closer look at the difference between GR and $\ae$-theory, although  in the above we already mentioned that the results from these two theories are quite similar. 
   To see these more clearly, we first note that the GR counterparts of $F$ and $B$ are given by
   \bqn
  \lb{FBGR}
  F^{GR} = 1-\frac{r_{MH}}{r}, \quad B^{GR} = 1.
  \eqn
  Thus,  the relative differences can be  defined as
  \bqn
  \lb{deltaFB}
 \Delta F \equiv \frac{F-F^{GR}}{F^{GR}}, \quad \Delta B \equiv \frac{B-B^{GR}}{B^{GR}}.
  \eqn
 Again, considering the representative case $c_2=9 \times 10^{-7}$, $c_{14}=2 \times 10^{-7}$ and $c_{13}=0$, we plot out the differences  $\Delta F$ and $\Delta B$ in Fig. \ref{DeltaFB}, from which 
 we find that in the range  $\xi \in (10^{-12}, 1)$ we have ${\cal{O}}(\Delta F) \lesssim 10^{-9}$. On the other hand, in the range $\xi \in (1, 10^{12})$,  we have ${\cal{O}}(\Delta F) \lesssim 10^{-5}$. 
 Similarly, in the range $\xi \in (10^{-12}, 1)$,  we have ${\cal{O}}(\Delta B) \lesssim 10^{-8}$. In addition, in the range  $\xi \in  (1, 10^{7})$ we have ${\cal{O}}(\Delta B) \lesssim 10^{-3}$. Thus, 
 we confirm that $F$ and $B$ are indeed quite  {close to}  their GR limits.

\begin{figure}[tbp]
\centering
\includegraphics[width=1\columnwidth]{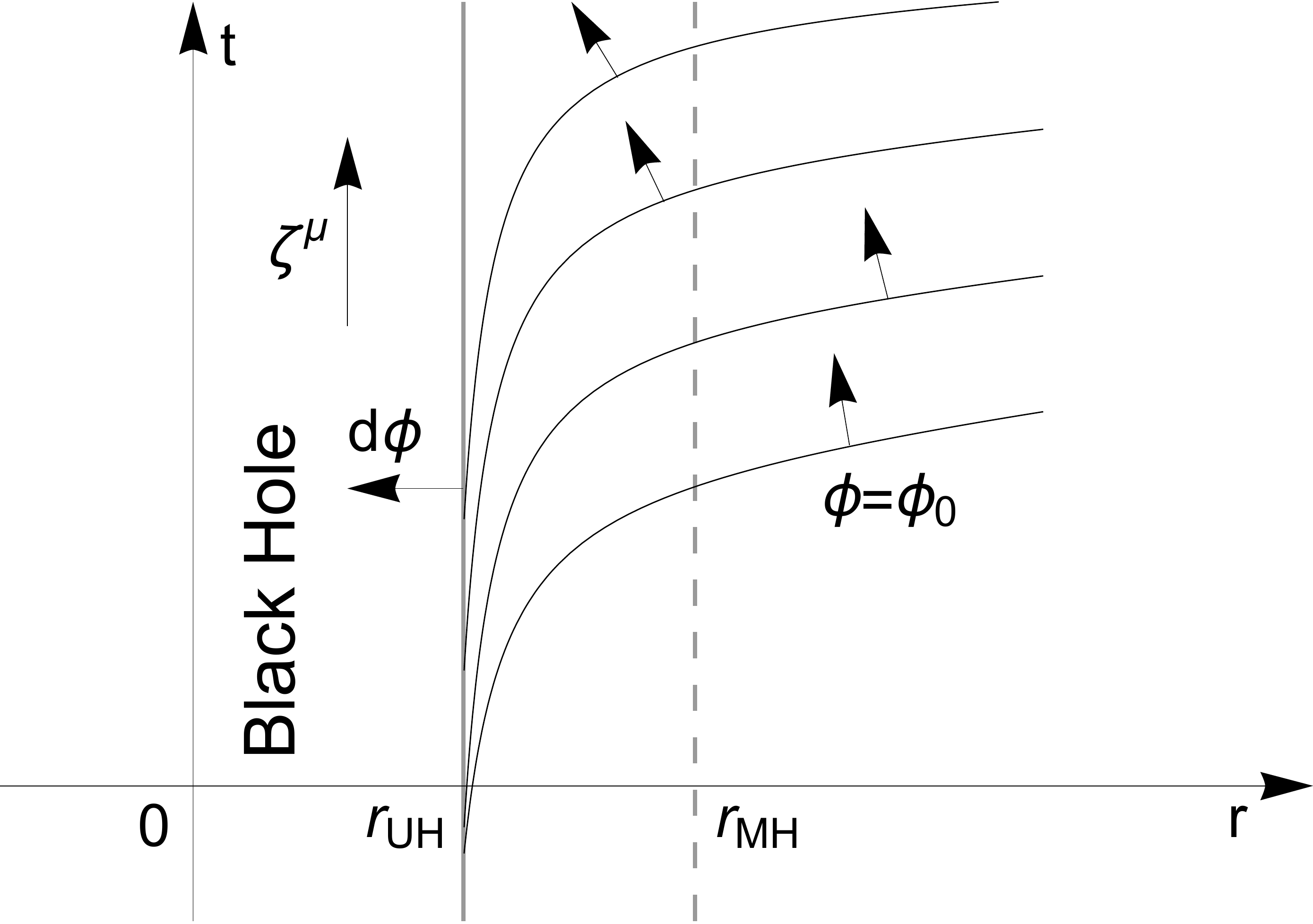}
\caption{Illustration of the bending of the $\phi$ = constant surfaces, and the existence of the UH  in a spherically symmetric static spacetime, 
where $\phi$ denotes the  globally timelike scalar  field, and $t$  is the Painlev\'{e}-Gullstrand-like coordinates, which covers the whole spacetime   \cite{LSW16}. Particles 
move always along the increasing direction of $\phi$. 
The Killing vector $\zeta^{\mu} = \delta^{\mu}_{v}$ always points upward at each point of the plane. The vertical dashed  line is   the location of the metric (Killing) horizon, 
$r=r_{MH}$. The UH, denoted by the vertical solid  line, is  located at $r = r_{UH}$, which is always inside the MH.} 
\label{fig9}
\end{figure}

\subsection{Universal Horizons}

 In theories with the broken LI,  the dispersion relation of a massive particle contains generically high-order momentum terms \cite{Wang17},
\bq
\lb{eq5.12}
E^{2} = m^{2} +  c_{k}^2k^{2}\left(1 + \sum^{2(z-1)}_{n=1}{a_{n}\left(\frac{k}{M_{*}}\right)^{n}}\right),
\eq
from which we can see that both of the group and phase velocities  
become unbounded as $k \rightarrow \infty$,
where $E$ and $k$ are the energy and momentum of the particle considered, and $c_k$ and $a_n$'s are coefficients, depending on the
species of the particle, while $M_{*}$ is the suppression energy  scale of the higher-dimensional operators. Note that there must be no confusion between  $c_k$ here and the 
four coupling constants $c_i$'s of the theory.
As an immediate result,    the causal
structure of the spacetimes in such theories is quite different from that given in GR,  where the light cone at a given point $p$ plays a fundamental 
role in determining the causal relationship of $p$ to other events \cite{GLLSW}.  In a UV complete theory, the above relationship is expected even in the 
gravitational sector. One of such examples is  { the  healthy extension  \cite{BPSa,BPSb} of Ho\v{r}ava gravity \cite{Horava,Wang17}, a possible UV extension of 
the khronometric theory (the HO $\ae$-theory \cite{Jacobson10,Jacobson14}).}

However, once LI is broken, the causal  structure will be dramatically changed. 
For example, in the Newtonian theory,  time is absolute and the speeds of signals are not limited. Then, the
causal structure of  a given point $p$ is uniquely determined by the time difference, $\Delta{t} \equiv t_{p} - t_{q}$, between the two events.  
 In particular, if $\Delta{t} > 0$, the event $q$ is to the past of $p$; if $\Delta{t} < 0$, it  is to the future; and if $\Delta{t} = 0$, the two events are
simultaneous.  In  theories with breaking  LI,  a similar situation occurs.

To provide a proper description of BHs in such theories,  UHs  were proposed \cite{BS11,Enrico11}, which represent 
 the absolute causal boundaries.   Particles even with infinitely large speeds    would just move on these
boundaries and  cannot escape to infinity.  The main idea is as follows. In a given spacetime,  a globally timelike scalar field  $\phi$ may exist \cite{LACW}.
In the spherically symmetric  case,  this globally timelike scalar field can be identified to the HO aether field $u_{\mu}$ via the relation (\ref{eq2.13}).
Then, similar to the 
Newtonian theory, this field defines  globally an absolute time, and all particles are assumed to move along the increasing direction of the timelike scalar field, so the causality is 
well defined. In such a spacetime, there may exist a surface at which the HO aether field $u_{\mu}$ is orthogonal to the timelike Killing vector,
$\zeta\; (\equiv \partial_v)$. Given that all particles move along the increasing direction of the HO aether field,   it is clear that a particle must cross this surface and move inward, 
once it arrives at it, no matter how large  its speed  is. This is a one-way membrane, and particles even with infinitely large speeds cannot escape from it, once they are 
inside it [cf. Fig. \ref{fig9}]. So, it acts as an absolute horizon to all particles (with any speed),  which is often called the UH \cite{BS11,Enrico11,Wang17}.  At the horizon, 
 as can be seen from Fig. \ref{fig9}, we have \cite{LGSW15},
\bq
\lb{eq5.12}
\left.\zeta \cdot u\right|_{r = r_{UH}}   = -\left. \frac{1}{2A}\left(1 + J\right)\right|_{r = r_{UH}} = 0,
\eq
where $J \equiv FA^2$. Therefore, the location of an UH is exactly the crossing point between the curve of $J(r)$ and the horizontal constant line $J = -1$,
as one can see from Figs. \ref{FAB} and \ref{FAB2}. From these figures we can also see that they are always located inside S0Hs, as expected. In addition, the curve $J(r)$ is oscillating
rapidly, and crosses the  horizontal line $J = -1$ back and forth infinite times. Therefore, in each case we have infinite number of $r_{UH-i}\; (i = 1, 2, ...)$. In this case, 
the UH is defined as the largest value of  $r_{UH-i}\; (i = 1, 2, ...)$.  In Table \ref{table4}, we show the locations of the first eight UHs for each case,  listed   in Tables 
\ref{table2} and \ref{table3}. It is interesting to note that the formation of multi-roots of UHs was  {first noticed in \cite{Enrico11}, and later} observed in gravitational collapse 
\cite{Bhattacharjee2018}.

\subsection{Other Observational Quantities}

 Another observationally interesting quantity is the  ISCO, which is the root of the equation, 
\bqn
\label{rISCO}
2 r {F'(r)}^2 -  F \left[3F'(r) +rF''(r)\right]=0.
\eqn
Note that in GR we have $ r_{ISCO}/r_{H}=3$ \cite{Paul2015}. Due to the tiny differences between the Schwarzschild solutions and the ones considered here, as shown in Fig. \ref{DeltaFB}, it is
expected that  $r_{ISCO}$'s in these cases  are also quite  {close to}  its GR limit. As a matter of fact, we find that this is indeed the case, and the differences in all the cases considered above appear only after six digits, that is, $\left|r_{ISCO} -  r_{ISCO}^{GR}\right| \le 10^{-6}$, as shown explicitly in  Tables \ref{table5} and  \ref{table6}.

In  Table \ref{table5}, we also show several other physical quantities. These  include the Lorentz gamma factor $\gamma_{ff}$,  the gravitational radius $r_g$, the orbital frequency of the ISCO 
$\omega_{ISCO}$, the maximum redshift $z_{max}$ of a photon emitted by a source orbiting the ISCO (measured at the infinity), 
and the impact parameter $b_{ph}$ of the  circular photon orbit (CPO), which are defined, respectively, by  \cite{Enrico11},
\bqn
\label{physical}
\gamma_{ff} &=& \left. \left( A+\frac{1}{4 A} \right)\right|_{r=r_{MH}}, \\
r_g &=& \left. - r_{S0H} \frac{d F}{d \xi} \right|_{\xi \to 0},\\
\omega_{ISCO} &=& \left. \sqrt{\frac{dF/dr}{2 r}} \right|_{r=r_{ISCO}}, \\
z_{max} &=& \left. \frac{1+\omega_{ISCO} r F^{-1/2}}{\sqrt{F-\omega_{ISCO}^2 r^2}} \right|_{r=r_{ISCO}}-1, ~~~~~~~~\\
b_{ph} &=& \left. \frac{r}{\sqrt{F}} \right|_{r=r_{ph}},~~~~
\eqn
where the radius $r_{ph}$ of the CPO  is defined as
\bqn
\label{rph}
\left. \left(2 F- r \frac{dF}{dr} \right) \right|_{r=r_{ph}} = 0.
\eqn

As pointed previously, these quantities are quite  {close to}  their relativistic limits, since they depend only on the spacetimes described by $F$ and $B$. As shown in Fig. \ref{DeltaFB}, 
the differences of these spacetimes between $\ae$-theory and GR are very small. To see this more clearly, let us introduce the quantities,
\bqn
\label{aeandGR}
\Delta r_{ISCO} &\equiv& \frac{r_{ISCO}}{r_{MH}}-\left(\frac{r_{ISCO}}{r_{MH}}\right)^{GR}, \nb\\
\Delta \omega_{ISCO} &\equiv& r_g \omega_{ISCO}-\left(r_g \omega_{ISCO}\right)^{GR}, \nb\\
\Delta z_{max} &\equiv& z_{max}-\left(z_{max}\right)^{GR}, \nb\\\
\Delta b_{ph} &\equiv& \frac{b_{ph}}{r_g}-\left(\frac{b_{ph}}{r_g}\right)^{GR}, 
\eqn
where the GR limits of $ r_{ISCO}/r_{MH}$,  $r_g \omega_{ISCO}$, $z_{max}$ and $b_{ph}/r_g$ are,  respectively,  $3, \; 2\times 6^{-3/2}$, $3/\sqrt{2}-1$ and $3 \sqrt{3}/2$. 
As can be seen from Table \ref{table6}, all of these quantities are fairly  {close to}  their GR limits. 

Therefore, we   conclude that it is quite difficult to distinguish GR and $\ae$-theory through the considerations of the 
physical quantities $r_{ISCO}$, $\omega_{ISCO}$, $z_{max}$ or $b_{ph}$, as far as the cases considered in this paper are concerned. 
Thus, it would be very
interesting to look for  other choices of $\{c_2, c_{13}, c_{14}\}$ (if there exist), which could result in distinguishable values in these observational  quantities.

\begin{table*}
	\caption{ $r_{UH-i}$'s for different cases listed in Tables \ref{table2} and \ref{table3}. Note that here we just show first eight UHs  of Eq.(\ref{eq5.12}) for each case.}  
	\label{table4}   
	\begin{tabular}{|c|c|c|c|c|c|c|c|c|} 
		\hline  
		$c_S^2$ & $r_{MH}/r_{UH-1}$  & $r_{MH}/r_{UH-2}$ &  $r_{MH}/r_{UH-3}$  & $r_{MH}/r_{UH-4}$  &$r_{MH}/r_{UH-5}$ & $r_{MH}/r_{UH-6}$ &  $r_{MH}/r_{UH-7}$ &  $r_{MH}/r_{UH-8}$ \\  
		\hline
		1.0049596 & 1.40913534  & 9.12519836 & 68.6766490 & 524.111256 & 4006.80012 & 30638.7274 & 234291.582 & 1791613.54
		\\  
		\hline
		1.3999982  &  1.39634652 & 6.27835216 & 33.1700700 & 178.825436 &  967.454326 & 5237.31538 & 28355.5481 & 153524.182 \\  
		\hline
		4.4999935  & 1.36429738  & 2.74101697 & 6.42094860 & 15.6447753 & 38.6164383  & 95.7784255 & 238.000717 & 591.850964  \\  
		\hline
		4.4999994  & 1.36429738  & 2.74101595 & 6.42094387 & 15.6447581  & 38.6163818 & 95.7782501 & 238.000194 & 591.849442  \\  
		\hline
		4.4999999  & 1.36429738  & 2.74101584 & 6.42094340 & 15.6447564 & 38.6163762  & 95.7782326 & 238.000141 & 591.849289  \\  
		\hline
		44.999939  &  1.33939835 & 1.56980254 & 1.91857535 & 2.41278107 & 3.08953425  & 4.00154026 & 5.22142096 & 6.84725395  \\  
		\hline
		449.93921  & 1.33429146  & 1.39226716 & 1.46010811 & 1.53855402 & 1.62835485  & 1.73026559 & 1.84507183 & 1.97362062  \\  
		\hline
	\end{tabular}
\end{table*} 


\begin{table*}
	\caption{ The quantities $r_{S0H}$, $\gamma_{ff}$,  $ r_{ISCO}$,  $\omega_{ISCO}$, $z_{max}$ and $b_{ph}$ for different cases listed in Tables \ref{table2} and \ref{table3} . }  
	\label{table5}   
	\begin{tabular}{|c|c|c|c|c|c|c|} 
		\hline  
		$c_S^2$ & $r_{MH}/r_{S0H}$  & $\gamma_{ff}$ &  $ r_{ISCO}/r_{MH}$  & $r_g \omega_{ISCO}$  & $z_{max}$ & $b_{ph}/r_g$  \\  
		\hline
		1.0049596 & 1.00058469  & 1.62614814 & 3.00000083 & 0.13608278  & 1.12132046 &  2.59807604 \\  
		\hline
		1.3999982  & 1.03812045   & 1.63971715  & 3.00000002 & 0.13608276  & 1.12132035  & 2.59807621  \\  
		\hline
		4.4999935  & 1.14507287   & 1.67376648  & 3.00000000 & 0.13608276  & 1.12132034  & 2.59807621  \\  
		\hline
		4.4999994  & 1.14507298   & 1.67376647  & 3.00000000 & 0.13608276  & 1.12132034  & 2.59807621  \\  
		\hline
		4.4999999  & 1.14507299   & 1.67376647  & 3.00000000 & 0.13608276  & 1.12132034  & 2.59807621  \\  
		\hline
		44.999939  & 1.26296693   & 1.69777578  & 3.00000000 & 0.13608276  & 1.12132034  & 2.59807621  \\  
		\hline
		449.93921  & 1.30916545   & 1.70149318  & 3.00000000 & 0.13608276  & 1.12132034  & 2.59807621  \\  
		\hline
	\end{tabular}
\end{table*}

\begin{table*}
	\caption{ $ \Delta r_{ISCO}$,  $ \Delta \omega_{ISCO}$, $\Delta z_{max}$ and $ \Delta b_{ph}$ for different cases listed in Tables \ref{table2} and \ref{table3} . } 
	\label{table6}   
	\begin{tabular}{|c|c|c|c|c|} 
		\hline  
		$c_S^2$ & $ \Delta r_{ISCO}$  & $ \Delta \omega_{ISCO}$  & $\Delta z_{max}$ & $ \Delta b_{ph}$  \\  
		\hline
		1.0049596  & $8.3 \times 10^{-7}$ & $1.3 \times 10^{-8}$ & $1.2 \times 10^{-7}$ & $-1.7 \times 10^{-7}$ \\  
		\hline
		1.3999982  & $1.8 \times 10^{-8}$  & $2.2 \times 10^{-10}$  & $2.0 \times 10^{-9}$  & $-3.2 \times 10^{-9}$ \\  
		\hline
		4.4999935  & $4.0 \times 10^{-9}$  & $1.5 \times 10^{-12}$  & $4.9 \times 10^{-11}$  & $-4.2 \times 10^{-10}$  \\  
		\hline
		4.4999994  & $4.0 \times 10^{-10}$  & $1.5 \times 10^{-11}$  & $-7.2 \times 10^{-11}$  &  $-3.2 \times 10^{-10}$  \\  
		\hline
		4.4999999  & $4.0 \times 10^{-11}$  & $2.3 \times 10^{-11}$  &  $-1.2 \times 10^{-10}$ & $-4.5 \times 10^{-10}$ \\  
		\hline
		44.999939  & $1.5 \times 10^{-10}$  & $9.6 \times 10^{-11}$  & $-5.6 \times 10^{-10}$  & $-1.9 \times 10^{-9}$ \\  
		\hline
		449.93921  & $1.1 \times 10^{-9}$  & $1.1 \times 10^{-11}$  & $-4.5 \times 10^{-10}$  & $-1.1 \times 10^{-9}$  \\  
		\hline
	\end{tabular}
\end{table*}

\section{Conclusions}
 \renewcommand{\theequation}{6.\arabic{equation}} \setcounter{equation}{0}
 
In this paper, we have systematically studied static spherically symmetric spacetimes in the framework of Einstein-aether theory, by paying particular attention to black holes that have regular 
S0Hs. In $\ae$-theory, a timelike vector - the aether, exists over the whole spacetime. As a result, in contrast to GR, now there are three gravitational modes, referred to
as,  respectively, the spin-0, spin-1 and spin-2 gravitons. 

To avoid the vacuum gravi-\v{C}erenkov radiation, all these modes must propagate with speeds greater than or at least equal to the 
speed of light  \cite{EMS05}. However, in the spherically symmetric spacetimes, only the spin-0 mode is relevant in the gravitational sector \cite{Jacobson}, and the boundaries of BHs
 are defined by this mode, which are the 
null surfaces with respect to the metric $g^{(S)}_{\mu\nu}$ defined  {in Eq.(\ref{eq1.3}),} the  so-called S0Hs.   Since now
$c_S \ge c$, where $c_S$ is the speed of the spin-0 mode, the S0Hs   are always inside or at most coincide with the metric (Killing) horizons. 
Then, in order to cover spacetimes both inside and outside  the MHs, working in the Eddington-Finkelstein coordinates (\ref{eq2.14}) is one of the natural choices.

In the process of gravitational radiations of compact objects, all of these three fundamental modes will be emitted, and the GW forms 
and energy loss rate should  be different from that of GR. In particular, to the leading order, both monopole and dipole emissions will co-exist with the quadrupole emission 
\cite{Foster07,Yagi13,Yagi14,HYY15,Kai19,Zhao19,Zhang20}. 
Despite of all these, it is remarkable that the theory still remains as a viable theory, and satisfies  all the constraints,
both theoretical and observational \cite{OMW18}, including the recent detection of the GW,  GW170817, observed by the LIGO/Virgo collaboration  
 \cite{GW170817}, which imposed the severe constraint on the  speed of the spin-2 gravitational mode, $- 3\times 10^{-15} < c_T -1 < 7\times 10^{-16}$. Consequently, it is one of few theories that violate
 Lorentz symmetry and meantime  is still consistent with all the observations carried out so far \cite{OMW18,Berti18a}.
 
Spherically symmetric static BHs in $\ae$-theory have been extensively studied both analytically  
\cite{Eling2006-1, Oost2019, Per12, Dingq15, Ding16, Kai19b, Ding19, Gao2013, Chan2020, Leon2019, Leon2020} 
and numerically \cite{Eling2006-2, Eling2007, Tamaki2008, BS11,Enrico11, Enrico2016, Zhu2019}, and various solutions have been obtained. Unfortunately, all these
solutions have been ruled out by current observations  \cite{OMW18}.

 Therefore, as a first step, in this paper we have investigated spherically symmetric static BHs in $\ae$-theory that satisfy all the
 observational constraints found lately in \cite{OMW18} in detail, and presented  various numerical new BH solutions.
 In particular, we have first shown  explicitly that among the five non-trivial field equations,  only three of them are independent. More important, 
 the two second-order differential equations given by  Eqs.(\ref{eq2.22a}) and (\ref{eq2.22b}) for  the two functions $F(r)$ and $A(r)$  are independent of the function $B(r)$, where $F(r)$ and $B(r)$ 
 are the metric coefficients  of the Eddington-Finkelstein metric (\ref{eq2.14}), and $A(r)$ describes the aether field, as shown by Eq.(\ref{eq2.15}).
Thus,   one  can first solve Eqs.(\ref{eq2.22a}) and (\ref{eq2.22b}) to find $F(r)$ and $A(r)$, and then from the third independent equation to find $B(r)$.
Another remarkable feature is that the function $B(r)$ can be obtained from the constraint (\ref{eq2.23a}), and  is given simply by  the algebraic expression of $F,\; A$ and their derivatives, 
as shown explicitly by Eq.(\ref{Bcv}). This not only saves the computational labor, but also makes the calculations more accurate, as pointed out explicitly in  \cite{Enrico11}, 
solving the first-order differential equation (\ref{eq2.22c}) for $B(r)$ can  ``potentially be affected by numerical inaccuracies when evaluated very  {close to}  the horizon".

Then, {\it now solving the (vacuum) field equations of spherically symmetric static spacetimes in $\ae$-theory simply reduces to solve
the two second-order differential equations (\ref{eq2.22a}) and (\ref{eq2.22b}).} This will considerably simplify the mathematical computations, which is very important, especially considering the fact
that the field equations involved are extremely complicated, as one can see from Eqs.(\ref{eq2.22a})-(\ref{eq2.23a}) and (\ref{fns}) - (\ref{nns}).  {Then, in the case $c_{13} = c_{14} = 0$ we have been
able to solve these equations  explicitly, and obtained a three-parameter family of exact solutions, which in general  depends on the coupling constant $c_2$. However,   requiring that the solutions 
be asymptotically flat, we have found that the solutions become independent of $c_2$, and the corresponding metric reduces
precisely to the Schwarzschild BH solution with a non-trivially coupling aether field given by Eq.(\ref{eqA6}), which is always timelike even in the region inside the BH. }

To simplify the problem further,  we have also taken the advantage of  the field redefinitions that are allowed by the internal symmetry of $\ae$-theory, first discovered by Foster in \cite{Foster05},
and later were used frequently, including   the works of \cite{Eling2006-2,Tamaki2008,Enrico11}. The advantage of the field redefinitions is that it allows us to choose the free parameter $\sigma$ involved in the 
field redefinitions, so that 
the S0H of the redefined metric $\tilde{g}_{\mu\nu}$ will coincide with its MH. This will reduce the four-dimensional space of the initial conditions,  spanned by $\tilde F_H, \; \tilde F'_H,\; 
\tilde A_H,\; \tilde A'_H$, to one-dimension, spanned only by $\tilde A_H$, if the initial conditions are imposed on the S0H. In Sec. III.D.  we have shown step by step how one can do it. In addition, in this same subsection we have also shown that the field equations 
are invariant under the rescaling $r \rightarrow C r$. In fact, introducing the dimensionless coordinate $\xi \equiv r_{S0H}/r$, the relevant four field equations take the scaling-invariant forms of Eqs.(\ref{eq2.36a})
- (\ref{eq2.36d}), which are all independent of $r_{S0H}$. Thus, when integrating these equations, without loss of generality, one can assign any value to $r_{S0H}$.

We would like also to note that in Section III.C we worked out the relations in detail among the fields ($g_{\mu\nu}, u^{\mu}, c_i)$, ($\hat g_{\mu\nu}, \hat u^{\mu}, \hat c_i)$ 
and ($\tilde g_{\mu\nu}, \tilde u^{\mu}, \tilde c_i)$, and clarified several subtle points. In particular, the redefined metric $\hat g_{\mu\nu}$  through Eqs.(\ref{eq2.2}) 
and (\ref{eq2.3}) does not take the standard form in the Eddington-Finkelstein coordinates, as shown explicitly by Eq.(\ref{eq2.16b}). Instead, only after a proper  coordinate transformation 
given by Eqs.(\ref{eq2.17}) and (\ref{C0Cr}), the resulting metric $\tilde g_{\mu\nu}$ takes the standard form, as given by Eq.(\ref{eq2.19}). Then, the field equations 
for ($\tilde g_{\mu\nu}, \tilde u^{\mu}, \tilde c_i)$  take the same forms as the ones for ($g_{\mu\nu}, u^{\mu}, c_i)$. 
Therefore, when we solved the field equations in terms of the redefined fields, they are the ones  of ($\tilde g_{\mu\nu},\; \tilde u^{\mu}$),
 not the ones for  ($\hat g_{\mu\nu}, \; \hat u^{\mu}$).

After clarifying all these subtle points,  in Sec. IV, we have worked out the detail on how to  carry out explicitly  our numerical analysis. In particular,   to monitor 
 the numerical errors of our code, we have  introduced the quantity $\tilde{\cal{C}}$ through Eq.(\ref{scC}), which is essentially  Eq.(\ref{eq2.30c}).
 Theoretically, it vanishes identically. But, due to numerical errors, it is expected that $\tilde{\cal{C}}$ has non-zero values, and the amplitude of it will provide a good indication on
 the numerical errors that our numerical code could produce.

 To show  further the accuracy of our numerical code, we have first reproduced the BH solutions obtained in  \cite{Eling2006-2,Enrico11}, but with an accuracy that are at least two orders higher 
  [cf. Table \ref{table1}]. It should be noted that  all these BH solutions have been ruled out by the current observations \cite{OMW18}. So, after checking our numerical code, 
in Sec. IV.B, we considered various new BH solutions  that satisfy all the observational constraints \cite{OMW18}, and presented  them in Tables \ref{table2} and \ref{table3}, as well as in Figs.
 \ref{FABtilde3}-\ref{FABtilde2}. 
 
  Then,  in Sec. V, we have presented   the  physical metric $g_{\mu\nu}$ and $\ae$-field $u^{\mu}$  for these viable new BH solutions obtained in Section IV. Before presenting the results, we have
  first shown that 
  the physical fields, ${g}_{\mu\nu}$ and ${u}^{\mu}$,  are also asymptotically flat, provided that the effective fields $\tilde g_{\mu\nu}$ and $\tilde u^{\mu}$ are [cf.
  Eqs.(\ref{normal}) and (\ref{normal2})]. Then, the physical BH solutions were plotted out in Figs. \ref{FAB} and \ref{FAB2}. Among several interesting features, we would like to point out 
  the different locations of the metric and spin-0 horizons for the physical metric ${g}_{\mu\nu}$, denoted by full solid circles and pentagrams, respectively.
  
   Another interesting point is that all these physical BH solutions are quite similar to the Schwarzschild one. In Fig. \ref{DeltaFB} we have shown the differences for the case  $c_2=9 \times 10^{-7}$, 
  $c_{14}=2 \times 10^{-7}$ and $c_{13}=0$, but similar results also hold for the other cases, listed in Tables \ref{table2} and \ref{table3}.
  
  In this section, we have also identified the locations of  the UHs of these solutions and several other observationally interesting  quantities, which include 
the ISCO $r_{ISCO}$,  the Lorentz gamma factor $\gamma_{ff}$,  the gravitational radius $r_g$, the orbital frequency $\omega_{ISCO}$ of the ISCO, the maximum redshift $z_{max}$ of a photon emitted by a source orbiting the ISCO (measured at the infinity), the radii $r_{ph}$ of the CPO, and the impact parameter 
$b_{ph}$ of the CPO. All of them are given in Table \ref{table4}-\ref{table5}. In Table \ref{table6} we also calculated the differences of these quantities obtained in $\ae$-theory and GR.
Looking at these results, we conclude  that it's very hard to distinguish GR and $\ae$-theory through these quantities, as far as the cases considered in this paper are concerned.
We would also like to note that for each BH solution, there are infinite number of UHs, $r = r_{UH-i}, \; (i = 1, 2, 3, ...)$,  {which was also observed in \cite{Enrico11}}. 
In Table \ref{table4} we have listed the first eight of them, and the largest one is
usually defined as the UH of the BH. In contrast, there are only one S0H and one MH for each solution. These features are also found in the gravitational collapse of 
a massless scalar field in $\ae$-theory \cite{Bhattacharjee2018}.
  
  An immediate implication of the above results  is that the QNMs of these BHs  for a test field, scalar, vector or tensor \cite{KZ07a}, will be quite similar to these given 
  in GR. Our preliminary results on such studies  indicate that this is indeed the case. However, we expect that there should be significant differences from GR, 
  when we consider the  metric perturbations of these  BH solutions - the gravitational spectra of perturbations  \cite{KZ07b}, as now the BH boundaries are the locations of the S0Hs, not 
  the locations of the MHs. This should be specially true for the cases with large speeds $c_S$ of the spin-0 modes, as in these cases the S0Hs are significantly different from the 
  MHs, and located deeply inside  them. Thus, imposing the non-out-going radiation on the S0Hs will be quite different from imposing the non-out-going radiation on the corresponding 
  MHs.   We wish to report our results along this direction soon in another occasion.

\begin{acknowledgments}

We would like very much to express our gratitude to Ted Jacobson, Ken Yagi, and Nico Yunes, for their valuable comments and suggestions, which lead us to improve the paper considerably.
This work is partially supported by the National Natural Science Foundation of China (NNSFC) under Grant Nos. 11675145, 11805166, 11975203,  11773028, 11633001, and 11653002.
\end{acknowledgments}

\section*{Appendix A: The coefficients of $f_n, a_n, b_n$ and $n_n$}
\renewcommand{\theequation}{A.\arabic{equation}} \setcounter{equation}{0}

In this appendix, we shall provide the explicit expressions of the coefficients of $f_n, a_n, b_n$ and $n_n$, encountered in the Einstein-aether field equations in the spherically symmetric spacetimes, for 
which the metric is written in the  Eddington-Finkelstein  coordinates (\ref{eq2.14}), with the aether field taking the form of Eq.(\ref{eq2.15}). In particular, the coefficients of $f_n, a_n$ and $b_n$ appearing in 
Eqs.(\ref{eq2.22a}) - (\ref{eq2.22c}) are given by,
\begin{widetext}
\bqn
\lb{fns}
f_0&=&-4 \left(c_2+c_{13}\right) \left(c_2+c_{13}-\left(c_2+1\right) c_{14}\right) r A(r) A'(r) \nb\\
&&-\left(c_{14} c_2^2-c_2^2+c_{13}^2+\left(c_2+1\right) c_{14}^2-\left(c_2+2\right) c_{13} c_{14}\right) r^2 A'(r)^2 \nb\\
&&-4 \left(\left(c_{14}+1\right) c_{13}^2+2 \left(c_2+1\right) c_{13}+\left(c_2-2\right) c_{14} c_{13}+c_2 \left(c_2+2\right)-2 \left(c_2^2+3 c_2+1\right) c_{14}\right) r A(r)^4 F'(r) \nb\\
&&+2 \left(c_2^2+\left(2-3 c_2\right) c_{14} c_2-9 c_{13} c_{14} c_2+\left(c_2+1\right) c_{14}^2-c_{13}^2 \left(4 c_{14}+1\right)\right) r^2 A(r)^3 A'(r) F'(r) \nb\\
&&-\left(5 c_{14} c_2^2-c_2^2+\left(c_2+1\right) c_{14}^2+\left(7 c_2-2\right) c_{13} c_{14}+c_{13}^2 \left(4 c_{14}+1\right)\right) r^2 A(r)^6 F'(r)^2 \nb\\
&&-2 \left(c_2+c_{13}\right) c_{14} \left(-c_2+c_{13}+c_{14}-4\right) r^3 A(r)^5 A'(r) F'(r)^2+\left(c_2+c_{13}\right) c_{14} \left(c_2-c_{13}+c_{14}\right) r^3 A(r)^8 F'(r)^3 \nb\\
&&-\left(c_2+c_{13}\right) A(r)^2 \left(-c_{14} \left(c_2-c_{13}+c_{14}\right) r^3 A'(r)^2 F'(r)+4 c_2 \left(c_{14}-1\right)+2 c_{13} c_{14}\right),\nb\\
f_1&=&-8 \left(c_2+c_{13}\right) \left(c_{14} c_2+c_2+c_{13}+c_{14}\right) r A(r)^3 A'(r) \nb\\
&&+4 \left(c_2^2-3 c_{13} c_{14} c_2+2 c_{14} c_2+\left(c_2+1\right) c_{14}^2-c_{13}^2 \left(c_{14}+1\right)\right) r^2 A(r)^2 A'(r)^2 \nb\\
&&-4 \left(\left(1-2 c_{14}\right) c_{13}^2+\left(-c_{14} c_2+2 c_2+c_{14}+4\right) c_{13}+c_2 \left(c_2+4\right)+\left(5 c_2^2+9 c_2+4\right) c_{14}\right) r A(r)^6 F'(r) \nb\\
&&+2 \left(c_{14} c_2^2+3 c_2^2-3 \left(c_2+1\right) c_{14}^2+\left(11 c_2+6\right) c_{13} c_{14}+c_{13}^2 \left(4 c_{14}-3\right)\right) r^2 A(r)^5 A'(r) F'(r) \nb\\
&&+2 \left(c_2^2+\left(3 c_2+2\right) c_{14} c_2+3 c_{13} c_{14} c_2+\left(c_2+1\right) c_{14}^2+c_{13}^2 \left(2 c_{14}-1\right)\right) r^2 A(r)^8 F'(r)^2 \nb\\
&&+2 \left(c_2+c_{13}\right) c_{14} \left(c_2-c_{13}+c_{14}\right) r^3 A(r)^7 A'(r) F'(r)^2 \nb\\
&&-2 \left(c_2+c_{13}\right) c_{14} A(r)^4 \left(\left(-c_2+c_{13}+c_{14}-4\right) r^3 A'(r)^2 F'(r)-4 \left(2 c_2+c_{13}+1\right)\right),\nb\\
f_2&=&2 \left(c_{14} c_2^2+3 c_2^2-3 \left(c_2+1\right) c_{14}^2+\left(11 c_2+6\right) c_{13} c_{14}+c_{13}^2 \left(4 c_{14}-3\right)\right) r^2 A(r)^4 A'(r)^2 \nb\\
&&+4 \left(-\left(c_{14}-1\right) c_{13}^2+2 \left(c_2-1\right) c_{13}+\left(c_2+4\right) c_{14} c_{13}+\left(c_2-2\right) c_2+\left(4 c_2^2+8 c_2+2\right) c_{14}\right) r A(r)^8 F'(r) \nb\\
&&+6 \left(\left(c_{14}+1\right) c_2^2+c_{14} \left(-c_{13}+c_{14}+2\right) c_2-c_{13}^2+c_{14}^2\right) r^2 A(r)^7 A'(r) F'(r) \nb\\
&&+\left(-\left(c_{14}-1\right) c_2^2+\left(c_{13}-c_{14}\right) c_{14} c_2-\left(c_{13}-c_{14}\right){}^2\right) r^2 A(r)^{10} F'(r)^2 \nb\\
&&+\left(-c_2-c_{13}\right) A(r)^6 \left(-c_{14} \left(c_2-c_{13}+c_{14}\right) r^3 A'(r)^2 F'(r)+8 c_2+4 \left(6 c_2+3 c_{13}+4\right) c_{14}\right),\nb\\
f_3&=&8 \left(c_2+c_{13}\right) \left(c_{14} c_2+c_2+c_{13}+c_{14}\right) r A(r)^7 A'(r)+8 \left(2 c_2^2+3 c_{13} c_2+c_2+c_{13}^2+c_{13}\right) c_{14} A(r)^8 \nb\\
&&+4 \left(c_2^2-3 c_{13} c_{14} c_2+2 c_{14} c_2+\left(c_2+1\right) c_{14}^2-c_{13}^2 \left(c_{14}+1\right)\right) r^2 A(r)^6 A'(r)^2 \nb\\
&&+4 \left(c_2+c_{13}\right) \left(c_2+c_{13}-\left(c_2+1\right) c_{14}\right) r A(r)^{10} F'(r) \nb\\
&&+2 \left(c_2-c_{13}+c_{14}\right) \left(c_2+c_{13}-\left(c_2+1\right) c_{14}\right) r^2 A(r)^9 A'(r) F'(r),\nb\\
f_4&=&4 \left(c_2+c_{13}\right) \left(c_2+c_{13}-\left(c_2+1\right) c_{14}\right) r A(r)^9 A'(r)-2 \left(c_2+c_{13}\right) \left(2 c_2 \left(c_{14}-1\right)+c_{13} c_{14}\right) A(r)^{10} \nb\\
&&+\left(-c_{14} c_2^2+c_2^2-c_{13}^2-\left(c_2+1\right) c_{14}^2+\left(c_2+2\right) c_{13} c_{14}\right) r^2 A(r)^8 A'(r)^2,
\eqn
\bqn
\lb{ans}
a_0&=&4 \left(-\left(c_{14}-1\right) c_{13}^2+\left(-c_{14} c_2+2 c_2+2 c_{14}-2\right) c_{13}+\left(c_2-2\right) c_2+2 \left(c_2^2+3 c_2+1\right) c_{14}\right) r A(r)^2 A'(r) \nb\\
&&+\left(c_{13}^2+\left(5 c_2 c_{14}+8\right) c_{13}-\left(c_2+1\right) c_{14}^2-\left(c_2-8\right) c_2-\left(5 c_2^2+18 c_2+8\right) c_{14}\right) r^2 A(r) A'(r)^2 \nb\\
&&+\left(c_2+c_{13}\right) c_{14} \left(c_2-c_{13}+c_{14}\right) r^3 A'(r)^3+4 \left(c_2+c_{13}\right) \left(c_{14} c_2+c_2+c_{13}+c_{14}\right) r A(r)^5 F'(r) \nb\\
&&-2 \left(\left(2 c_{14}-1\right) c_{13}^2+\left(\left(3 c_2-2\right) c_{14}+4\right) c_{13}-\left(c_2+1\right) c_{14}^2+c_2 \left(c_2+4\right)+\left(3 c_2^2+4 c_2+4\right) c_{14}\right) r^2 A(r)^4 A'(r) F'(r) \nb\\
&&+\left(-c_2^2-\left(c_2+2\right) c_{14} c_2+c_{13} c_{14} c_2+c_{13}^2-\left(c_2+1\right) c_{14}^2\right) r^2 A(r)^7 F'(r)^2 \nb\\
&&+\left(c_2+c_{13}\right) c_{14} \left(c_2-c_{13}+c_{14}\right) r^3 A(r)^6 A'(r) F'(r)^2 \nb\\
&&-2 \left(c_2+c_{13}\right) A(r)^3 \left(c_{14} \left(-c_2+c_{13}+c_{14}-4\right) r^3 A'(r)^2 F'(r)+2 c_2+2 c_2 c_{14}+c_{13} c_{14}+4\right),\nb\\
%
a_1&=&4 \left(\left(2 c_{14}+1\right) c_{13}^2+\left(3 c_{14} c_2+2 c_2+c_{14}-4\right) c_{13}+\left(c_2-4\right) c_2-\left(3 c_2^2+7 c_2+4\right) c_{14}\right) r A(r)^4 A'(r) \nb\\
&&+\left(-\left(8 c_{14}-3\right) c_{13}^2-\left(\left(23 c_2+14\right) c_{14}-8\right) c_{13}+3 \left(c_2+1\right) c_{14}^2-c_2 \left(3 c_2-8\right)-\left(c_2^2-8 c_2-8\right) c_{14}\right) r^2 A(r)^3 A'(r)^2 \nb\\
&&-2 \left(c_2+c_{13}\right) c_{14} \left(-c_2+c_{13}+c_{14}-4\right) r^3 A(r)^2 A'(r)^3-8 \left(c_2+1\right) \left(c_2+c_{13}\right) c_{14} r A(r)^7 F'(r) \nb\\
&&+4 \left(\left(c_{14}+1\right) c_{13}^2+2 \left(c_2 c_{14}-1\right) c_{13}-\left(c_2+1\right) c_{14}^2-c_2 \left(c_2+2\right)+\left(c_2^2+2 c_2+2\right) c_{14}\right) r^2 A(r)^6 A'(r) F'(r) \nb\\
&&+\left(c_{14} c_2^2-c_2^2+c_{13}^2+\left(c_2+1\right) c_{14}^2-\left(c_2+2\right) c_{13} c_{14}\right) r^2 A(r)^9 F'(r)^2 \nb\\
&&-2 \left(c_2+c_{13}\right) A(r)^5 \left(-c_{14} \left(c_2-c_{13}+c_{14}\right) r^3 A'(r)^2 F'(r)-2 c_2-\left(6 c_2+3 c_{13}+4\right) c_{14}\right),\nb\\
%
a_2&=&-2 \left(c_2+c_{13}\right) \left(\left(6 c_2+3 c_{13}+4\right) c_{14}-2 \left(c_2+2\right)\right) A(r)^7 \nb\\
&&-4 \left(\left(c_{14}+1\right) c_{13}^2+\left(c_2 \left(3 c_{14}+2\right)+2\right) c_{13}+c_2 \left(c_2+2\right)-2 \left(2 c_2+1\right) c_{14}\right) r A(r)^6 A'(r) \nb\\
&&+\left(-3 c_2^2+\left(5 c_2-6\right) c_{14} c_2+19 c_{13} c_{14} c_2-3 \left(c_2+1\right) c_{14}^2+c_{13}^2 \left(8 c_{14}+3\right)\right) r^2 A(r)^5 A'(r)^2 \nb\\
&&+\left(c_2+c_{13}\right) c_{14} \left(c_2-c_{13}+c_{14}\right) r^3 A(r)^4 A'(r)^3-4 \left(c_2+c_{13}\right) \left(c_2+c_{13}-\left(c_2+1\right) c_{14}\right) r A(r)^9 F'(r) \nb\\
&&-2 \left(c_2-c_{13}+c_{14}\right) \left(c_2+c_{13}-\left(c_2+1\right) c_{14}\right) r^2 A(r)^8 A'(r) F'(r),\nb\\
%
a_3&=&2 \left(c_2+c_{13}\right) \left(2 c_2 \left(c_{14}-1\right)+c_{13} c_{14}\right) A(r)^9-4 \left(c_2+c_{13}\right) \left(c_2+c_{13}-\left(c_2+1\right) c_{14}\right) r A(r)^8 A'(r) \nb\\
&&+\left(c_{14} c_2^2-c_2^2+c_{13}^2+\left(c_2+1\right) c_{14}^2-\left(c_2+2\right) c_{13} c_{14}\right) r^2 A(r)^7 A'(r)^2,\nb\\
\eqn
and
\bqn
\lb{bns}
b_0&=&4 \left(c_2+1\right) \left(c_2+c_{13}\right) c_{14} A(r)^2-4 \left(c_2+c_{13}\right){}^2 c_{14} r A(r) A'(r) \nb\\
&&+\left(c_2+c_{13}\right) c_{14} \left(c_2-c_{13}+c_{14}\right) r^2 A'(r)^2-4 \left(c_2+c_{13}\right){}^2 c_{14} r A(r)^4 F'(r) \nb\\
&&-2 \left(c_2+c_{13}\right) c_{14} \left(-c_2+c_{13}+c_{14}-4\right) r^2 A(r)^3 A'(r) F'(r) \nb\\
&&+\left(c_2+c_{13}\right) c_{14} \left(c_2-c_{13}+c_{14}\right) r^2 A(r)^6 F'(r)^2,\nb\\
%
b_1&=&-2 \left(c_2+c_{13}\right) c_{14} \left(-c_2+c_{13}+c_{14}-4\right) r^2 A(r)^2 A'(r)^2-8 \left(c_2+1\right) \left(c_2+c_{13}\right) c_{14} A(r)^4 \nb\\
&&+2 \left(c_2+c_{13}\right) c_{14} \left(c_2-c_{13}+c_{14}\right) r^2 A(r)^5 A'(r) F'(r)+4 \left(c_2+c_{13}\right){}^2 c_{14} r A(r)^6 F'(r),\nb\\
%
b_2&=&4 \left(c_2+c_{13}\right){}^2 c_{14} r A(r)^5 A'(r)+4 \left(c_2+1\right) \left(c_2+c_{13}\right) c_{14} A(r)^6 \nb\\
&&+\left(c_2+c_{13}\right) c_{14} \left(c_2-c_{13}+c_{14}\right) r^2 A(r)^4 A'(r)^2.
\eqn

On the other hand, the coefficients $n_n$'s appearing in Eq.(\ref{eq2.23a}) are given by
\bqn
\lb{nns}
n_0&=&\frac{c_2 A'(r)}{2 r A(r)^3 B(r)^3}-\frac{c_2}{2 r^2 A(r)^2 B(r)^3}-\frac{c_{13}}{4 r^2 A(r)^2 B(r)^3}-\frac{1}{r^2 B(r)} \nb\\
&&-\frac{\left(c_2+c_{13}+c_{14}\right) A'(r) F'(r)}{4 A(r) B(r)^3}-\frac{\left(c_2+c_{13}-c_{14}\right) A'(r)^2}{8 A(r)^4 B(r)^3}+\frac{\left(c_2+2\right) F'(r)}{2 r B(r)^3} \nb\\
&&-\frac{\left(c_2+c_{13}-c_{14}\right) A(r)^2 F'(r)^2}{8 B(r)^3},\nb\\
%
%
n_1&=&\frac{\left(-c_2-c_{13}-c_{14}\right) A'(r)^2}{4 A(r)^2 B(r)^3}-\frac{c_2 A(r)^2 F'(r)}{2 r B(r)^3}+\frac{2 c_2+c_{13}+2}{2 r^2 B(r)^3} \nb\\
&&+\frac{\left(-c_2-c_{13}+c_{14}\right) A(r) A'(r) F'(r)}{4 B(r)^3},\nb\\
%
n_2&=&-\frac{c_2 A(r) A'(r)}{2 r B(r)^3}+\frac{\left(-c_2-c_{13}+c_{14}\right) A'(r)^2}{8 B(r)^3}+\frac{\left(-2 c_2-c_{13}\right) A(r)^2}{4 r^2 B(r)^3}.
\eqn

When $c_S=1$, i.e. $c_2=\left(-2 c_{13}+2 c_{14}- c_{13}^2 c_{14}\right)/\left(2-4 c_{14}+3 c_{13} c_{14}\right)$, the   coefficients $f_0, a_0, b_0$ and $n_0$ reduce to
\bqn
\lb{fab0Special}
f_0&=&  \left[ r A(r)^2 F'(r)+\frac{c_{14} c_{13}+2 c_{13}-2 c_{14}}{2 \left(c_{13}-1\right) c_{14}} \right] b_0,\nb\\
a_0&=&   \frac{2 \left(c_{13}-1\right) c_{14} r A'(r)+\left(c_{13}-2\right) \left(c_{14}-2\right) A(r)}{c_{13} \left(c_{14} \left(2 r A(r)^2 F'(r)+1\right)+2\right)-2 c_{14} \left(r A(r)^2 F'(r)+1\right)} f_0,\nb\\
b_0&=&\frac{1}{\left(\left(3 c_{13}-4\right) c_{14}+2\right){}^2} \left\{- 2 \left(c_{13}-1\right){}^2 c_{14}^2 \left(4 c_{14} c_{13}^2+\left(-3 c_{14}^2-4 c_{14}+4\right) c_{13}+4 \left(c_{14}-1\right) c_{14}\right) r^2 A'(r)^2 \right. \nb\\
&&\left.-4 \left(c_{13}-1\right){}^2 c_{14}^2 r A(r) A'(r) \left[\left(4 c_{14} c_{13}^2+\left(3 c_{14}^2-16 c_{14}+4\right) c_{13}-4 \left(c_{14}^2-4 c_{14}+2\right)\right) r A(r)^2 F'(r) \right.\right. \nb\\
&&\left.\left.+4 \left(c_{13}-1\right){}^2 c_{14}\right] -2 \left(c_{13}-1\right){}^2 c_{14}^2 A(r)^2 \left[ \left(4 c_{14} c_{13}^2+\left(-3 c_{14}^2-4 c_{14}+4\right) c_{13}  \right.\right.\right.\nb\\ 
&&\left.\left.\left.+4 \left(c_{14}-1\right) c_{14}\right) r^2 A(r)^4 F'(r)^2 +8 \left(c_{13}-1\right){}^2 c_{14} r A(r)^2 F'(r)+4 \left(c_{13}-1\right) \left(\left(c_{13}-2\right) c_{14}+2\right)\right] \right\},
\eqn
and
\bqn
\lb{n0Special}
n_0&=&\frac{c_{14} \left(-2 c_{13}^2+\left(3 c_{14}+4\right) c_{13}-4 c_{14}\right) A(r)^2 F'(r)^2}{8 \left(\left(3 c_{13}-4\right) c_{14}+2\right) B(r)^3} \nb\\
&&+F'(r)\left[ \left(c_{14} \left(-2 c_{13}^2-3 c_{14} c_{13}+4 c_{13}+4 c_{14}-4\right) r A'(r)-2 \left(c_{14} c_{13}^2+\left(2-6 c_{14}\right) c_{13}+6 c_{14}-4\right) A(r)\right)\right] \nb\\
&&\times\left[4 \left(\left(3 c_{13}-4\right) c_{14}+2\right) r A(r) B(r)^3\right]^{-1}     +  \left[8 \left(\left(3 c_{13}-4\right) c_{14}+2\right) r^2 A(r)^4 B(r)^3\right]^{-1} \nb\\
&&\times\left[c_{14} \left(-2 c_{13}^2+\left(3 c_{14}+4\right) c_{13}-4 c_{14}\right) r^2 A'(r)^2-4 \left(c_{14} c_{13}^2+2 c_{13}-2 c_{14}\right) r A(r) A'(r) \right.\nb\\
&&\left.-8 \left(\left(3 c_{13}-4\right) c_{14}+2\right) A(r)^4 B(r)^2+\left(-2 c_{14} c_{13}^2+\left(8 c_{14}+4\right) c_{13}-8 c_{14}\right) A(r)^2\right]. 
\eqn

\end{widetext}

 {It should be noted that, due to the complexities of the expressions given in Eqs.(\ref{fns}) - (\ref{n0Special}), we  extract these coefficients directly from our Mathematica code.
In addition, they are further tested by the exact  solutions   presented in \cite{Eling2006-1,Per12}, as well as by the numerical solutions presented in
\cite{Eling2006-2,Enrico11}. In the latter, we find that there are no differences between our numerical solutions and the ones presented in \cite{Eling2006-2,Enrico11}, within the errors allowed by the numerical codes.}


\end{document}